\newif\ifconfver
\confverfalse     

\newif\ifplainver  
\plainvertrue

\ifplainver
    \confverfalse   
\fi

\ifconfver
     \documentclass[10pt,twocolumn,twoside,table]{IEEEtran}
\else
    \ifplainver
        \documentclass[11pt]{article}
        \usepackage{fullpage}
    \else
        \documentclass[11pt,draftcls,onecolumn]{IEEEtran}
    \fi
\fi
\usepackage{calc,amsfonts,amssymb,amsmath,bm,url,color,theorem,graphicx,cite,shortcuts_OPT, bbm}
\usepackage{psfrag,subfigure,float,hyperref,microtype}
\usepackage{algorithm}
\usepackage{algorithmic}

\hypersetup{
  colorlinks   = true, 
  urlcolor     = blue, 
  linkcolor    = blue, 
  citecolor    = blue   
}

\definecolor{orange}{RGB}{255,107,0}
\def\blue{\textcolor{blue}}

\newtheorem{Lemma}{Lemma}
\newtheorem{Prop}{Proposition}
\newtheorem{Theorem}{Theorem}
\newtheorem{Assumption}{Assumption}
\newtheorem{Def}{Definition}
\newtheorem{Corollary}{Corollary}

\newtheorem{Observation}{Observation}
\newtheorem{Exa}{Example}

\usepackage{enumitem}
\setlist{leftmargin=4.5mm}

\usepackage[table]{xcolor}
\usepackage{pgfplots}
\usetikzlibrary{arrows,shapes,calc,tikzmark,backgrounds,matrix,decorations.markings}
\usepgfplotslibrary{fillbetween}

\pgfplotsset{compat=1.3}

\usepackage{relsize}
\tikzset{fontscale/.style = {font=\relsize{#1}}
    }

\definecolor{lavander}{cmyk}{0,0.48,0,0}
\definecolor{violet}{cmyk}{0.79,0.88,0,0}
\definecolor{burntorange}{cmyk}{0,0.52,1,0}

\definecolor{asuorange}{rgb}{1,0.699,0.0625}
\definecolor{asured}{rgb}{0.598,0,0.199}
\definecolor{asuborder}{rgb}{0.953,0.484,0}
\definecolor{asugrey}{rgb}{0.309,0.332,0.340}
\definecolor{asublue}{rgb}{0,0.555,0.836}
\definecolor{asugold}{rgb}{1,0.777,0.008}

\tikzset{USA map/.cd,
state/.style={fill, draw=white, ultra thick},
HI/.style={}, AK/.style={}, FL/.style={}, NH/.style={}, MI/.style={}, MI/.style={}, VT/.style={}, ME/.style={}, RI/.style={}, NY/.style={}, PA/.style={}, NJ/.style={}, DE/.style={}, MD/.style={}, VA/.style={}, WV/.style={}, OH/.style={}, IN/.style={}, IL/.style={}, CT/.style={}, WI/.style={}, NC/.style={}, DC/.style={}, MA/.style={}, TN/.style={}, AR/.style={}, MO/.style={}, GA/.style={}, SC/.style={}, KY/.style={}, AL/.style={}, LA/.style={}, MS/.style={}, IA/.style={}, MN/.style={}, OK/.style={}, TX/.style={}, NM/.style={}, KS/.style={}, NE/.style={}, SD/.style={}, ND/.style={}, WY/.style={}, MT/.style={}, CO/.style={}, ID/.style={}, UT/.style={}, AZ/.style={}, NV/.style={}, OR/.style={}, WA/.style={}, CA/.style={}}

\tikzset{
every state/.style={USA map/state/.style={#1}},
HI/.style={USA map/HI/.style={#1}}, AK/.style={USA map/AK/.style={#1}}, FL/.style={USA map/FL/.style={#1}}, NH/.style={USA map/NH/.style={#1}}, MI/.style={USA map/MI/.style={#1}}, VT/.style={USA map/VT/.style={#1}}, ME/.style={USA map/ME/.style={#1}}, RI/.style={USA map/RI/.style={#1}}, NY/.style={USA map/NY/.style={#1}}, PA/.style={USA map/PA/.style={#1}}, NJ/.style={USA map/NJ/.style={#1}}, DE/.style={USA map/DE/.style={#1}}, MD/.style={USA map/MD/.style={#1}}, VA/.style={USA map/VA/.style={#1}}, WV/.style={USA map/WV/.style={#1}}, OH/.style={USA map/OH/.style={#1}}, IN/.style={USA map/IN/.style={#1}}, IL/.style={USA map/IL/.style={#1}}, CT/.style={USA map/CT/.style={#1}}, WI/.style={USA map/WI/.style={#1}}, NC/.style={USA map/NC/.style={#1}}, DC/.style={USA map/DC/.style={#1}}, MA/.style={USA map/MA/.style={#1}}, TN/.style={USA map/TN/.style={#1}}, AR/.style={USA map/AR/.style={#1}}, MO/.style={USA map/MO/.style={#1}}, GA/.style={USA map/GA/.style={#1}}, SC/.style={USA map/SC/.style={#1}}, KY/.style={USA map/KY/.style={#1}}, AL/.style={USA map/AL/.style={#1}}, LA/.style={USA map/LA/.style={#1}}, MS/.style={USA map/MS/.style={#1}}, IA/.style={USA map/IA/.style={#1}}, MN/.style={USA map/MN/.style={#1}}, OK/.style={USA map/OK/.style={#1}}, TX/.style={USA map/TX/.style={#1}}, NM/.style={USA map/NM/.style={#1}}, KS/.style={USA map/KS/.style={#1}}, NE/.style={USA map/NE/.style={#1}}, SD/.style={USA map/SD/.style={#1}}, ND/.style={USA map/ND/.style={#1}}, WY/.style={USA map/WY/.style={#1}}, MT/.style={USA map/MT/.style={#1}}, CO/.style={USA map/CO/.style={#1}}, ID/.style={USA map/ID/.style={#1}}, UT/.style={USA map/UT/.style={#1}}, AZ/.style={USA map/AZ/.style={#1}}, NV/.style={USA map/NV/.style={#1}}, OR/.style={USA map/OR/.style={#1}}, WA/.style={USA map/WA/.style={#1}}, CA/.style={USA map/CA/.style={#1}}
}

\newcommand{\USA}[1][]{
    \begin{scope}[y=0.80pt,x=0.80pt,yscale=-1, inner sep=0pt, outer sep=0pt,
    #1
    ]

    \path[USA map/state, USA map/HI, local bounding box=HI] (233.0875,519.3095) -- (235.0274,515.7529) -- (237.2907,515.4296) --
      (237.6140,516.2379) -- (235.5124,519.3095) -- (233.0875,519.3095) --
      cycle(243.2722,515.5913) -- (249.4153,518.1778) -- (251.5169,517.8545) --
      (253.1335,513.9747) -- (252.4869,510.5798) -- (248.2837,510.0948) --
      (244.2421,511.8731) -- (243.2722,515.5913) -- cycle(273.9878,525.6143) --
      (277.7060,531.1107) -- (280.1309,530.7874) -- (281.2625,530.3024) --
      (282.7175,531.5957) -- (286.4357,531.4341) -- (287.4057,529.9791) --
      (284.4958,528.2009) -- (282.5558,524.4826) -- (280.4542,520.9261) --
      (274.6344,523.8360) -- (273.9878,525.6143) -- cycle(294.1954,534.5056) --
      (295.4887,532.5657) -- (300.1769,533.5357) -- (300.8236,533.0507) --
      (306.9667,533.6973) -- (306.6434,534.9906) -- (304.0568,536.4456) --
      (299.6919,536.1222) -- (294.1954,534.5056) -- cycle(299.5303,539.6788) --
      (301.4702,543.5587) -- (304.5418,542.4270) -- (304.8651,540.8104) --
      (303.2485,538.7088) -- (299.5303,538.3855) -- (299.5303,539.6788) --
      cycle(306.4817,538.5472) -- (308.7450,535.6373) -- (313.4331,538.0622) --
      (317.7980,539.1938) -- (322.1628,541.9421) -- (322.1628,543.8820) --
      (318.6063,545.6603) -- (313.7565,546.6302) -- (311.3315,545.1753) --
      (306.4817,538.5472) -- cycle(323.1328,554.0666) -- (324.7494,552.7734) --
      (328.1443,554.3900) -- (335.7424,557.9465) -- (339.1373,560.0481) --
      (340.7539,562.4730) -- (342.6938,566.8379) -- (346.7353,569.4245) --
      (346.4120,570.7178) -- (342.5321,573.9510) -- (338.3290,575.4059) --
      (336.8740,574.7593) -- (333.8024,576.5375) -- (331.3775,579.7708) --
      (329.1143,582.6807) -- (327.3360,582.5190) -- (323.7794,579.9324) --
      (323.4561,575.4059) -- (324.1028,572.9810) -- (322.4862,567.3229) --
      (320.3846,565.5446) -- (320.2229,562.9580) -- (322.4862,561.9880) --
      (324.5878,558.9165) -- (325.0727,557.9465) -- (323.4561,556.1682) --
      (323.1328,554.0666) -- cycle;

    \path[USA map/state, USA map/AK, local bounding box=AK] (158.0767,453.6750) -- (157.7534,539.0322) -- (159.3700,540.0021) --
      (162.4416,540.1638) -- (163.8965,539.0322) -- (166.4831,539.0322) --
      (166.6447,541.9420) -- (173.5962,548.7318) -- (174.0812,551.3184) --
      (177.4760,549.3785) -- (178.1227,549.2168) -- (178.4460,546.1452) --
      (179.9010,544.5286) -- (181.0326,544.3670) -- (182.9725,542.9120) --
      (186.0441,545.0136) -- (186.6907,547.9235) -- (188.6307,549.0551) --
      (189.7623,551.4801) -- (193.6422,553.2583) -- (197.0371,559.2398) --
      (199.7853,563.1197) -- (202.0486,565.8679) -- (203.5035,569.5861) --
      (208.5150,571.3644) -- (213.6882,573.4660) -- (214.6581,577.8308) --
      (215.1431,580.9024) -- (214.1732,584.2973) -- (212.3949,586.5605) --
      (210.7783,585.7522) -- (209.3233,582.6807) -- (206.5751,581.2257) --
      (204.7968,580.0941) -- (203.9885,580.9024) -- (205.4434,583.6507) --
      (205.6051,587.3689) -- (204.4735,587.8538) -- (202.5335,585.9139) --
      (200.4320,584.6206) -- (200.9169,586.2372) -- (202.2102,588.0155) --
      (201.4019,588.8238) .. controls (201.4019,588.8238) and (200.5936,588.5005) ..
      (200.1086,587.8538) .. controls (199.6236,587.2072) and (198.0070,584.4590) ..
      (198.0070,584.4590) -- (197.0371,582.1957) .. controls (197.0371,582.1957) and
      (196.7137,583.4890) .. (196.0671,583.1657) .. controls (195.4204,582.8423) and
      (194.7738,581.7107) .. (194.7738,581.7107) -- (196.5521,579.7708) --
      (195.0971,578.3158) -- (195.0971,573.3043) -- (194.2888,573.3043) --
      (193.4805,576.6992) -- (192.3489,577.1842) -- (191.3789,573.4660) --
      (190.7323,569.7478) -- (189.9240,569.2628) -- (190.2473,574.9209) --
      (190.2473,576.0526) -- (188.7923,574.7593) -- (185.2358,568.7778) --
      (183.1342,568.2928) -- (182.4876,564.5746) -- (180.8709,561.6647) --
      (179.2543,560.5331) -- (179.2543,558.2698) -- (181.3559,556.9765) --
      (180.8709,556.6532) -- (178.2844,557.2999) -- (174.8895,554.8750) --
      (172.3029,551.9650) -- (167.4531,549.3785) -- (163.4115,546.7919) --
      (164.7048,543.5587) -- (164.7048,541.9421) -- (162.9265,543.5587) --
      (160.0166,544.6903) -- (156.2984,543.5587) -- (150.6403,541.1338) --
      (145.1438,541.1338) -- (144.4972,541.6187) -- (138.0307,537.7389) --
      (135.9291,537.4155) -- (133.1809,531.5957) -- (129.6243,531.9191) --
      (126.0678,533.3740) -- (126.5528,537.9005) -- (127.6844,534.9906) --
      (128.6544,535.3139) -- (127.1994,539.6788) -- (130.4326,536.9306) --
      (131.0793,538.5472) -- (127.1994,542.9120) -- (125.9061,542.5887) --
      (125.4211,540.6488) -- (124.1279,539.8405) -- (122.8346,540.9721) --
      (120.0863,539.1938) -- (117.0148,541.2954) -- (115.2365,543.3970) --
      (111.8416,545.4986) -- (107.1534,545.3369) -- (106.6684,543.2353) --
      (110.3866,542.5887) -- (110.3866,541.2954) -- (108.1234,540.6488) --
      (109.0934,538.2238) -- (111.3566,534.3440) -- (111.3566,532.5657) --
      (111.5183,531.7574) -- (115.8831,529.4941) -- (116.8531,530.7874) --
      (119.6013,530.7874) -- (118.3081,528.2009) -- (114.5898,527.8775) --
      (109.5783,530.6258) -- (107.1534,534.0206) -- (105.3752,536.6072) --
      (104.2435,538.8705) -- (100.0403,540.3254) -- (96.9688,542.9120) --
      (96.6454,544.5286) -- (98.9087,545.4986) -- (99.7170,547.6002) --
      (96.9688,550.8334) -- (90.5023,555.0366) -- (82.7426,559.2398) --
      (80.6410,560.3714) -- (75.3062,561.5031) -- (69.9713,563.7663) --
      (71.7496,565.0596) -- (70.2947,566.5146) -- (69.8097,567.6462) --
      (67.0614,566.6762) -- (63.8282,566.8379) -- (63.0199,569.1011) --
      (62.0499,569.1011) -- (62.3733,566.6762) -- (58.8167,567.9695) --
      (55.9068,568.9395) -- (52.5119,567.6462) -- (49.6020,569.5861) --
      (46.3688,569.5861) -- (44.2672,570.8794) -- (42.6506,571.6877) --
      (40.5490,571.3644) -- (37.9624,570.2328) -- (35.6992,570.8794) --
      (34.7292,571.8494) -- (33.1126,570.7178) -- (33.1126,568.7778) --
      (36.1841,567.4845) -- (42.4889,568.1312) -- (46.8538,566.5146) --
      (48.9554,564.4130) -- (51.8653,563.7663) -- (53.6436,562.9580) --
      (56.3918,563.1197) -- (58.0084,564.4130) -- (58.9784,564.0896) --
      (61.2416,561.3414) -- (64.3132,560.3714) -- (67.7081,559.7248) --
      (69.0014,559.4015) -- (69.6480,559.8864) -- (70.4563,559.8864) --
      (71.7496,556.1682) -- (75.7911,554.7133) -- (77.7311,550.9951) --
      (79.9943,546.4686) -- (81.6110,545.0136) -- (81.9343,542.4270) --
      (80.3177,543.7203) -- (76.9228,544.3670) -- (76.2761,541.9421) --
      (74.9828,541.6187) -- (74.0129,542.5887) -- (73.8512,545.4986) --
      (72.3963,545.3369) -- (70.9413,539.5171) -- (69.6480,540.8104) --
      (68.5164,540.3254) -- (68.1931,538.3855) -- (64.1515,538.5472) --
      (62.0499,539.6788) -- (59.4634,539.3555) -- (60.9183,537.9005) --
      (61.4033,535.3139) -- (60.7566,533.3740) -- (62.2116,532.4040) --
      (63.5049,532.2424) -- (62.8582,530.4641) -- (62.8582,526.0993) --
      (61.8883,525.1293) -- (61.0800,526.5842) -- (54.9368,526.5842) --
      (53.4819,525.2909) -- (52.8352,521.4111) -- (50.7337,517.8545) --
      (50.7337,516.8846) -- (52.8352,516.0763) -- (52.9969,513.9747) --
      (54.1285,512.8430) -- (53.3202,512.3581) -- (52.0269,512.8430) --
      (50.8953,510.0948) -- (51.8653,505.0833) -- (56.3918,501.8501) --
      (58.9784,500.2335) -- (60.9183,496.5153) -- (63.6666,495.2220) --
      (66.2531,496.3536) -- (66.5765,498.7785) -- (69.0014,498.4552) --
      (72.2346,496.0303) -- (73.8512,496.6769) -- (74.8212,497.3236) --
      (76.4378,497.3236) -- (78.7010,496.0303) -- (79.5094,491.6654) .. controls
      (79.5094,491.6654) and (79.8327,488.7555) .. (80.4793,488.2705) .. controls
      (81.1260,487.7855) and (81.4493,487.3006) .. (81.4493,487.3006) --
      (80.3177,485.3606) -- (77.7311,486.1689) -- (74.4978,486.9772) --
      (72.5579,486.4923) -- (69.0014,484.7140) -- (63.9899,484.5523) --
      (60.4333,480.8341) -- (60.9183,476.9542) -- (61.5650,474.5293) --
      (59.4634,472.7511) -- (57.5234,469.0328) -- (58.0084,468.2245) --
      (64.7982,467.7396) -- (66.8998,467.7396) -- (67.8697,468.7095) --
      (68.5164,468.7095) -- (68.3547,467.0929) -- (72.2346,466.4463) --
      (74.8212,466.7696) -- (76.2761,467.9012) -- (74.8212,470.0028) --
      (74.3362,471.4578) -- (77.0844,473.0744) -- (82.0959,474.8526) --
      (83.8742,473.8827) -- (81.6110,469.5178) -- (80.6410,466.2846) --
      (81.6110,465.4763) -- (78.2161,463.5364) -- (77.7311,462.4047) --
      (78.2161,460.7881) -- (77.4078,456.9083) -- (74.4978,452.2201) --
      (72.0729,448.0169) -- (74.9828,446.0769) -- (78.2161,446.0769) --
      (79.9943,446.7236) -- (84.1975,446.5619) -- (87.9157,443.0054) --
      (89.0474,439.9338) -- (92.7656,437.5089) -- (94.3822,438.4789) --
      (97.1304,437.8322) -- (100.8486,435.7306) -- (101.9803,435.5690) --
      (102.9502,436.3773) -- (107.4767,436.2156) -- (110.2250,433.1441) --
      (111.3566,433.1441) -- (114.9132,435.5690) -- (116.8531,437.6706) --
      (116.3681,438.8022) -- (117.0148,439.9338) -- (118.6314,438.3172) --
      (122.5112,438.6405) -- (122.8346,442.3587) -- (124.7745,443.8137) --
      (131.8876,444.4603) -- (138.1924,448.6635) -- (139.6473,447.6936) --
      (144.8205,450.2801) -- (146.9221,449.6335) -- (148.8620,448.8252) --
      (153.7119,450.7651) -- (158.0767,453.6750) -- cycle(42.9739,482.6124) --
      (45.0755,487.9472) -- (44.9138,488.9172) -- (42.0039,488.5938) --
      (40.2257,484.5523) -- (38.4474,483.0974) -- (36.0225,483.0974) --
      (35.8608,480.5108) -- (37.6391,478.0859) -- (38.7707,480.5108) --
      (40.2257,481.9657) -- (42.9739,482.6124) -- cycle(40.3873,516.0763) --
      (44.1055,516.8846) -- (47.8237,517.8545) -- (48.6321,518.8245) --
      (47.0154,522.5427) -- (43.9439,522.3810) -- (40.5490,518.8245) --
      (40.3873,516.0763) -- cycle(19.6947,502.0117) -- (20.8263,504.5983) --
      (21.9580,506.2149) -- (20.8263,507.0232) -- (18.7247,503.9517) --
      (18.7247,502.0117) -- (19.6947,502.0117) -- cycle(5.9535,575.0826) --
      (9.3484,572.8193) -- (12.7433,571.8494) -- (15.3298,572.1727) --
      (15.8148,573.7893) -- (17.7548,574.2743) -- (19.6947,572.3344) --
      (19.3714,570.7178) -- (22.1196,570.0711) -- (25.0295,572.6577) --
      (23.8979,574.4360) -- (19.5330,575.5676) -- (16.7848,575.0826) --
      (13.0666,573.9510) -- (8.7017,575.4059) -- (7.0851,575.7292) --
      (5.9535,575.0826) -- cycle(54.9368,570.5561) -- (56.5535,572.4960) --
      (58.6550,570.8794) -- (57.2001,569.5861) -- (54.9368,570.5561) --
      cycle(57.8467,573.6276) -- (58.9784,571.3644) -- (61.0800,571.6877) --
      (60.2717,573.6276) -- (57.8467,573.6276) -- cycle(81.4493,571.6877) --
      (82.9042,573.4660) -- (83.8742,572.3344) -- (83.0659,570.3944) --
      (81.4493,571.6877) -- cycle(90.1790,559.2398) -- (91.3106,565.0596) --
      (94.2205,565.8679) -- (99.2320,562.9580) -- (103.5969,560.3714) --
      (101.9803,557.9465) -- (102.4652,555.5216) -- (100.3636,556.8149) --
      (97.4538,556.0066) -- (99.0704,554.8750) -- (101.0103,555.6833) --
      (104.8902,553.9050) -- (105.3751,552.4500) -- (102.9502,551.6417) --
      (103.7585,549.7018) -- (101.0103,551.6417) -- (96.3221,555.1983) --
      (91.4723,558.1082) -- (90.1790,559.2398) -- cycle(132.5342,539.3555) --
      (134.9592,537.9005) -- (133.9892,536.1222) -- (132.2109,537.0922) --
      (132.5342,539.3555) -- cycle;

    \path[USA map/state, USA map/FL, local bounding box=FL] (759.8167,439.1428) -- (762.0824,446.4614) -- (765.8121,456.2037) --
      (771.1468,465.5800) -- (774.8650,471.8847) -- (779.7149,477.3812) --
      (783.7564,481.0994) -- (785.3730,484.0093) -- (784.2414,485.3025) --
      (783.4330,486.5958) -- (786.3429,494.0322) -- (789.2528,496.9421) --
      (791.8394,502.2769) -- (795.3959,508.0967) -- (799.9224,516.3413) --
      (801.2157,523.9394) -- (801.7007,535.9023) -- (802.3473,537.6805) --
      (802.0240,541.0754) -- (799.5991,542.3687) -- (799.9224,544.3086) --
      (799.2758,546.2485) -- (799.5991,548.6734) -- (800.0841,550.6134) --
      (797.3358,553.8466) -- (794.2643,555.3015) -- (790.3844,555.4632) --
      (788.9295,557.0798) -- (786.5046,558.0497) -- (785.2113,557.5648) --
      (784.0797,556.5948) -- (783.7564,553.6849) -- (782.9481,550.2900) --
      (779.5532,545.1169) -- (775.9967,542.8537) -- (772.1168,542.5303) --
      (771.3085,543.8236) -- (768.2370,539.4588) -- (767.5903,535.9023) --
      (765.0037,531.8608) -- (763.2255,530.7291) -- (761.6089,532.8307) --
      (759.8306,532.5074) -- (757.7290,527.4959) -- (754.8191,523.6161) --
      (751.9092,518.2813) -- (749.3227,515.2097) -- (745.7662,511.4915) --
      (747.8677,509.0666) -- (751.1009,503.5702) -- (750.9393,501.9536) --
      (746.4128,500.9836) -- (744.7962,501.6302) -- (745.1195,502.2769) --
      (747.7061,503.2468) -- (746.2511,507.7733) -- (745.4428,508.2583) --
      (743.6646,504.2168) -- (742.3713,499.3670) -- (742.0480,496.6188) --
      (743.5029,491.9306) -- (743.5029,482.3927) -- (740.4314,478.6745) --
      (739.1381,475.6029) -- (733.9649,474.3096) -- (732.0250,473.6630) --
      (730.4084,471.0764) -- (727.0135,469.4598) -- (725.8819,466.0649) --
      (723.1337,465.0950) -- (720.7088,461.3768) -- (716.5056,459.9219) --
      (713.5957,458.4669) -- (711.0092,458.4669) -- (706.9676,459.2752) --
      (706.8060,461.2151) -- (707.6143,462.1851) -- (707.1293,463.3167) --
      (704.0578,463.1551) -- (700.3396,466.7116) -- (696.7830,468.6515) --
      (692.9032,468.6515) -- (689.6700,469.9448) -- (689.3466,467.1966) --
      (687.7300,465.2566) -- (684.8202,464.1250) -- (683.2036,462.6701) --
      (675.1205,458.7902) -- (667.5225,457.0120) -- (663.1577,457.6586) --
      (657.1762,458.1436) -- (651.1948,460.2452) -- (647.7155,460.8581) --
      (647.4776,452.8084) -- (644.8910,450.8685) -- (643.1128,449.0902) --
      (643.4361,446.0186) -- (653.6207,444.7254) -- (679.1631,441.8155) --
      (685.9529,441.1688) -- (691.3889,441.4491) -- (693.9754,445.3290) --
      (695.4304,446.7839) -- (703.5285,447.2991) -- (714.3483,446.6525) --
      (735.8607,445.3592) -- (741.3064,444.6848) -- (746.4140,444.8893) --
      (746.8408,447.7992) -- (749.0738,448.6075) -- (749.3087,443.9775) --
      (747.7805,439.8046) -- (749.0889,438.3647) -- (754.6436,438.8195) --
      (759.8167,439.1428) -- cycle(772.3621,571.5479) -- (774.7870,570.9012) --
      (776.0803,570.6588) -- (777.5353,568.3147) -- (779.8793,566.6980) --
      (781.1726,567.1830) -- (782.8701,567.5064) -- (783.2742,568.5571) --
      (779.7985,569.7696) -- (775.5953,571.2246) -- (773.2512,572.4370) --
      (772.3621,571.5479) -- cycle(785.8608,566.5364) -- (787.0733,567.5872) --
      (789.8215,565.4856) -- (795.1563,561.2824) -- (798.8745,557.4025) --
      (801.3803,550.7744) -- (802.3502,549.0770) -- (802.5119,545.6821) --
      (801.7844,546.1671) -- (800.8145,548.9962) -- (799.3595,553.6035) --
      (796.1263,558.8575) -- (791.7614,563.0607) -- (788.3666,565.0006) --
      (785.8608,566.5364) -- cycle;

    \path[USA map/state, USA map/NH, local bounding box=NH] (880.7990,142.4248) -- (881.6680,141.3483) -- (882.7582,138.0572) --
      (880.2152,137.1438) -- (879.7302,134.0722) -- (875.8503,132.9406) --
      (875.5270,130.1923) -- (868.2523,106.7515) -- (863.6508,92.2085) --
      (862.7538,92.2034) -- (862.1071,93.8200) -- (861.4605,93.3351) --
      (860.4905,92.3651) -- (859.0356,94.3050) -- (858.9871,99.3371) --
      (859.2987,105.0043) -- (861.2386,107.7525) -- (861.2386,111.7941) --
      (857.5204,116.8568) -- (854.9339,117.9885) -- (854.9339,119.1201) --
      (856.0655,120.8984) -- (856.0655,129.4664) -- (855.2572,138.6811) --
      (855.0955,143.5309) -- (856.0655,144.8242) -- (855.9038,149.3507) --
      (855.4188,151.1289) -- (856.3876,151.8382) -- (873.1753,147.4136) --
      (875.3502,146.8112) -- (877.1938,144.0378) -- (880.7990,142.4247) -- cycle;

    \begin{scope}
    \path[USA map/state, USA map/MI, local bounding box=MI] (697.8601,177.2369) -- (694.6269,168.9922)
        -- (692.3636,159.9392) -- (689.9387,156.7060) -- (687.3521,154.9277) --
        (685.7355,156.0594) -- (681.8557,157.8376) -- (679.9158,162.8491) --
        (677.1675,166.5673) -- (676.0359,167.2139) -- (674.5810,166.5673) .. controls
        (674.5810,166.5673) and (671.9944,165.1123) .. (672.1561,164.4657) .. controls
        (672.3177,163.8191) and (672.6410,159.4542) .. (672.6410,159.4542) --
        (676.0359,158.1609) -- (676.8442,154.7661) -- (677.4908,152.1795) --
        (679.9158,150.5629) -- (679.5924,140.5400) -- (677.9758,138.2767) --
        (676.6825,137.4684) -- (675.8742,135.3668) -- (676.6825,134.5585) --
        (678.2991,134.8818) -- (678.4608,133.2652) -- (676.0359,131.0020) --
        (674.7426,128.4154) -- (672.1561,128.4154) -- (667.6296,126.9605) --
        (662.1331,123.5656) -- (659.3849,123.5656) -- (658.7382,124.2123) --
        (657.7683,123.7273) -- (654.6967,121.4640) -- (651.7868,123.2423) --
        (648.8769,125.5055) -- (649.2003,129.0621) -- (650.1702,129.3854) --
        (652.2718,129.8704) -- (652.7568,130.6787) -- (650.1702,131.4870) --
        (647.5837,131.8103) -- (646.1287,133.5886) -- (645.8054,135.6901) --
        (646.1287,137.3067) -- (646.4520,142.8032) -- (642.8955,144.9048) --
        (642.2489,144.7431) -- (642.2489,140.5400) -- (643.5421,138.1151) --
        (644.1888,135.6901) -- (643.3805,134.8818) -- (641.4406,135.6901) --
        (640.4706,139.8933) -- (637.7224,141.0249) -- (635.9441,142.9649) --
        (635.7824,143.9348) -- (636.4291,144.7431) -- (635.7824,147.3297) --
        (633.5192,147.8147) -- (633.5192,148.9463) -- (634.3275,151.3712) --
        (633.1959,157.5143) -- (631.5793,161.5558) -- (632.2259,166.2440) --
        (632.7109,167.3756) -- (631.9026,169.8005) -- (631.5793,170.6088) --
        (631.2560,173.3570) -- (634.8125,179.3385) -- (637.7224,185.8049) --
        (639.1773,190.6547) -- (638.3690,195.3429) -- (637.3991,201.3243) --
        (634.9741,206.4974) -- (634.6508,209.2457) -- (631.3920,212.3308) --
        (635.8006,212.1688) -- (657.2191,209.9055) -- (664.4969,208.9184) --
        (664.5933,210.5848) -- (671.4452,209.3723) -- (681.7433,207.8692) --
        (685.5975,207.4083) -- (685.7356,206.8207) -- (685.8972,205.3658) --
        (687.9988,201.6476) -- (689.9994,199.9098) -- (689.7771,194.8579) --
        (691.3741,193.2609) -- (692.4647,192.9179) -- (692.6870,189.3614) --
        (694.2227,186.3303) -- (695.2735,186.9365) -- (695.4352,187.5832) --
        (696.2435,187.7448) -- (698.1834,186.7749) -- (697.8601,177.2369) -- cycle;

      \path[USA map/state, USA map/MI, local bounding box=MI2] (581.6193,82.0590) -- (583.4483,80.0014) --
        (585.6202,79.2012) -- (590.9929,75.3146) -- (593.2791,74.7431) --
        (593.7363,75.2003) -- (588.5923,80.3443) -- (585.2773,82.2876) --
        (583.2197,83.2021) -- (581.6193,82.0590) -- cycle(667.7937,114.1872) --
        (668.4403,116.6929) -- (671.6736,116.8546) -- (672.9668,115.6421) .. controls
        (672.9668,115.6421) and (672.8860,114.1872) .. (672.5627,114.0255) .. controls
        (672.2394,113.8639) and (670.9461,112.1664) .. (670.9461,112.1664) --
        (668.7637,112.4089) -- (667.1470,112.5706) -- (666.8237,113.7022) --
        (667.7937,114.1872) -- cycle(567.4921,111.2132) -- (568.2084,110.6328) --
        (570.9566,109.8245) -- (574.5131,107.5612) -- (574.5131,106.5913) --
        (575.1598,105.9446) -- (581.1412,104.9747) -- (583.5661,103.0347) --
        (587.9310,100.9331) -- (588.0926,99.6399) -- (590.0325,96.7300) --
        (591.8108,95.9217) -- (593.1041,94.1434) -- (595.3673,91.8802) --
        (599.7322,89.4553) -- (604.4203,88.9703) -- (605.5519,90.1019) --
        (605.2286,91.0719) -- (601.5104,92.0418) -- (600.0555,95.1134) --
        (597.7922,95.9217) -- (597.3073,98.3466) -- (594.8824,101.5798) --
        (594.5590,104.1664) -- (595.3673,104.6513) -- (596.3373,103.5197) --
        (599.8938,100.6098) -- (601.1871,101.9031) -- (603.4504,101.9031) --
        (606.6836,102.8731) -- (608.1385,104.0047) -- (609.5934,107.0762) --
        (612.3417,109.8245) -- (616.2215,109.6628) -- (617.6765,108.6928) --
        (619.2931,109.9861) -- (620.9097,110.4711) -- (622.2030,109.6628) --
        (623.3346,109.6628) -- (624.9512,108.6928) -- (628.9927,105.1363) --
        (632.3876,104.0047) -- (639.0157,103.6814) -- (643.5421,101.7414) --
        (646.1287,100.4482) -- (647.5837,100.6098) -- (647.5837,106.2679) --
        (648.0687,106.5913) -- (650.9785,107.3996) -- (652.9185,106.9146) --
        (659.0616,105.2980) -- (660.1932,104.1664) -- (661.6481,104.6513) --
        (661.6481,111.6027) -- (664.8813,114.6743) -- (666.1746,115.3209) --
        (667.4679,116.2909) -- (666.1746,116.6142) -- (665.3663,116.2909) --
        (661.6481,115.8059) -- (659.5465,116.4526) -- (657.2833,116.2909) --
        (654.0501,117.7458) -- (652.2718,117.7458) -- (646.4520,116.4526) --
        (641.2789,116.6142) -- (639.3390,119.2008) -- (632.3876,119.8474) --
        (629.9627,120.6557) -- (628.8311,123.7273) -- (627.5378,124.8589) --
        (627.0528,124.6972) -- (625.5978,123.0806) -- (621.0714,125.5055) --
        (620.4247,125.5055) -- (619.2931,123.8889) -- (618.4848,124.0506) --
        (616.5449,128.4154) -- (615.5749,132.4569) -- (612.3938,139.4577) --
        (611.2170,138.4235) -- (609.8453,137.3922) -- (607.9045,127.1041) --
        (604.3600,125.7341) -- (602.3074,123.4479) -- (590.1871,120.7044) --
        (587.3318,119.6747) -- (579.1014,117.5020) -- (571.2114,116.3589) --
        (567.4921,111.2132) -- cycle;

    \end{scope}
    \path[USA map/state, USA map/VT, local bounding box=VT] (844.4842,154.0579) -- (844.8009,148.7123) -- (841.9101,137.9281) --
      (841.2635,137.6048) -- (838.3536,136.3115) -- (839.1619,133.4016) --
      (838.3536,131.3000) -- (835.6536,126.6600) -- (836.6235,122.7802) --
      (835.8152,117.6070) -- (833.3903,111.1406) -- (832.5847,106.2181) --
      (859.0041,99.4863) -- (859.3128,105.0085) -- (861.2291,107.7507) --
      (861.2291,111.7922) -- (857.5219,116.8502) -- (854.9353,117.9929) --
      (854.9243,119.1135) -- (856.2343,120.6326) -- (855.9234,128.7305) --
      (855.3139,137.9894) -- (855.0860,143.5463) -- (856.0560,144.8396) --
      (855.8943,149.4103) -- (855.4093,151.1002) -- (856.4235,151.8274) --
      (848.9860,153.3341) -- (844.4842,154.0579) -- cycle;

    \path[USA map/state, USA map/ME, local bounding box=ME] (922.8398,78.8307) -- (924.7797,80.9323) -- (927.0429,84.6505) --
      (927.0429,86.5904) -- (924.9413,91.2786) -- (923.0014,91.9252) --
      (919.6065,94.9968) -- (914.7567,100.4932) .. controls (914.7567,100.4932) and
      (914.1101,100.4932) .. (913.4635,100.4932) .. controls (912.8168,100.4932) and
      (912.4935,98.3916) .. (912.4935,98.3916) -- (910.7152,98.5533) --
      (909.7453,100.0082) -- (907.3204,101.4632) -- (906.3504,102.9181) --
      (907.9670,104.3731) -- (907.4820,105.0197) -- (906.9970,107.7679) --
      (905.0571,107.6063) -- (905.0571,105.9897) -- (904.7338,104.6964) --
      (903.2789,105.0197) -- (901.5006,101.7865) -- (899.3990,103.0798) --
      (900.6923,104.5347) -- (901.0156,105.6664) -- (900.2073,106.9596) --
      (900.5306,110.0312) -- (900.6923,111.6478) -- (899.0757,114.2344) --
      (896.1658,114.7193) -- (895.8425,117.6292) -- (890.5077,120.7008) --
      (889.2144,121.1858) -- (887.5978,119.7308) -- (884.5262,123.2873) --
      (885.4962,126.5206) -- (884.0412,127.8138) -- (883.8796,132.1787) --
      (882.7563,138.4380) -- (880.2941,137.2821) -- (879.8091,134.2105) --
      (875.9292,133.0789) -- (875.6059,130.3306) -- (868.3311,106.8898) --
      (863.6326,92.2501) -- (865.0531,92.1319) -- (866.5669,92.5418) --
      (866.5669,89.9553) -- (867.8752,85.4588) -- (870.4618,80.7706) --
      (871.9167,76.7291) -- (869.9768,74.3042) -- (869.9768,68.3228) --
      (870.7851,67.3528) -- (871.5934,64.6046) -- (871.4317,63.1497) --
      (871.2701,58.2998) -- (873.0483,53.4500) -- (875.9582,44.5587) --
      (878.0598,40.3555) -- (879.3531,40.3555) -- (880.6464,40.5172) --
      (880.6464,41.6488) -- (881.9397,43.9121) -- (884.6879,44.5587) --
      (885.4962,43.7504) -- (885.4962,42.7804) -- (889.5377,39.8705) --
      (891.3160,38.0923) -- (892.7709,38.2539) -- (898.7523,40.6788) --
      (900.6923,41.6488) -- (909.7453,71.5560) -- (915.7267,71.5560) --
      (916.5350,73.4959) -- (916.6967,78.3457) -- (919.6066,80.6090) --
      (920.4149,80.6090) -- (920.5765,80.1240) -- (920.0915,78.9924) --
      (922.8398,78.8307) -- cycle(901.9080,108.9783) -- (903.4438,107.4425) --
      (904.8179,108.4933) -- (905.3837,110.9182) -- (903.6863,111.8073) --
      (901.9080,108.9782) -- cycle(908.6169,103.0776) -- (910.3952,104.9367) ..
      controls (910.3952,104.9367) and (911.6885,105.0175) .. (911.6885,104.6942) ..
      controls (911.6885,104.3709) and (911.9310,102.6735) .. (911.9310,102.6735) --
      (912.8201,101.8652) -- (912.0118,100.0869) -- (909.9911,100.8144) --
      (908.6169,103.0776) -- cycle;

    \path[USA map/state, USA map/RI, local bounding box=RI] (874.0700,178.8954) -- (870.3742,163.9394) -- (876.6435,162.0942) --
      (878.8346,164.0214) -- (882.1411,168.3420) -- (884.8290,172.7441) --
      (881.8297,174.3689) -- (880.5364,174.2072) -- (879.4048,175.9855) --
      (876.9799,177.9254) -- (874.0700,178.8954) -- cycle;

    \path[USA map/state, USA map/NY, local bounding box=NY] (830.3794,188.7456) -- (829.2478,187.7756) -- (826.6612,187.6140) --
      (824.3980,185.6741) -- (822.7674,179.5449) -- (819.3089,179.6354) --
      (816.8652,176.9272) -- (797.4799,181.3092) -- (754.4781,190.0389) --
      (746.9485,191.2669) -- (746.2103,184.7985) -- (747.6384,183.6731) --
      (748.9317,182.5415) -- (749.9017,180.9249) -- (751.6799,179.7933) --
      (753.6198,178.0150) -- (754.1048,176.3984) -- (756.2064,173.6502) --
      (757.3380,172.6802) -- (757.1764,171.7103) -- (755.8831,168.6387) --
      (754.1048,168.4770) -- (752.1649,162.3339) -- (755.0748,160.5557) --
      (759.4396,159.1007) -- (763.4811,157.8074) -- (766.7143,157.3225) --
      (773.0191,157.1608) -- (774.9590,158.4541) -- (776.5756,158.6158) --
      (778.6772,157.3225) -- (781.2638,156.1908) -- (786.4369,155.7059) --
      (788.5385,153.9276) -- (790.3168,150.6944) -- (791.9334,148.7545) --
      (794.0350,148.7545) -- (795.9749,147.6228) -- (796.1365,145.3596) --
      (794.6816,143.2580) -- (794.3583,141.8031) -- (795.4899,139.7015) --
      (795.4899,138.2465) -- (793.7116,138.2465) -- (791.9334,137.4382) --
      (791.1251,136.3066) -- (790.9634,133.7200) -- (796.7832,128.2236) --
      (797.4298,127.4153) -- (798.8848,124.5054) -- (801.7947,119.9789) --
      (804.5429,116.2607) -- (806.6445,113.8358) -- (809.0596,112.0102) --
      (812.1409,110.7643) -- (817.6374,109.4710) -- (820.8706,109.6326) --
      (825.3971,108.1777) -- (832.9623,106.1065) -- (833.4821,111.0862) --
      (835.9070,117.5526) -- (836.7153,122.7258) -- (835.7453,126.6056) --
      (838.3319,131.1321) -- (839.1402,133.2337) -- (838.3319,136.1436) --
      (841.2418,137.4369) -- (841.8884,137.7602) -- (844.9600,148.7532) --
      (844.4237,153.8128) -- (843.9387,164.6441) -- (844.7470,170.1406) --
      (845.5553,173.6971) -- (847.0103,180.9719) -- (847.0103,189.0549) --
      (845.8787,191.3182) -- (847.7180,193.3109) -- (848.5145,194.9894) --
      (846.5746,196.7676) -- (846.8979,198.0609) -- (848.1912,197.7376) --
      (849.6462,196.4443) -- (851.9094,193.8577) -- (853.0410,193.2111) --
      (854.6576,193.8577) -- (856.9209,194.0194) -- (864.8422,190.1396) --
      (867.7521,187.3913) -- (869.0454,185.9364) -- (873.2486,187.5530) --
      (869.8537,191.1095) -- (865.9739,194.0194) -- (858.8608,199.3542) --
      (856.2742,200.3242) -- (850.4545,202.2641) -- (846.4130,203.3957) --
      (845.2382,202.8628) -- (844.9942,199.1743) -- (845.4792,196.4260) --
      (845.3175,194.3244) -- (842.5040,192.6254) -- (837.9775,191.6555) --
      (834.0976,190.5238) -- (830.3794,188.7456) -- cycle;

    \path[USA map/state, USA map/PA, local bounding box=PA] (825.1237,224.6920) -- (826.4321,224.4211) -- (828.7616,223.1678) --
      (829.9735,220.6847) -- (831.5901,218.4215) -- (834.8233,215.3499) --
      (834.8233,214.5416) -- (832.3984,212.9250) -- (828.8419,210.5001) --
      (827.8719,207.9135) -- (825.1237,207.5902) -- (824.9620,206.4586) --
      (824.1537,203.7103) -- (826.4170,202.5787) -- (826.5787,200.1538) --
      (825.2854,198.8605) -- (825.4470,197.2439) -- (827.3870,194.1724) --
      (827.3870,191.1008) -- (830.0846,188.4549) -- (829.1643,187.7799) --
      (826.6402,187.5870) -- (824.3457,185.6471) -- (822.7958,179.5310) --
      (819.2912,179.6316) -- (816.8360,176.9282) -- (798.7450,181.1260) --
      (755.7432,189.8557) -- (746.8519,191.3106) -- (746.2312,184.7892) --
      (740.8687,189.8569) -- (739.5754,190.3419) -- (735.3731,193.3508) --
      (738.2839,212.4882) -- (740.7655,222.2176) -- (744.3373,241.4791) --
      (747.6066,240.8414) -- (759.5502,239.3389) -- (797.4768,231.6737) --
      (812.3531,228.8504) -- (820.6534,227.2280) -- (820.9205,226.9895) --
      (823.0221,225.3729) -- (825.1237,224.6920) -- cycle;

    \path[USA map/state, USA map/NJ, local bounding box=NJ] (829.6794,188.4602) -- (827.3569,191.1944) -- (827.3569,194.2660) --
      (825.4169,197.3375) -- (825.2553,198.9542) -- (826.5486,200.2474) --
      (826.3869,202.6724) -- (824.1237,203.8040) -- (824.9320,206.5522) --
      (825.0936,207.6838) -- (827.8419,208.0072) -- (828.8118,210.5937) --
      (832.3684,213.0187) -- (834.7933,214.6353) -- (834.7933,215.4436) --
      (831.8101,218.1401) -- (830.1934,220.4034) -- (828.7385,223.1516) --
      (826.4752,224.4449) -- (826.0128,226.0474) -- (825.7703,227.2598) --
      (825.1611,229.8666) -- (826.2533,232.1108) -- (829.4865,235.0206) --
      (834.3364,237.2839) -- (838.3779,237.9305) -- (838.5395,239.3855) --
      (837.7312,240.3554) -- (838.0545,243.1037) -- (838.8628,243.1037) --
      (840.9644,240.6788) -- (841.7727,235.8289) -- (844.5210,231.7874) --
      (847.5925,225.3210) -- (848.7241,219.8246) -- (848.0775,218.6929) --
      (847.9158,209.3166) -- (846.2992,205.9218) -- (845.1676,206.7301) --
      (842.4194,207.0534) -- (841.9344,206.5684) -- (843.0660,205.5984) --
      (845.1676,203.6585) -- (845.2307,202.5647) -- (844.8463,199.1308) --
      (845.4197,196.3826) -- (845.3022,194.4136) -- (842.4947,192.6632) --
      (837.4025,191.4875) -- (833.2651,190.1059) -- (829.6795,188.4602) -- cycle;

    \path[USA map/state, USA map/DE, local bounding box=DE] (825.6261,228.2791) -- (825.9944,226.1322) -- (826.3695,224.4412) --
      (824.7465,224.8389) -- (823.1310,225.3065) -- (820.9248,227.0708) --
      (822.6449,232.1137) -- (824.9081,237.7718) -- (827.0097,247.4714) --
      (828.6263,253.7762) -- (833.6378,253.6145) -- (839.7799,252.4339) --
      (837.5157,245.0476) -- (836.5457,245.5326) -- (832.9892,243.1077) --
      (831.2109,238.4195) -- (829.2710,234.8630) -- (826.1239,231.9927) --
      (825.2597,229.8946) -- (825.6261,228.2791) -- cycle;

    \path[USA map/state, USA map/MD, local bounding box=MD] (839.7917,252.4148) -- (833.7832,253.6186) -- (828.6403,253.7361) --
      (826.7967,246.8137) -- (824.8719,237.6444) -- (822.2993,231.4560) --
      (821.0109,227.0576) -- (813.5049,228.6800) -- (798.6287,231.5033) --
      (761.1773,239.0542) -- (762.3086,244.0659) -- (763.2785,249.7240) --
      (763.6018,249.4007) -- (765.7034,246.9758) -- (767.9667,244.3581) --
      (770.3916,243.7425) -- (771.8466,242.2876) -- (773.6248,239.7010) --
      (774.9181,240.3477) -- (777.8280,240.0243) -- (780.4146,237.9228) --
      (782.4215,236.4695) -- (784.2667,235.9845) -- (785.9110,237.1145) --
      (788.8209,238.5694) -- (790.7609,240.3477) -- (791.9733,241.8835) --
      (796.0957,243.5809) -- (796.0957,246.4908) -- (801.5921,247.7841) --
      (802.7366,248.3260) -- (804.1485,246.2977) -- (807.0304,248.2679) --
      (805.7523,250.7498) -- (804.9870,254.7355) -- (803.2087,257.3220) --
      (803.2087,259.4236) -- (803.8554,261.2019) -- (808.9193,262.5576) --
      (813.2304,262.4959) -- (816.3020,263.4659) -- (818.4035,263.7892) --
      (819.3735,261.6876) -- (817.9186,259.5860) -- (817.9186,257.8077) --
      (815.4937,255.7062) -- (813.3921,250.2097) -- (814.6854,244.8749) --
      (814.5237,242.7733) -- (813.2304,241.4800) .. controls (813.2304,241.4800) and
      (814.6854,239.8634) .. (814.6854,239.2168) .. controls (814.6854,238.5701) and
      (815.1703,237.1152) .. (815.1703,237.1152) -- (817.1103,235.8219) --
      (819.0502,234.2053) -- (819.5352,235.1753) -- (818.0802,236.7919) --
      (816.7869,240.5101) -- (817.1103,241.6417) -- (818.8885,241.9650) --
      (819.3735,247.4615) -- (817.2719,248.4314) -- (817.5952,251.9880) --
      (818.0802,251.8263) -- (819.2118,249.8864) -- (820.8285,251.6646) --
      (819.2118,252.9579) -- (818.8885,256.3528) -- (821.4751,259.7477) --
      (825.3549,260.2327) -- (826.9716,259.4244) -- (830.2081,263.6073) --
      (831.5665,264.1436) -- (838.2201,261.3466) -- (840.2277,257.3228) --
      (839.7917,252.4148) -- cycle(823.8222,261.4435) -- (824.9538,263.9492) --
      (825.1155,265.7275) -- (826.2471,267.5866) .. controls (826.2471,267.5866) and
      (827.1362,266.6975) .. (827.1362,266.3741) .. controls (827.1362,266.0508) and
      (826.4087,263.3026) .. (826.4087,263.3026) -- (825.6813,260.9585) --
      (823.8222,261.4435) -- cycle;

    \path[USA map/state, USA map/VA, local bounding box=VA] (831.6389,266.0689) -- (831.4949,264.1219) -- (837.9484,261.5720) --
      (837.1780,264.7899) -- (834.2580,268.5690) -- (833.8399,273.1548) --
      (834.3017,276.5452) -- (832.4737,281.5234) -- (830.3094,283.4395) --
      (828.8391,278.7987) -- (829.2850,273.3496) -- (830.8720,269.1665) --
      (831.6389,266.0689) -- cycle(834.9790,294.3703) -- (776.8049,306.9457) --
      (739.3779,312.2248) -- (732.6996,311.8496) -- (730.1143,313.7760) --
      (722.7752,313.9967) -- (714.3931,314.9743) -- (703.4781,316.5890) --
      (713.9475,310.9778) -- (713.9344,308.9028) -- (715.4545,306.7567) --
      (726.0083,295.2553) -- (729.9550,299.7327) -- (733.7380,300.6967) --
      (736.2815,299.5564) -- (738.5187,298.2452) -- (741.0553,299.5887) --
      (744.9695,298.1610) -- (746.8462,293.6046) -- (749.4471,294.1447) --
      (752.3024,292.0134) -- (754.1016,292.5070) -- (756.9288,288.8304) --
      (757.2771,286.7473) -- (756.3134,285.4718) -- (757.3162,283.6051) --
      (762.5905,271.3280) -- (763.2072,265.5929) -- (764.4361,265.0694) --
      (766.6147,267.5122) -- (770.5505,267.2111) -- (772.4797,259.6374) --
      (775.2737,259.0766) -- (776.3235,256.3355) -- (778.9033,253.9886) --
      (781.6751,248.2934) -- (781.7600,243.2259) -- (791.5815,247.0487) .. controls
      (792.2624,247.3891) and (792.4144,241.9996) .. (792.4144,241.9996) --
      (796.0670,243.5979) -- (796.1353,246.5361) -- (801.9195,247.8355) --
      (804.0525,249.0117) -- (805.7124,251.0674) -- (805.0578,254.7161) --
      (803.1104,257.3071) -- (803.2202,259.3662) -- (803.8092,261.2191) --
      (808.7880,262.4875) -- (813.2392,262.5274) -- (816.3081,263.4860) --
      (818.2516,263.7953) -- (818.9664,266.8838) -- (822.1568,267.2863) --
      (823.0249,268.4863) -- (822.5854,273.1764) -- (823.9601,274.2790) --
      (823.4812,276.2094) -- (824.7106,276.9991) -- (824.4888,278.3837) --
      (821.7948,278.2888) -- (821.8838,279.9044) -- (824.1648,281.4472) --
      (824.2863,282.8591) -- (826.0594,284.6445) -- (826.5512,287.1686) --
      (823.9982,288.5499) -- (825.5704,290.0442) -- (831.3714,288.3584) --
      (834.9790,294.3703) -- cycle;

    \path[USA map/state, USA map/WV, local bounding box=WV] (761.1855,238.9673) -- (762.2975,243.9118) -- (763.3810,249.9432) --
      (765.5113,247.3628) -- (767.7745,244.2913) -- (770.3129,243.6757) --
      (771.7678,242.2208) -- (773.5461,239.6342) -- (774.9911,240.2808) --
      (777.9010,239.9575) -- (780.4875,237.8559) -- (782.4944,236.4027) --
      (784.3397,235.9177) -- (785.6436,236.9342) -- (789.2868,238.7558) --
      (791.2268,240.5341) -- (792.6009,241.8273) -- (791.8392,247.3823) --
      (786.0042,244.8411) -- (781.7590,243.2190) -- (781.6579,248.3975) --
      (778.9102,253.9342) -- (776.3802,256.3609) -- (775.1881,259.1102) --
      (772.5445,259.6103) -- (771.6467,263.2122) -- (770.6034,267.1619) --
      (766.6352,267.5026) -- (764.3115,265.0638) -- (763.2403,265.6232) --
      (762.6076,271.0929) -- (761.2574,274.6274) -- (756.2990,285.5823) --
      (757.1956,286.7430) -- (756.9898,288.6516) -- (754.1811,292.5360) --
      (752.3726,291.9918) -- (749.4045,294.1515) -- (746.8622,293.5793) --
      (744.8629,298.1349) .. controls (744.8629,298.1349) and (741.6036,299.5651) ..
      (740.9400,299.5026) .. controls (740.7795,299.4875) and (738.4709,298.2535) ..
      (738.4709,298.2535) -- (736.1344,299.6329) -- (733.7246,300.6773) --
      (729.9799,299.7881) -- (728.8585,298.6199) -- (726.6663,295.5965) --
      (723.5237,293.6084) -- (721.8121,289.9851) -- (717.5273,286.5169) --
      (716.8806,284.2537) -- (714.2940,282.7987) -- (713.4857,281.1821) --
      (713.2432,275.9282) -- (715.4257,275.8474) -- (717.3656,275.0391) --
      (717.5273,272.2908) -- (719.1439,270.8359) -- (719.3055,265.8244) --
      (720.2755,261.9445) -- (721.5688,261.2979) -- (722.8620,262.4295) --
      (723.3470,264.2078) -- (725.1253,263.2378) -- (725.6103,261.6212) --
      (724.4787,259.8430) -- (724.4787,257.4180) -- (725.4486,256.1248) --
      (727.7119,252.7299) -- (729.0052,251.2749) -- (731.1068,251.7599) --
      (733.3700,250.1433) -- (736.4415,246.7484) -- (738.7048,242.8686) --
      (739.0281,237.2105) -- (739.5131,232.1990) -- (739.5131,227.5108) --
      (738.3815,224.4393) -- (739.3514,222.9843) -- (740.6349,221.6910) --
      (744.1262,241.5181) -- (748.7572,240.7670) -- (761.1855,238.9673) -- cycle;

    \path[USA map/state, USA map/OH, local bounding box=OH] (735.3250,193.3283) -- (729.2314,197.3817) -- (725.3516,199.6449) --
      (721.9567,203.3631) -- (717.9152,207.2430) -- (714.6820,208.0513) --
      (711.7721,208.5362) -- (706.2756,211.1228) -- (704.1741,211.2845) --
      (700.7792,208.2129) -- (695.6061,208.8596) -- (693.0195,207.4046) --
      (690.6384,206.0538) -- (685.7459,206.7572) -- (675.5612,208.3738) --
      (664.3544,210.5585) -- (665.6477,225.1888) -- (667.4259,238.9300) --
      (670.0125,262.3708) -- (670.5783,267.2020) -- (674.7007,267.0729) --
      (677.1256,266.2646) -- (680.4894,267.7678) -- (682.5598,272.1326) --
      (687.6988,272.1155) -- (689.5905,274.2342) -- (691.3517,274.1689) --
      (693.8901,272.8274) -- (696.3943,273.1989) -- (701.8155,273.6816) --
      (703.5425,271.5489) -- (705.8882,270.2557) -- (707.9587,269.5748) --
      (708.6053,272.3230) -- (710.3836,273.2930) -- (713.8593,275.6371) --
      (716.0417,275.5563) -- (717.3748,275.0638) -- (717.5595,272.3023) --
      (719.1449,270.8473) -- (719.2441,266.0546) .. controls (719.2441,266.0546) and
      (720.2680,261.9455) .. (720.2680,261.9455) -- (721.5673,261.3442) --
      (722.8887,262.4920) -- (723.4268,264.1890) -- (725.1459,263.1516) --
      (725.5849,261.6908) -- (724.4682,259.7878) -- (724.5345,257.4733) --
      (725.2835,256.4010) -- (727.4363,253.0946) -- (728.4865,251.5512) --
      (730.5881,252.0362) -- (732.8513,250.4196) -- (735.9229,247.0247) --
      (738.6944,242.9460) -- (739.0147,237.8905) -- (739.4997,232.8790) --
      (739.3229,227.5721) -- (738.3681,224.6773) -- (738.7193,223.4875) --
      (740.5237,221.7374) -- (738.2349,212.6901) -- (735.3250,193.3283) -- cycle;

    \path[USA map/state, USA map/IN, local bounding box=IN] (619.5695,299.9713) -- (619.6348,297.1127) -- (620.1198,292.5862) --
      (622.3831,289.6764) -- (624.1613,285.7965) -- (626.7479,281.5933) --
      (626.2629,275.7735) -- (624.4847,273.0253) -- (624.1613,269.7921) --
      (624.9697,264.2956) -- (624.4847,257.3442) -- (623.1914,241.3398) --
      (621.8981,225.9820) -- (620.9276,214.2620) -- (623.9987,215.1515) --
      (625.4536,216.1215) -- (626.5853,215.7982) -- (628.6868,213.8582) --
      (631.5164,212.2413) -- (636.6092,212.0792) -- (658.5951,209.8160) --
      (664.1708,209.2828) -- (665.6740,225.2390) -- (669.9253,262.0806) --
      (670.5238,267.8521) -- (670.1523,270.1154) -- (671.3802,271.9108) --
      (671.4766,273.2833) -- (668.9554,274.8828) -- (665.4159,276.4341) --
      (662.2138,276.9844) -- (661.6153,281.8514) -- (657.0406,285.1638) --
      (654.2442,289.1743) -- (654.5675,291.5510) -- (653.9862,293.0852) --
      (650.6597,293.0852) -- (649.0742,291.4686) -- (646.5809,292.7308) --
      (643.8979,294.2339) -- (644.0596,297.2884) -- (642.8658,297.5464) --
      (642.3979,296.5283) -- (640.2311,295.0251) -- (636.9807,296.3666) --
      (635.4294,299.3729) -- (633.9916,298.5646) -- (632.5366,296.9651) --
      (628.0723,297.4500) -- (622.4795,298.4200) -- (619.5696,299.9713) -- cycle;

    \path[USA map/state, USA map/IL, local bounding box=IL] (619.5415,300.3424) -- (619.5727,297.1127) -- (620.1400,292.4668) --
      (622.4726,289.5509) -- (624.3392,285.4751) -- (626.5722,281.4798) --
      (626.2007,276.2274) -- (624.1955,272.6848) -- (624.0991,269.3382) --
      (624.7940,264.0687) -- (623.9686,256.8903) -- (622.9022,241.1128) --
      (621.6089,226.0955) -- (620.6867,214.4563) -- (620.4141,213.5349) --
      (619.6058,210.9483) -- (618.3126,207.2301) -- (616.6960,205.4519) --
      (615.2410,202.8653) -- (615.0074,197.3764) -- (569.2110,199.9746) --
      (569.4396,202.3466) -- (571.7259,203.0324) -- (572.6403,204.1755) --
      (573.0976,206.0045) -- (576.9842,209.4339) -- (577.6701,211.7201) --
      (576.9842,215.1494) -- (575.1552,218.8074) -- (574.4693,221.3222) --
      (572.1831,223.1512) -- (570.3541,223.8371) -- (565.0958,225.2088) --
      (564.4099,227.0378) -- (563.7241,229.0954) -- (564.4099,230.4672) --
      (566.2389,232.0675) -- (566.0103,236.1827) -- (564.1813,237.7831) --
      (563.4954,239.3834) -- (563.4954,242.1269) -- (561.6665,242.5841) --
      (560.0661,243.7273) -- (559.8375,245.0990) -- (560.0661,247.1566) --
      (558.3514,248.4712) -- (557.3226,251.2718) -- (557.7799,254.9298) --
      (560.0661,262.2457) -- (567.3820,269.7902) -- (572.8690,273.4482) --
      (572.6403,277.7920) -- (573.5548,279.1638) -- (579.9563,279.6210) --
      (582.6997,280.9928) -- (582.0139,284.6507) -- (579.7277,290.5949) --
      (579.0418,293.7956) -- (581.3280,297.6822) -- (587.7294,302.9405) --
      (592.3019,303.6264) -- (594.3595,308.6561) -- (596.4171,311.8568) --
      (595.5026,314.8289) -- (597.1030,318.9441) -- (598.9319,321.0017) --
      (600.3460,320.1210) -- (601.2536,318.0462) -- (603.4667,316.2990) --
      (605.5982,315.6846) -- (608.2007,316.8644) -- (611.8277,318.2401) --
      (613.0167,317.9419) -- (613.2165,315.6834) -- (611.9292,313.2717) --
      (612.2334,310.8949) -- (614.0718,309.5475) -- (617.0944,308.7372) --
      (618.3553,308.2787) -- (617.7427,306.8918) -- (616.9513,304.5374) --
      (618.3839,303.5565) -- (619.5414,300.3424) -- cycle;

    \path[USA map/state, USA map/CT, local bounding box=CT] (874.0683,178.8629) -- (870.3909,163.9841) -- (865.6721,164.9044) --
      (844.4433,169.6475) -- (845.4435,172.8731) -- (846.8984,180.1479) --
      (847.0752,189.1148) -- (845.8552,191.2897) -- (847.7760,193.2220) --
      (852.0475,189.3164) -- (855.6040,186.0832) -- (857.5439,183.9816) --
      (858.3523,184.6282) -- (861.1005,183.1733) -- (866.2736,182.0417) --
      (874.0683,178.8629) -- cycle;

    \path[USA map/state, USA map/WI, local bounding box=WI] (615.0659,197.3687) -- (614.9992,194.2112) -- (613.8201,189.6847) --
      (613.1734,183.5417) -- (612.0418,181.1167) -- (613.0118,178.0452) --
      (613.8201,175.1353) -- (615.2750,172.5487) -- (614.6284,169.1539) --
      (613.9817,165.5973) -- (614.4667,163.8191) -- (616.4066,161.3942) --
      (616.5683,158.6459) -- (615.7600,157.3526) -- (616.4066,154.7661) --
      (615.9541,150.5954) -- (618.7024,144.9373) -- (621.6122,138.1475) --
      (621.7739,135.8843) -- (621.4506,134.9143) -- (620.6423,135.3993) --
      (616.4391,141.7041) -- (613.6909,145.7456) -- (611.7510,147.5238) --
      (610.9427,149.7871) -- (608.9877,150.5954) -- (607.8561,152.5353) --
      (606.4011,152.2120) -- (606.2395,150.4337) -- (607.5328,148.0088) --
      (609.6343,143.3207) -- (611.4126,141.7040) -- (612.4034,139.3462) --
      (609.8430,137.4449) -- (607.8682,127.0779) -- (604.3207,125.7359) --
      (602.3744,123.4276) -- (590.2447,120.7059) -- (587.3688,119.6939) --
      (579.1557,117.5266) -- (571.2378,116.3678) -- (567.4726,111.2372) --
      (566.7222,111.7912) -- (565.5243,111.6295) -- (564.8777,110.4979) --
      (563.5437,110.7944) -- (562.4120,110.9561) -- (560.6338,111.9261) --
      (559.6638,111.2794) -- (560.3105,109.3395) -- (562.2504,106.2679) --
      (563.3820,105.1363) -- (561.4421,103.6814) -- (559.3405,104.4897) --
      (556.4306,106.4296) -- (548.9942,109.6628) -- (546.0843,110.3094) --
      (543.1745,109.8245) -- (542.1927,108.9462) -- (540.0760,111.7814) --
      (539.8474,114.5249) -- (539.8474,122.9839) -- (538.7043,124.5843) --
      (533.4460,128.4708) -- (531.1597,134.4150) -- (531.6170,134.6437) --
      (534.1318,136.7013) -- (534.8177,139.9020) -- (532.9887,143.1027) --
      (532.9887,146.9893) -- (533.4460,153.6193) -- (536.4181,156.5914) --
      (539.8474,156.5914) -- (541.6764,159.7922) -- (545.1057,160.2494) --
      (548.9923,165.9650) -- (556.0796,170.0802) -- (558.1372,172.8236) --
      (559.0517,180.2539) -- (559.7376,183.5689) -- (562.0238,185.1693) --
      (562.2524,186.5410) -- (560.1948,189.9703) -- (560.4234,193.1711) --
      (562.9383,197.0576) -- (565.4531,198.2007) -- (568.4252,198.6580) --
      (569.7676,200.0381) -- (615.0659,197.3687) -- cycle;

    \path[USA map/state, USA map/NC, local bounding box=NC] (834.9815,294.3155) -- (837.0665,299.2329) -- (840.6231,305.6993) --
      (843.0480,308.1242) -- (843.6946,310.3875) -- (841.2697,310.5491) --
      (842.0780,311.1958) -- (841.7547,315.3989) -- (839.1681,316.6922) --
      (838.5215,318.7938) -- (837.2282,321.7037) -- (833.5100,323.3203) --
      (831.0851,322.9970) -- (829.6301,322.8353) -- (828.0135,321.5420) --
      (828.3369,322.8353) -- (828.3369,323.8053) -- (830.2768,323.8053) --
      (831.0851,325.0986) -- (829.1452,331.4033) -- (833.3483,331.4033) --
      (833.9950,333.0199) -- (836.2582,330.7567) -- (837.5515,330.2717) --
      (835.6116,333.8282) -- (832.5400,338.6781) -- (831.2468,338.6781) --
      (830.1151,338.1931) -- (827.3669,338.8397) -- (822.1938,341.2646) --
      (815.7273,346.5994) -- (812.3325,351.2876) -- (810.3926,357.7540) --
      (809.9076,360.1789) -- (805.2194,360.6639) -- (799.7663,362.0005) --
      (789.8199,353.7980) -- (777.2103,346.2000) -- (774.3004,345.3916) --
      (761.6909,346.8466) -- (757.4145,347.5967) -- (755.7979,344.3635) --
      (752.8275,342.2468) -- (736.3381,342.7318) -- (729.0634,343.5401) --
      (720.0104,348.0666) -- (713.8673,350.6532) -- (692.6897,353.2398) --
      (693.1898,349.1854) -- (694.9681,347.7305) -- (697.7163,347.0838) --
      (698.3630,343.3656) -- (702.5661,340.6174) -- (706.4460,339.1624) --
      (710.6492,335.6059) -- (715.0140,333.5043) -- (715.6606,330.4328) --
      (719.5405,326.5529) -- (720.1871,326.3913) .. controls (720.1871,326.3913) and
      (720.1871,327.5229) .. (720.9955,327.5229) .. controls (721.8038,327.5229) and
      (722.9354,327.8462) .. (722.9354,327.8462) -- (725.1986,324.2897) --
      (727.3002,323.6430) -- (729.5635,323.9664) -- (731.1801,320.4098) --
      (734.0900,317.8232) -- (734.5750,315.7217) -- (734.7625,312.0735) --
      (739.0390,312.0510) -- (746.2375,311.1952) -- (761.9948,308.9427) --
      (777.1308,306.8562) -- (798.7713,302.1368) -- (818.7546,297.8782) --
      (829.9316,295.4724) -- (834.9815,294.3156) -- cycle(839.2520,327.5221) --
      (841.8386,325.0164) -- (844.9909,322.4298) -- (846.5267,321.7831) --
      (846.6884,319.7624) -- (846.0417,313.6193) -- (844.5868,311.2752) --
      (843.9401,309.4161) -- (844.6676,309.1736) -- (847.4159,314.6701) --
      (847.8200,319.1157) -- (847.6584,322.5106) -- (844.2635,324.0464) --
      (841.4344,326.4713) -- (840.3028,327.6838) -- (839.2520,327.5221) -- cycle;

    \path[USA map/state, USA map/DC, local bounding box=DC] (805.8194,250.8438) -- (803.9612,249.0197) -- (802.7285,248.3334) --
      (804.1715,246.3109) -- (807.0606,248.2594) -- (805.8194,250.8438) -- cycle;

    \path[USA map/state, USA map/MA, local bounding box=MA] (899.6235,173.2539) -- (901.7954,172.5681) -- (902.2527,170.8534) --
      (903.2815,170.9677) -- (904.3103,173.2539) -- (903.0529,173.7112) --
      (899.1662,173.8255) -- (899.6235,173.2539) -- cycle(890.2499,174.0541) --
      (892.5362,171.4250) -- (894.1365,171.4250) -- (895.9655,172.9110) --
      (893.5650,173.9398) -- (891.3931,174.9686) -- (890.2499,174.0541) --
      cycle(855.4508,152.0659) -- (873.0977,147.4253) -- (875.3609,146.7786) --
      (877.2750,143.9829) -- (881.0118,142.3196) -- (883.9010,146.7324) --
      (881.4761,151.9056) -- (881.1528,153.3605) -- (883.0927,155.9471) --
      (884.2244,155.1388) -- (886.0026,155.1388) -- (888.2659,157.7253) --
      (892.1457,163.7068) -- (895.7023,164.1918) -- (897.9655,163.2218) --
      (899.7438,161.4435) -- (898.9355,158.6953) -- (896.8339,157.0787) --
      (895.3789,157.8870) -- (894.4090,156.5937) -- (894.8939,156.1087) --
      (896.9955,155.9471) -- (898.7738,156.7554) -- (900.7137,159.1803) --
      (901.6837,162.0902) -- (902.0070,164.5151) -- (897.8038,165.9700) --
      (893.9240,167.9099) -- (890.0441,172.4364) -- (888.1042,173.8914) --
      (888.1042,172.9214) -- (890.5291,171.4665) -- (891.0141,169.6882) --
      (890.2058,166.6167) -- (887.2959,168.0716) -- (886.4876,169.5266) --
      (886.9726,171.7898) -- (884.9063,172.7902) -- (882.1591,168.2631) --
      (878.7642,163.8983) -- (876.6937,162.0858) -- (870.1604,163.9620) --
      (865.0681,165.0128) -- (844.3929,169.6050) -- (843.7252,164.8371) --
      (844.3718,154.2484) -- (848.6611,153.3592) -- (855.4508,152.0659) -- cycle;

    \path[USA map/state, USA map/TN, local bounding box=TN] (696.6779,318.2541) -- (644.7848,323.2656) -- (629.0252,325.0439) --
      (624.4040,325.5566) -- (620.5357,325.5289) -- (620.3147,329.6297) --
      (612.1293,329.8937) -- (605.1779,330.5403) -- (597.0871,330.4165) --
      (595.6733,337.4894) -- (593.9771,342.9694) -- (590.6839,345.7202) --
      (589.3352,350.1013) -- (589.0118,352.6879) -- (584.9703,354.9511) --
      (586.4253,358.5076) -- (585.4553,362.8725) -- (584.4869,363.6621) --
      (692.6455,353.2546) -- (693.0487,349.2996) -- (694.8595,347.8093) --
      (697.6936,347.0598) -- (698.3656,343.3428) -- (702.4642,340.6379) --
      (706.5111,339.1438) -- (710.5947,335.5735) -- (715.0308,333.5480) --
      (715.5520,330.4807) -- (719.6166,326.4957) -- (720.1674,326.3815) .. controls
      (720.1674,326.3815) and (720.1986,327.5132) .. (721.0070,327.5132) .. controls
      (721.8153,327.5132) and (722.9469,327.8677) .. (722.9469,327.8677) --
      (725.2101,324.2799) -- (727.2805,323.6333) -- (729.5556,323.9285) --
      (731.1539,320.3956) -- (734.1092,317.7517) -- (734.5308,315.8126) --
      (734.8398,312.1015) -- (732.6932,311.9017) -- (730.0916,313.9300) --
      (723.0983,313.9591) -- (704.7390,316.3460) -- (696.6779,318.2542) -- cycle;

    \path[USA map/state, USA map/AR, local bounding box=AR] (593.8248,343.0530) -- (589.8449,343.7697) -- (584.7327,343.1356) --
      (585.1534,341.5336) -- (588.1332,338.9669) -- (589.0766,335.3106) --
      (587.2476,332.3385) -- (508.8300,334.8534) -- (510.4304,341.7121) --
      (510.4304,349.9425) -- (511.8021,360.9165) -- (512.0307,398.7534) --
      (514.3170,400.6967) -- (517.2891,399.3250) -- (520.0325,400.4681) --
      (520.7129,407.0414) -- (576.3341,405.9008) -- (577.4798,403.8104) --
      (577.1932,400.2609) -- (575.3675,397.2888) -- (576.9662,395.8036) --
      (575.3675,393.2921) -- (576.0517,390.7822) -- (577.4201,385.1768) --
      (579.9383,383.1142) -- (579.2524,380.8296) -- (582.9104,375.4578) --
      (585.6539,374.0894) -- (585.5404,372.5959) -- (585.1949,370.7702) --
      (588.0519,365.1715) -- (590.4549,363.9149) -- (590.8391,360.4873) --
      (592.6097,359.2456) -- (589.4662,358.7613) -- (588.1248,354.7509) --
      (590.9288,352.3742) -- (591.4791,350.3550) -- (592.7586,346.3083) --
      (593.8248,343.0530) -- cycle;

    \path[USA map/state, USA map/MO, local bounding box=MO] (558.4402,248.1132) -- (555.9203,245.0259) -- (554.7772,242.7397) --
      (490.4200,245.1402) -- (488.1337,245.2545) -- (489.3912,247.7694) --
      (489.1626,250.0556) -- (491.6774,253.9422) -- (494.7638,258.0574) --
      (497.8502,260.8009) -- (500.0114,261.0295) -- (501.5082,261.9440) --
      (501.5082,264.9161) -- (499.6792,266.5164) -- (499.2219,268.8027) --
      (501.2795,272.2320) -- (503.7944,275.2041) -- (506.3092,277.0331) --
      (507.6810,288.6928) -- (507.9951,324.7650) -- (508.2237,329.4518) --
      (508.6810,334.8353) -- (531.1140,333.9685) -- (554.3200,333.2826) --
      (575.1246,332.4816) -- (586.7794,332.2513) -- (588.9488,335.6773) --
      (588.2646,338.9848) -- (585.1773,341.3878) -- (584.6050,343.2252) --
      (589.9834,343.6824) -- (593.8784,342.9966) -- (595.5956,337.5029) --
      (596.2470,331.6461) -- (598.3450,329.0910) -- (600.9411,327.6041) --
      (600.9925,324.5538) -- (602.0085,322.6174) -- (600.3143,320.0736) --
      (598.9833,321.0579) -- (596.9907,318.8306) -- (595.7057,314.0716) --
      (596.5067,311.5534) -- (594.5626,308.1258) -- (592.7319,303.5500) --
      (587.9325,302.7506) -- (580.9637,297.1519) -- (579.2449,293.0383) --
      (580.0442,289.8376) -- (582.1035,283.7799) -- (582.5624,280.9163) --
      (580.6133,279.8850) -- (573.7579,279.0873) -- (572.7299,277.3752) --
      (572.6181,273.1448) -- (567.1312,269.7138) -- (560.1557,261.9423) --
      (557.8695,254.6264) -- (557.6392,250.4011) -- (558.4402,248.1132) -- cycle;

    \path[USA map/state, USA map/GA, local bounding box=GA] (672.2923,355.5518) -- (672.2923,357.7342) -- (672.4539,359.8358) --
      (673.1006,363.2307) -- (676.4955,371.1521) -- (678.9204,381.0134) --
      (680.3753,387.1565) -- (681.9919,392.0063) -- (683.4469,398.9577) --
      (685.5485,405.2625) -- (688.1350,408.6574) -- (688.6200,412.0522) --
      (690.5599,412.8605) -- (690.7216,414.9621) -- (688.9433,419.8119) --
      (688.4584,423.0452) -- (688.2967,424.9851) -- (689.9133,429.3499) --
      (690.2366,434.6847) -- (689.4283,437.1096) -- (690.0750,437.9179) --
      (691.5299,438.7262) -- (691.7346,441.9443) -- (693.9676,445.2939) --
      (696.2181,447.4559) -- (704.1395,447.6176) -- (714.9592,446.9709) --
      (736.4716,445.6777) -- (741.9173,445.0033) -- (746.4946,445.0310) --
      (746.6562,447.9409) -- (749.2428,448.7492) -- (749.5661,444.3843) --
      (747.9495,439.8578) -- (749.0811,438.2412) -- (754.9009,439.0495) --
      (759.8783,439.3673) -- (759.1029,433.0685) -- (761.3661,423.0456) --
      (762.8211,418.8424) -- (762.3361,416.2558) -- (765.6705,410.0115) --
      (765.1602,408.6599) -- (763.2468,409.3644) -- (760.6602,408.0711) --
      (760.0136,405.9695) -- (758.7203,402.4130) -- (756.4571,400.3114) --
      (753.8705,399.6648) -- (752.2539,394.8150) -- (749.3289,388.4800) --
      (745.1257,386.5400) -- (743.0241,384.6001) -- (741.7308,382.0135) --
      (739.6292,380.0736) -- (737.3660,378.7803) -- (735.1027,375.8704) --
      (732.0312,373.6072) -- (727.5047,371.8289) -- (727.0197,370.3740) --
      (724.5948,367.4641) -- (724.1098,366.0091) -- (720.7149,361.0386) --
      (717.1951,361.1378) -- (713.4401,358.7817) -- (712.0219,357.4884) --
      (711.6985,355.7102) -- (712.5693,353.7702) -- (714.7960,352.6601) --
      (714.1620,350.5629) -- (672.2923,355.5518) -- cycle;

    \path[USA map/state, USA map/SC, local bounding box=SC] (764.9433,408.1649) -- (763.1662,409.1344) -- (760.5796,407.8411) --
      (759.9330,405.7395) -- (758.6397,402.1830) -- (756.3765,400.0814) --
      (753.7899,399.4347) -- (752.1733,394.5849) -- (749.4251,388.6035) --
      (745.2219,386.6635) -- (743.1203,384.7236) -- (741.8270,382.1370) --
      (739.7254,380.1971) -- (737.4622,378.9038) -- (735.1989,375.9939) --
      (732.1274,373.7307) -- (727.6009,371.9524) -- (727.1159,370.4975) --
      (724.6910,367.5876) -- (724.2060,366.1326) -- (720.8111,360.9595) --
      (717.4162,361.1211) -- (713.3747,358.6962) -- (712.0814,357.4029) --
      (711.7581,355.6247) -- (712.5664,353.6848) -- (714.8297,352.7148) --
      (714.3189,350.4257) -- (720.0870,348.0891) -- (729.2025,343.5001) --
      (736.9772,342.6918) -- (753.0916,342.2693) -- (755.7298,344.1468) --
      (757.4089,347.5050) -- (761.7113,346.8950) -- (774.3208,345.4400) --
      (777.2307,346.2484) -- (789.8402,353.8464) -- (799.9483,361.9681) --
      (794.5272,367.4264) -- (791.9406,373.5695) -- (791.4556,379.8743) --
      (789.8390,380.6826) -- (788.7074,383.4308) -- (786.2825,384.0775) --
      (784.1809,387.6340) -- (781.4327,390.3822) -- (779.1694,393.7771) --
      (777.5528,394.5854) -- (773.9963,397.9803) -- (771.0864,398.1419) --
      (772.0564,401.3751) -- (767.0449,406.8716) -- (764.9433,408.1649) -- cycle;

    \path[USA map/state, USA map/KY, local bounding box=KY] (725.9944,295.2707) -- (723.7011,297.6724) -- (720.1229,301.6664) --
      (715.1983,307.1311) -- (713.9826,308.8469) -- (713.9201,310.9484) --
      (709.5402,313.1125) -- (703.8821,316.5074) -- (696.6502,318.3063) --
      (644.7823,323.2051) -- (629.0228,324.9834) -- (624.4016,325.4961) --
      (620.5332,325.4684) -- (620.3063,329.6887) -- (612.1269,329.8332) --
      (605.1755,330.4799) -- (597.1880,330.4197) -- (598.3958,329.0996) --
      (600.8953,327.5587) -- (601.1239,324.3580) -- (602.0384,322.5290) --
      (600.4316,319.9901) -- (601.2334,318.0833) -- (603.4967,316.3051) --
      (605.5983,315.6584) -- (608.3465,316.9517) -- (611.9030,318.2450) --
      (613.0347,317.9217) -- (613.1963,315.6584) -- (611.9030,313.2335) --
      (612.2264,310.9702) -- (614.1663,309.5153) -- (616.7529,308.8687) --
      (618.3695,308.2220) -- (617.5612,306.4437) -- (616.9145,304.5038) --
      (618.4211,303.5080) .. controls (618.4241,303.4709) and (619.6751,299.9857) ..
      (619.6594,299.8502) -- (622.7127,298.3715) -- (628.0324,297.4016) --
      (632.5265,296.9166) -- (633.9189,298.5440) -- (635.4472,299.4148) --
      (637.0380,296.3066) -- (640.2250,295.0240) -- (642.4301,296.5080) --
      (642.8407,297.5071) -- (644.0142,297.2430) -- (643.8525,294.2901) --
      (646.9834,292.5409) -- (649.1315,291.4674) -- (650.6609,293.1283) --
      (653.9790,293.0841) -- (654.5663,291.5128) -- (654.1988,289.2496) --
      (656.7994,285.2511) -- (661.5759,281.8132) -- (662.2819,276.9773) --
      (665.2069,276.5214) -- (668.9983,274.8757) -- (671.4417,273.1675) --
      (671.2433,271.6025) -- (670.1009,270.1476) -- (670.6667,267.1527) --
      (674.8516,267.0352) -- (677.1515,266.2894) -- (680.4989,267.7185) --
      (682.5530,272.0833) -- (687.6853,272.0941) -- (689.7363,274.3023) --
      (691.3517,274.1546) -- (693.9534,272.8765) -- (699.1905,273.4498) --
      (701.7654,273.6673) -- (703.4530,271.6111) -- (706.0709,270.1852) --
      (707.9527,269.4781) -- (708.5993,272.3147) -- (710.6428,273.3731) --
      (713.2855,275.4556) -- (713.4030,281.1288) -- (714.2113,282.7012) --
      (716.8010,284.2575) -- (717.5727,286.5520) -- (721.7325,289.9890) --
      (723.5379,293.6122) -- (725.9944,295.2707) -- cycle;

    \path[USA map/state, USA map/AL, local bounding box=AL] (631.3065,460.4157) -- (629.8159,446.0942) -- (627.0676,427.3416) --
      (627.2293,413.2771) -- (628.0376,382.2382) -- (627.8759,365.5872) --
      (628.0410,359.1681) -- (672.5255,355.5487) -- (672.3777,357.7311) --
      (672.5394,359.8327) -- (673.1860,363.2276) -- (676.5809,371.1489) --
      (679.0058,381.0102) -- (680.4607,387.1534) -- (682.0773,392.0032) --
      (683.5323,398.9546) -- (685.6339,405.2593) -- (688.2205,408.6542) --
      (688.7054,412.0491) -- (690.6454,412.8574) -- (690.8070,414.9590) --
      (689.0287,419.8088) -- (688.5438,423.0420) -- (688.3821,424.9820) --
      (689.9987,429.3468) -- (690.3220,434.6816) -- (689.5137,437.1065) --
      (690.1604,437.9148) -- (691.6153,438.7231) -- (691.9435,441.6119) --
      (686.3458,441.2584) -- (679.5561,441.9050) -- (654.0137,444.8149) --
      (643.6021,446.2217) -- (643.3807,449.0991) -- (645.1590,450.8774) --
      (647.7456,452.8173) -- (648.3264,460.7527) -- (642.7844,463.3256) --
      (640.0361,463.0023) -- (642.7844,461.0624) -- (642.7844,460.0924) --
      (639.7128,454.1110) -- (637.4496,453.4643) -- (635.9946,457.8291) --
      (634.7013,460.5774) -- (634.0547,460.4157) -- (631.3065,460.4157) -- cycle;

    \path[USA map/state, USA map/LA, local bounding box=LA] (607.9671,459.1612) -- (604.6824,455.9951) -- (605.6924,450.4949) --
      (605.0310,449.6018) -- (595.7693,450.6084) -- (570.7410,451.0673) --
      (570.0568,448.6726) -- (570.9696,440.2169) -- (574.2855,434.2711) --
      (579.3169,425.5800) -- (578.7428,423.1820) -- (579.9994,422.5012) --
      (580.4583,420.5487) -- (578.1721,418.4927) -- (578.0603,416.5503) --
      (576.2296,412.2048) -- (576.0826,405.8662) -- (520.6088,406.7902) --
      (520.6374,416.3637) -- (521.3233,425.7373) -- (522.0091,429.6238) --
      (524.5240,433.7390) -- (525.4385,438.7688) -- (529.7823,444.2557) --
      (530.0109,447.4564) -- (530.6968,448.1423) -- (530.0109,456.6013) --
      (527.0388,461.6310) -- (528.6392,463.6886) -- (527.9533,466.2035) --
      (527.2675,473.5194) -- (525.8957,476.7201) -- (526.0182,480.3365) --
      (530.7047,478.8164) -- (542.8180,479.0234) -- (553.1643,482.5799) --
      (559.6307,483.7116) -- (563.3489,482.2566) -- (566.5821,483.3882) --
      (569.8153,484.3582) -- (570.6236,482.2566) -- (567.3904,481.1250) --
      (564.8038,481.6100) -- (562.0556,479.9934) .. controls (562.0556,479.9934) and
      (562.2173,478.7001) .. (562.8639,478.5384) .. controls (563.5105,478.3768) and
      (565.9355,477.5685) .. (565.9355,477.5685) -- (567.7137,479.0234) --
      (569.4920,478.0534) -- (572.7252,478.7001) -- (574.1801,481.1250) --
      (574.5035,483.3882) -- (579.0299,483.7116) -- (580.8082,485.4898) --
      (579.9999,487.1064) -- (578.7066,487.9147) -- (580.3232,489.5313) --
      (588.7296,493.0879) -- (592.2861,491.7946) -- (593.2561,489.3697) --
      (595.8426,488.7230) -- (597.6209,487.2681) -- (598.9142,488.2381) --
      (599.7225,491.1479) -- (597.4592,491.9562) -- (598.1059,492.6029) --
      (601.5008,491.3096) -- (603.7640,487.9147) -- (604.5723,487.4298) --
      (602.4707,487.1064) -- (603.2790,485.4898) -- (603.1174,484.0349) --
      (605.2189,483.5499) -- (606.3506,482.2566) -- (606.9972,483.0649) .. controls
      (606.9972,483.0649) and (606.8355,486.1365) .. (607.6439,486.1365) .. controls
      (608.4522,486.1365) and (611.8470,486.7831) .. (611.8470,486.7831) --
      (615.8885,488.7230) -- (616.8585,490.1780) -- (619.7684,490.1780) --
      (620.9000,491.1479) -- (623.1633,488.0764) -- (623.1633,486.6214) --
      (621.8700,486.6214) -- (618.4751,483.8732) -- (612.6553,483.0649) --
      (609.4221,480.8017) -- (610.5537,478.0534) -- (612.8170,478.3768) --
      (612.9786,477.7301) -- (611.2004,476.7602) -- (611.2004,476.2752) --
      (614.4336,476.2752) -- (616.2119,473.2036) -- (614.9186,471.2637) --
      (614.5953,468.5155) -- (613.1403,468.6771) -- (611.2004,470.7787) --
      (610.5537,473.3653) -- (607.4822,472.7186) -- (606.5122,470.9404) --
      (608.2905,469.0005) -- (610.1938,465.5548) -- (609.1327,463.1426) --
      (607.9671,459.1612) -- cycle;

    \path[USA map/state, USA map/MS, local bounding box=MS] (631.5588,459.3446) -- (631.3046,460.6007) -- (626.1314,460.6007) --
      (624.6765,459.7924) -- (622.5749,459.4691) -- (615.7851,461.4090) --
      (614.0069,460.6007) -- (611.4203,464.8039) -- (610.3178,465.5819) --
      (609.1939,463.0939) -- (608.0508,459.2074) -- (604.6215,456.0066) --
      (605.7646,450.4621) -- (605.0787,449.5476) -- (603.2498,449.7762) --
      (595.3318,450.6496) -- (570.7853,451.0230) -- (570.0156,448.7976) --
      (570.8890,440.4208) -- (574.0058,434.7480) -- (579.2329,425.6031) --
      (578.7871,423.1705) -- (580.0240,422.5142) -- (580.4599,420.5948) --
      (578.1424,418.5158) -- (578.0273,416.3743) -- (576.1915,412.2532) --
      (576.0825,406.2905) -- (577.4101,403.8095) -- (577.1868,400.3937) --
      (575.4173,397.3111) -- (576.9437,395.8289) -- (575.3731,393.3294) --
      (575.8303,391.6772) -- (577.4077,385.1508) -- (579.8937,383.1145) --
      (579.2520,380.7475) -- (582.9100,375.4450) -- (585.7419,374.0885) --
      (585.5209,372.4134) -- (585.2328,370.7323) -- (588.1088,365.1646) --
      (590.4545,363.9331) -- (590.6062,363.0401) -- (627.9496,359.1589) --
      (628.1345,365.4422) -- (628.2962,382.0933) -- (627.4879,413.1322) --
      (627.3262,427.1966) -- (630.0744,445.9493) -- (631.5588,459.3446) -- cycle;

    \path[USA map/state, USA map/IA, local bounding box=IA] (569.1915,199.5843) -- (569.4559,202.3705) -- (571.6796,202.9478) --
      (572.6336,204.1731) -- (573.1336,206.0285) -- (576.9264,209.3871) --
      (577.6123,211.7786) -- (576.9380,215.2031) -- (575.3556,218.4351) --
      (574.5563,221.1768) -- (572.3836,222.7789) -- (570.6680,223.3513) --
      (565.0890,225.2115) -- (563.6976,229.0602) -- (564.4262,230.4319) --
      (566.2667,232.1145) -- (565.9838,236.1508) -- (564.2206,237.6887) --
      (563.4492,239.3318) -- (563.5764,242.1081) -- (561.6901,242.5654) --
      (560.0647,243.6703) -- (559.7859,245.0229) -- (560.0647,247.1378) --
      (558.5137,248.2539) -- (556.0431,245.1206) -- (554.7806,242.6707) --
      (489.0447,245.1856) -- (488.1267,245.3510) -- (486.0743,240.8351) --
      (485.8457,234.2050) -- (484.2453,230.0898) -- (483.5595,224.8315) --
      (481.2732,221.1735) -- (480.3588,216.3724) -- (477.6153,208.8279) --
      (476.4722,203.4552) -- (475.1004,201.2833) -- (473.5001,198.5399) --
      (475.4541,193.6960) -- (476.8258,187.9805) -- (474.0823,185.9229) --
      (473.6251,183.1794) -- (474.5396,180.6645) -- (476.2542,180.6645) --
      (558.9082,179.3951) -- (559.7425,183.5782) -- (561.9947,185.1392) --
      (562.2514,186.5622) -- (560.2219,189.9516) -- (560.4123,193.1571) --
      (562.9271,196.9553) -- (565.4539,198.2489) -- (568.5332,198.7519) --
      (569.1915,199.5843) -- cycle;

    \path[USA map/state, USA map/MN, local bounding box=MN] (475.2378,128.8244) -- (474.7806,120.3653) -- (472.9516,113.0494) --
      (471.1226,99.5607) -- (470.6654,89.7299) -- (468.8364,86.3006) --
      (467.2360,81.2709) -- (467.2360,70.9829) -- (467.9219,67.0963) --
      (466.1009,61.6446) -- (496.2334,61.6799) -- (496.5567,53.4352) --
      (497.2033,53.2735) -- (499.4666,53.7585) -- (501.4065,54.5668) --
      (502.2148,60.0633) -- (503.6697,66.2064) -- (505.2863,67.8230) --
      (510.1362,67.8230) -- (510.4595,69.2779) -- (516.7642,69.6012) --
      (516.7642,71.7028) -- (521.6141,71.7028) -- (521.9374,70.4095) --
      (523.0690,69.2779) -- (525.3322,68.6313) -- (526.6255,69.6012) --
      (529.5354,69.6012) -- (533.4153,72.1878) -- (538.7501,74.6127) --
      (541.1750,75.0977) -- (541.6599,74.1277) -- (543.1149,73.6428) --
      (543.5999,76.5526) -- (546.1864,77.8459) -- (546.6714,77.3609) --
      (547.9647,77.5226) -- (547.9647,79.6242) -- (550.5513,80.5942) --
      (553.6228,80.5942) -- (555.2394,79.7859) -- (558.4726,76.5526) --
      (561.0592,76.0677) -- (561.8675,77.8459) -- (562.3525,79.1392) --
      (563.3224,79.1392) -- (564.2924,78.3309) -- (573.1837,78.0076) --
      (574.9620,81.0791) -- (575.6086,81.0791) -- (576.3223,79.9949) --
      (580.7622,79.6242) -- (580.1501,81.9037) -- (576.2113,83.7408) --
      (566.9656,87.8019) -- (562.1908,89.8088) -- (559.1193,92.3954) --
      (556.6944,95.9519) -- (554.4311,99.8318) -- (552.6529,100.6401) --
      (548.1264,105.6515) -- (546.8331,105.8132) -- (542.5053,108.5703) --
      (540.0424,111.7754) -- (539.8138,114.9668) -- (539.9082,123.0102) --
      (538.5322,124.6989) -- (533.4506,128.4589) -- (531.2205,134.4413) --
      (534.0923,136.6750) -- (534.7722,139.9020) -- (532.9169,143.1409) --
      (533.0877,146.8889) -- (533.4566,153.6193) -- (536.4848,156.6213) --
      (539.8138,156.6213) -- (541.7050,159.7539) -- (545.0841,160.2572) --
      (548.9433,165.9287) -- (556.0306,170.0454) -- (558.1737,172.9205) --
      (558.8449,179.3600) -- (477.6334,180.5048) -- (477.2955,144.8280) --
      (476.8382,141.8559) -- (472.7230,138.4265) -- (471.5799,136.5976) --
      (471.5799,134.9972) -- (473.6375,133.3968) -- (475.0092,132.0251) --
      (475.2379,128.8244) -- cycle;

    \path[USA map/state, USA map/OK, local bounding box=OK] (380.3431,320.8215) -- (363.6589,319.5482) -- (362.7787,330.5006) --
      (383.2441,331.6575) -- (415.2997,332.9611) -- (412.9651,357.3797) --
      (412.5078,375.2123) -- (412.7364,376.8126) -- (417.0803,380.4706) --
      (419.1379,381.6137) -- (419.8237,381.3851) -- (420.5096,379.3275) --
      (421.8813,381.1565) -- (423.9389,381.1565) -- (423.9389,379.7847) --
      (426.6824,381.1565) -- (426.2252,385.0430) -- (430.3404,385.2717) --
      (432.8552,386.4148) -- (436.9704,387.1007) -- (439.4853,388.9296) --
      (441.7715,386.8720) -- (445.2009,387.5579) -- (447.7157,390.9872) --
      (448.6302,390.9872) -- (448.6302,393.2735) -- (450.9164,393.9593) --
      (453.2026,391.6731) -- (455.0316,392.3590) -- (457.5465,392.3590) --
      (458.4610,394.8738) -- (464.7620,396.9528) -- (466.1338,396.2669) --
      (467.9628,392.1517) -- (469.1059,392.1517) -- (470.2490,394.2093) --
      (474.3642,394.8952) -- (478.0221,396.2669) -- (480.9942,397.1814) --
      (482.8232,396.2669) -- (483.5091,393.7521) -- (487.8529,393.7521) --
      (489.9105,394.6666) -- (492.6540,392.6090) -- (493.7971,392.6090) --
      (494.4830,394.2093) -- (498.5982,394.2093) -- (500.1985,392.1517) --
      (502.0275,392.6090) -- (504.0851,395.1238) -- (507.2858,396.9528) --
      (510.4866,397.8673) -- (512.4277,398.9862) -- (512.0386,361.7692) --
      (510.6668,350.7952) -- (510.5063,341.9229) -- (509.0665,335.3852) --
      (508.2883,328.2055) -- (508.2202,324.3893) -- (496.0833,324.7081) --
      (449.6733,324.2508) -- (404.6344,322.1932) -- (380.3432,320.8215) -- cycle;

    \path[USA map/state, USA map/TX, local bounding box=TX] (361.4642,330.5736) -- (384.1550,331.6595) -- (415.2477,332.8026) --
      (412.9131,356.2584) -- (412.6163,374.4120) -- (412.6844,376.4938) --
      (417.0283,380.3122) -- (419.0149,381.7593) -- (420.1991,381.1997) --
      (420.5725,379.3819) -- (421.7128,381.1856) -- (423.8245,381.2295) --
      (423.8215,379.7824) -- (425.4914,380.7496) -- (426.6301,381.1585) --
      (426.2708,385.1261) -- (430.3590,385.2197) -- (433.2843,386.4168) --
      (437.2391,386.9422) -- (439.6205,389.0212) -- (441.7446,386.9450) --
      (445.4695,387.5599) -- (447.6904,390.7849) -- (448.7654,391.1058) --
      (448.6049,393.0711) -- (450.8185,393.8634) -- (453.1487,391.8086) --
      (455.2817,392.4235) -- (457.5111,392.4590) -- (458.4441,394.8944) --
      (464.7722,397.0089) -- (466.3653,396.2420) -- (467.8547,392.0643) --
      (468.1955,392.0643) -- (469.1020,392.1458) -- (470.3310,394.2145) --
      (474.2609,394.8798) -- (477.5979,396.0026) -- (481.0235,397.1986) --
      (482.8641,396.2236) -- (483.5779,393.7088) -- (488.0311,393.7530) --
      (489.8398,394.6837) -- (492.6391,392.5772) -- (493.7427,392.6214) --
      (494.5937,394.2265) -- (498.6485,394.2265) -- (500.1673,392.1979) --
      (502.0347,392.6051) -- (503.9807,395.0084) -- (507.5013,397.0526) --
      (510.3601,397.8624) -- (511.8737,398.6622) -- (514.3204,400.6595) --
      (517.3634,399.3317) -- (520.0545,400.4706) -- (520.6183,406.5766) --
      (520.5785,416.2787) -- (521.2644,425.8127) -- (521.9666,429.4179) --
      (524.6419,433.8377) -- (525.5401,438.7884) -- (529.7560,444.3265) --
      (529.9520,447.4714) -- (530.6984,448.2572) -- (529.9683,456.6373) --
      (527.0962,461.6438) -- (528.6292,463.7967) -- (527.9991,466.1348) --
      (527.3296,473.5391) -- (525.8252,476.8771) -- (526.1201,480.3794) --
      (520.4552,481.9646) -- (510.5940,486.4911) -- (509.6240,488.4310) --
      (507.0374,490.3710) -- (504.9358,491.8259) -- (503.6425,492.6342) --
      (497.9844,497.9690) -- (495.2362,500.0706) -- (489.9014,503.3038) --
      (484.2433,505.7287) -- (477.9385,509.1236) -- (476.1603,510.5785) --
      (470.3405,514.1350) -- (466.9456,514.7817) -- (463.0658,520.2781) --
      (459.0243,520.6015) -- (458.0543,522.5414) -- (460.3176,524.4813) --
      (458.8626,529.9778) -- (457.5693,534.5043) -- (456.4377,538.3841) --
      (455.6294,542.9106) -- (456.4377,545.3355) -- (458.2160,552.2869) --
      (459.1859,558.4300) -- (460.9642,561.1782) -- (459.9942,562.6332) --
      (456.9227,564.5731) -- (451.2646,560.6933) -- (445.7681,559.5616) --
      (444.4748,560.0466) -- (441.2416,559.4000) -- (437.0384,556.3284) --
      (431.8653,555.1968) -- (424.2673,551.8019) -- (422.1657,547.9221) --
      (420.8724,541.4557) -- (417.6392,539.5157) -- (416.9925,537.2525) --
      (417.6392,536.6059) -- (417.9625,533.2110) -- (416.6692,532.5643) --
      (416.0226,531.5944) -- (417.3159,527.2295) -- (415.6993,524.9663) --
      (412.4660,523.6730) -- (409.0712,519.3082) -- (405.5146,512.6801) --
      (401.3115,510.0935) -- (401.4731,508.1536) -- (396.1383,495.8674) --
      (395.3300,491.6642) -- (393.5518,489.7243) -- (393.3901,488.2694) --
      (387.4087,482.9346) -- (384.8221,479.8630) -- (384.8221,478.7314) --
      (382.2355,476.6298) -- (375.4458,475.4982) -- (368.0094,474.8516) --
      (364.9379,472.5883) -- (360.4114,474.3666) -- (356.8548,475.8215) --
      (354.5916,479.0547) -- (353.6216,482.7729) -- (349.2568,488.9160) --
      (346.8319,491.3409) -- (344.2453,490.3710) -- (342.4671,489.2393) --
      (340.5271,488.5927) -- (336.6473,486.3295) -- (336.6473,485.6828) --
      (334.8690,483.7429) -- (329.6959,481.6413) -- (322.2595,473.8816) --
      (319.9963,469.1934) -- (319.9963,461.1104) -- (316.7631,454.6440) --
      (316.2781,451.8958) -- (314.6615,450.9258) -- (313.5298,448.8242) --
      (308.5184,446.7226) -- (307.2251,445.1060) -- (300.1120,437.1847) --
      (298.8187,433.9515) -- (294.1306,431.6882) -- (292.6756,427.3233) --
      (290.0890,424.4135) -- (288.1491,423.9285) -- (287.4999,419.2509) --
      (295.5018,419.9368) -- (324.5368,422.6802) -- (353.5718,424.2806) --
      (355.8054,404.8187) -- (359.6919,349.2637) -- (361.2923,330.5164) --
      (362.6641,330.5450)(461.6934,560.2077) -- (461.1276,553.0947) --
      (458.3794,545.9007) -- (457.8135,538.8685) -- (459.3493,530.6238) --
      (462.6634,523.7532) -- (466.1391,518.3375) -- (469.2915,514.7810) --
      (469.9381,515.0235) -- (465.1691,521.6516) -- (460.8043,528.1989) --
      (458.7835,534.8270) -- (458.4602,540.0001) -- (459.3493,546.1432) --
      (461.9359,553.3372) -- (462.4209,558.5103) -- (462.5825,559.9653) --
      (461.6934,560.2077) -- cycle;

    \path[USA map/state, USA map/NM, local bounding box=NM] (288.1526,424.0131) -- (287.3771,419.2650) -- (296.0209,419.7904) --
      (326.1927,422.7363) -- (353.4608,424.4262) -- (355.6761,405.7188) --
      (359.5335,349.8428) -- (361.2711,330.4536) -- (362.8428,330.5821) --
      (363.6683,319.4187) -- (259.6638,308.7828) -- (242.1664,429.2176) --
      (257.6271,431.2067) -- (258.9204,421.1838) -- (288.1525,424.0131) -- cycle;

    \path[USA map/state, USA map/KS, local bounding box=KS] (507.8806,324.3803) -- (495.2623,324.5847) -- (449.1732,324.1275) --
      (404.6158,322.0699) -- (379.9860,320.8124) -- (383.8798,256.2175) --
      (405.9633,256.8926) -- (446.2524,257.7340) -- (490.5536,258.7216) --
      (495.6493,258.7216) -- (497.8337,260.8840) -- (499.8513,260.8626) --
      (501.4916,261.8751) -- (501.4291,264.8843) -- (499.6001,266.6097) --
      (499.2679,268.8419) -- (501.1110,272.2442) -- (504.0633,275.4393) --
      (506.3907,277.0537) -- (507.6915,288.2945) -- (507.8806,324.3803) -- cycle;

    \path[USA map/state, USA map/NE, local bounding box=NE] (486.0979,240.7006) -- (489.3285,247.7205) -- (489.1999,250.0230) --
      (492.6591,255.5169) -- (495.3784,258.6692) -- (490.3289,258.6692) --
      (446.8463,257.7305) -- (406.0595,256.8402) -- (383.8072,256.0564) --
      (384.8800,234.7285) -- (352.5618,231.8083) -- (356.9056,187.7984) --
      (372.4519,188.8272) -- (392.5707,189.9703) -- (410.4033,191.1134) --
      (434.1801,192.2566) -- (444.9253,191.7993) -- (446.9829,194.0855) --
      (451.7840,197.0576) -- (452.9271,197.9721) -- (457.2709,196.6004) --
      (461.1575,196.1431) -- (463.9010,195.9145) -- (465.7300,197.2863) --
      (469.7874,198.8866) -- (472.7595,200.4870) -- (473.2167,202.0873) --
      (474.1312,204.1449) -- (475.9602,204.1449) -- (476.7582,204.1911) --
      (477.6524,208.8730) -- (480.5727,217.3409) -- (481.1452,221.0976) --
      (483.6687,224.8718) -- (484.2383,229.9860) -- (485.8455,234.2263) --
      (486.0979,240.7006) -- cycle;

    \path[USA map/state, USA map/SD, local bounding box=SD] (476.4469,204.0247) -- (476.3995,203.4438) -- (473.5038,198.5983) --
      (475.3640,193.8862) -- (476.8567,187.9997) -- (474.0748,185.9200) --
      (473.6897,183.1765) -- (474.4821,180.6222) -- (477.6706,180.6374) --
      (477.5475,175.6312) -- (477.2142,145.4570) -- (476.5965,141.6894) --
      (472.5242,138.3585) -- (471.5415,136.6815) -- (471.4790,135.0727) --
      (473.5012,133.5433) -- (475.0334,131.8776) -- (475.2783,129.2208) --
      (417.0212,127.6205) -- (362.2220,124.1714) -- (356.8968,187.8626) --
      (371.4870,188.7664) -- (391.4369,189.9720) -- (409.1799,190.9006) --
      (432.9567,192.2042) -- (444.9394,191.7795) -- (446.9057,194.0247) --
      (452.1003,197.2781) -- (452.8642,198.0008) -- (457.4057,196.5480) --
      (463.9462,195.9331) -- (465.6215,197.2694) -- (469.8260,198.8655) --
      (472.7711,200.5013) -- (473.1701,201.9851) -- (474.2096,204.2260) --
      (476.4469,204.0246) -- cycle;

    \path[USA map/state, USA map/ND, local bounding box=ND] (475.3053,128.9185) -- (474.6904,120.4848) -- (473.0134,113.6689) --
      (471.1219,100.6446) -- (470.6647,89.6576) -- (468.9252,86.5805) --
      (467.1686,81.3861) -- (467.1998,70.9418) -- (467.8232,67.1177) --
      (465.9891,61.6500) -- (437.3468,61.0859) -- (418.7559,60.4393) --
      (392.2436,59.1460) -- (369.2972,57.0121) -- (362.3040,124.1890) --
      (417.2362,127.5326) -- (475.3052,128.9185) -- cycle;

    \path[USA map/state, USA map/WY, local bounding box=WY] (360.3767,143.2759) -- (253.6341,129.8188) -- (239.5506,218.2768) --
      (352.8152,231.8623) -- (360.3767,143.2759) -- cycle;

    \path[USA map/state, USA map/MT, local bounding box=MT] (369.2095,56.9691) -- (338.5352,54.1613) -- (309.2746,50.6048) --
      (280.0141,46.5633) -- (247.6820,41.2285) -- (229.2527,37.8336) --
      (196.5291,30.9009) -- (192.0500,52.2484) -- (195.4794,59.7929) --
      (194.1076,64.3654) -- (195.9366,68.9378) -- (199.1374,70.3096) --
      (203.7582,81.0790) -- (206.4533,84.2555) -- (206.9105,85.3987) --
      (210.3399,86.5418) -- (210.7971,88.5994) -- (203.7098,106.2033) --
      (203.7098,108.7182) -- (206.2247,111.9189) -- (207.1391,111.9189) --
      (211.9402,108.9468) -- (212.6261,107.8037) -- (214.2264,108.4896) --
      (213.9978,113.7479) -- (216.7413,126.3221) -- (219.7134,128.8370) --
      (220.6279,129.5228) -- (222.4569,131.8091) -- (221.9996,135.2384) --
      (222.6855,138.6677) -- (223.8286,139.5822) -- (226.1148,137.2960) --
      (228.8583,137.2960) -- (232.0590,138.8964) -- (234.5739,137.9819) --
      (238.6891,137.9819) -- (242.3470,139.5822) -- (245.0905,139.1250) --
      (245.5477,136.1529) -- (248.5198,135.4670) -- (249.8916,136.8388) --
      (250.3488,140.0395) -- (251.7747,140.8741) -- (253.6616,129.8394) --
      (360.4073,143.2683) -- (369.2095,56.9691) -- cycle;

    \path[USA map/state, USA map/CO, local bounding box=CO] (380.0324,320.9646) -- (384.9357,234.6396) -- (271.5471,221.9956) --
      (259.3333,309.9348) -- (380.0324,320.9646) -- cycle;

    \path[USA map/state, USA map/ID, local bounding box=ID] (148.4788,176.4839) -- (157.2497,141.2632) -- (158.6214,137.0337) --
      (161.1363,131.0895) -- (159.8788,128.8033) -- (157.3640,128.9176) --
      (156.5638,127.8888) -- (157.0211,126.7457) -- (157.3640,123.6593) --
      (161.8221,118.1723) -- (163.6511,117.7151) -- (164.7942,116.5720) --
      (165.3658,113.3713) -- (166.2803,112.6854) -- (170.1668,106.8555) --
      (174.0534,102.5117) -- (174.2821,98.7394) -- (170.8527,96.1103) --
      (169.3172,91.7093) -- (182.9421,28.3676) -- (196.4597,30.8957) --
      (192.0516,52.2787) -- (195.6119,59.7641) -- (194.0308,64.4249) --
      (196.0007,69.0661) -- (199.1389,70.3213) -- (202.9742,79.8779) --
      (206.4869,84.3151) -- (206.9942,85.4582) -- (210.3351,86.6013) --
      (210.7040,88.6984) -- (203.7330,106.0745) -- (203.5678,108.6404) --
      (206.1989,111.9621) -- (207.1040,111.9132) -- (212.0153,108.8876) --
      (212.6927,107.7926) -- (214.2550,108.4515) -- (213.9766,113.8052) --
      (216.7158,126.3879) -- (220.6337,129.5658) -- (222.3148,131.7313) --
      (221.5982,135.8151) -- (222.6644,138.6226) -- (223.7261,139.7138) --
      (226.2054,137.3624) -- (229.0535,137.4113) -- (231.9728,138.7465) --
      (234.7528,138.0646) -- (238.5471,137.9041) -- (242.5260,139.5045) --
      (245.2694,139.2077) -- (245.7662,136.1704) -- (248.6988,135.4056) --
      (249.9589,136.9215) -- (250.3999,139.8664) -- (251.8242,141.0797) --
      (243.4382,194.6883) .. controls (243.4382,194.6883) and (155.4722,177.9877) ..
      (148.4788,176.4840) -- cycle;

    \path[USA map/state, USA map/UT, local bounding box=UT] (259.4984,310.1051) -- (175.7493,298.2328) -- (196.3369,185.6915) --
      (243.1173,194.4366) -- (241.6325,205.0670) -- (239.3208,218.2397) --
      (247.1285,219.1681) -- (263.5350,220.9729) -- (271.7460,221.8285) --
      (259.4984,310.1051) -- cycle;

    \path[USA map/state, USA map/AZ, local bounding box=AZ] (144.9112,382.6291) -- (142.2842,384.7874) -- (141.9609,386.2424) --
      (142.4459,387.2123) -- (161.3601,397.8819) -- (173.4847,405.4800) --
      (188.1958,414.0480) -- (205.0084,424.0709) -- (217.2946,426.4958) --
      (242.2458,429.2007) -- (259.5014,310.0737) -- (175.7658,298.1564) --
      (172.6734,314.5689) -- (171.0671,314.5842) -- (169.3524,317.2133) --
      (166.8376,317.0990) -- (165.5802,314.3556) -- (162.8367,314.0126) --
      (161.9222,312.8695) -- (161.0077,312.8695) -- (160.0932,313.4411) --
      (158.1499,314.4699) -- (158.0356,321.4429) -- (157.8070,323.1575) --
      (157.2354,335.7318) -- (155.7494,337.9037) -- (155.1778,341.2187) --
      (157.9213,346.1341) -- (159.1787,351.9640) -- (159.9789,352.9928) --
      (161.0077,353.5643) -- (160.8934,355.8505) -- (159.2930,357.2223) --
      (155.8637,358.9370) -- (153.9204,360.8803) -- (152.4344,364.5382) --
      (151.8628,369.4536) -- (149.0050,372.1971) -- (146.9474,372.8829) --
      (147.0831,373.7128) -- (146.6259,375.4275) -- (147.0831,376.2277) --
      (150.7411,376.7992) -- (150.1695,379.5427) -- (148.6835,381.7146) --
      (144.9112,382.6291) -- cycle;

    \path[USA map/state, USA map/NV, local bounding box=NV] (196.3927,185.5755) -- (172.7538,314.3983) -- (170.9216,314.7474) --
      (169.3488,317.1536) -- (166.9759,317.1643) -- (165.5039,314.4208) --
      (162.8855,314.0424) -- (162.1145,312.9348) -- (161.0767,312.8808) --
      (158.2983,314.5251) -- (157.9881,321.3106) -- (157.6260,327.0877) --
      (157.2774,335.6805) -- (155.8303,337.7697) -- (153.3914,336.6957) --
      (84.3115,232.4945) -- (103.3006,164.9096) -- (196.3927,185.5756) -- cycle;

    \path[USA map/state, USA map/OR, local bounding box=OR] (148.7218,175.5315) -- (157.5715,140.7300) -- (158.6223,136.5005) --
      (160.9767,130.8773) -- (160.3612,129.7144) -- (157.8463,129.6682) --
      (156.5647,127.9975) -- (157.0220,126.5334) -- (157.5254,123.2865) --
      (161.9835,117.7996) -- (163.8125,116.7004) -- (164.9556,115.5573) --
      (166.4417,111.9917) -- (170.4887,106.3223) -- (174.0543,102.4599) --
      (174.2830,99.0086) -- (171.0141,96.5399) -- (169.2307,91.8973) --
      (156.5669,88.2853) -- (141.4778,84.7417) -- (126.0458,84.8560) --
      (125.5886,83.4842) -- (120.1016,85.5418) -- (115.6435,84.9703) --
      (113.2430,83.3699) -- (111.9855,84.0558) -- (107.2988,83.8272) --
      (105.5841,82.4554) -- (100.3258,80.3978) -- (99.5256,80.5121) --
      (95.1818,79.0261) -- (93.2385,80.8551) -- (87.0657,80.5121) --
      (81.1215,76.3969) -- (81.8073,75.5968) -- (82.0360,67.8236) --
      (79.7497,63.9370) -- (75.6345,63.3654) -- (74.9487,60.8506) --
      (72.5947,60.3840) -- (66.7962,62.4428) -- (64.5330,68.9092) --
      (61.2998,78.9322) -- (58.0665,85.3986) -- (53.0551,99.4631) --
      (46.5887,113.0425) -- (38.5056,125.6521) -- (36.5657,128.5619) --
      (35.7574,137.1299) -- (36.1435,149.2102) -- (148.7218,175.5315) -- cycle;

    \path[USA map/state, USA map/WA, local bounding box=WA] (102.0732,7.6118) -- (106.4381,9.0667) -- (116.1377,11.8149) --
      (124.7057,13.7549) -- (144.7516,19.4130) -- (167.7074,25.0711) --
      (182.9305,28.2783) -- (169.2981,91.8641) -- (156.8531,88.3388) --
      (141.3451,84.7681) -- (126.1158,84.8013) -- (125.6603,83.4566) --
      (120.0611,85.6359) -- (115.4656,84.8992) -- (113.3187,83.3151) --
      (112.0054,83.9731) -- (107.2698,83.8329) -- (105.5714,82.4832) --
      (100.3084,80.3709) -- (99.5734,80.5178) -- (95.1843,78.9934) --
      (93.2910,80.8108) -- (87.0251,80.5120) -- (81.0994,76.3863) --
      (81.8784,75.4536) -- (81.9996,67.7761) -- (79.7176,63.9364) --
      (75.6024,63.3294) -- (74.9250,60.8188) -- (72.6494,60.3618) --
      (69.0945,61.5924) -- (66.8313,58.3732) -- (67.1546,55.4633) --
      (69.9028,55.1400) -- (71.5194,51.0984) -- (68.9328,49.9668) --
      (69.0945,46.2486) -- (73.4593,45.6020) -- (70.7111,42.8538) --
      (69.2562,35.7407) -- (69.9028,32.8308) -- (69.9028,24.9094) --
      (68.1245,21.6762) -- (70.3878,12.2999) -- (72.4894,12.7849) --
      (74.9143,15.6948) -- (77.6625,18.2814) -- (80.8957,20.2213) --
      (85.4222,22.3229) -- (88.4938,22.9695) -- (91.4036,24.4245) --
      (94.7985,25.3944) -- (97.0618,25.2328) -- (97.0618,22.8079) --
      (98.3550,21.6762) -- (100.4566,20.3830) -- (100.7800,21.5146) --
      (101.1033,23.2928) -- (98.8400,23.7778) -- (98.5167,25.8794) --
      (100.2950,27.3344) -- (101.4266,29.7593) -- (102.0732,31.6992) --
      (103.5282,31.5375) -- (103.6898,30.2442) -- (102.7199,28.9510) --
      (102.2349,25.7177) -- (103.0432,23.9395) -- (102.3966,22.4845) --
      (102.3966,20.2213) -- (104.1748,16.6648) -- (103.0432,14.0782) --
      (100.6183,9.2284) -- (100.9416,8.4201) -- (102.0732,7.6118) --
      cycle(92.6165,13.5907) -- (94.6373,13.4291) -- (95.1223,14.8032) --
      (96.6581,13.1866) -- (99.0022,13.1866) -- (99.8105,14.7224) --
      (98.2747,16.4198) -- (98.9213,17.2281) -- (98.1939,19.2489) --
      (96.8197,19.6530) .. controls (96.8197,19.6530) and (95.9306,19.7339) ..
      (95.9306,19.4105) .. controls (95.9306,19.0872) and (97.3856,16.8240) ..
      (97.3856,16.8240) -- (95.6881,16.2581) -- (95.3648,17.7131) --
      (94.6373,18.3597) -- (93.1015,16.0965) -- (92.6165,13.5907) -- cycle;

    \path[USA map/state, USA map/CA, local bounding box=CA] (144.6944,382.1981) -- (148.6345,381.7095) -- (150.1206,379.6981) --
      (150.6651,376.7570) -- (147.1136,376.1669) -- (146.5994,375.4986) --
      (147.0769,373.4663) -- (146.9176,372.8767) -- (148.8402,372.2571) --
      (151.8830,369.4244) -- (152.4645,364.4293) -- (153.8444,361.0272) --
      (155.7877,358.8609) -- (159.3066,357.2712) -- (160.9610,355.6664) --
      (161.0297,353.5576) -- (160.0363,352.9776) -- (159.0132,351.9048) --
      (157.8580,346.0564) -- (155.1728,341.2263) -- (155.7386,337.7213) --
      (153.3190,336.6920) -- (84.2577,232.5136) -- (103.1598,164.9121) --
      (36.0799,149.2141) -- (34.5730,153.9474) -- (34.4114,161.3838) --
      (29.2382,173.1850) -- (26.1667,175.7715) -- (25.8434,176.9032) --
      (24.0651,177.7115) -- (22.6102,181.9146) -- (21.8019,185.1478) --
      (24.5501,189.3510) -- (26.1667,193.5542) -- (27.2983,197.1107) --
      (26.9750,203.5771) -- (25.1967,206.6487) -- (24.5501,212.4685) --
      (23.5801,216.1867) -- (25.3584,220.0665) -- (28.1066,224.5930) --
      (30.3699,229.4428) -- (31.6632,233.4843) -- (31.3398,236.7175) --
      (31.0165,237.2025) -- (31.0165,239.3041) -- (36.6746,245.6089) --
      (36.1896,248.0338) -- (35.5430,250.2970) -- (34.8964,252.2369) --
      (35.0580,260.4816) -- (37.1596,264.1998) -- (39.0995,266.7864) --
      (41.8478,267.2714) -- (42.8177,270.0196) -- (41.6861,273.5761) --
      (39.5845,275.1927) -- (38.4529,275.1927) -- (37.6446,279.0726) --
      (38.1296,281.9825) -- (41.3628,286.3473) -- (42.9794,291.6821) --
      (44.4343,296.3702) -- (45.7276,299.4418) -- (49.1225,305.2616) --
      (50.5774,307.8481) -- (51.0624,310.7580) -- (52.6790,311.7280) --
      (52.6790,314.1529) -- (51.8707,316.0928) -- (50.0924,323.2059) --
      (49.6075,325.1458) -- (52.0324,327.8940) -- (56.2355,328.3790) --
      (60.7620,330.1573) -- (64.6419,332.2589) -- (67.5518,332.2589) --
      (70.4617,335.3304) -- (73.0482,340.1802) -- (74.1799,342.4435) --
      (78.0597,344.5451) -- (82.9095,345.3534) -- (84.3645,347.4550) --
      (85.0111,350.6882) -- (83.5562,351.3348) -- (83.8795,352.3048) --
      (87.1127,353.1131) -- (89.8609,353.2747) -- (93.0208,351.5879) --
      (96.9007,355.7911) -- (97.7090,358.0543) -- (100.2955,362.2575) --
      (100.6189,365.4907) -- (100.6189,374.8670) -- (101.1038,376.6453) --
      (111.1268,378.1002) -- (130.8494,380.8484) -- (144.6944,382.1981) --
      cycle(56.5592,338.4814) -- (57.8525,340.0172) -- (57.6908,341.3105) --
      (54.4576,341.2297) -- (53.8918,340.0173) -- (53.2452,338.5623) --
      (56.5592,338.4815) -- cycle(58.4992,338.4814) -- (59.7116,337.8348) --
      (63.2682,339.9364) -- (66.3397,341.1488) -- (65.4506,341.7955) --
      (60.9241,341.5530) -- (59.3075,339.9364) -- (58.4992,338.4814) --
      cycle(79.1918,358.2849) -- (80.9700,360.6290) -- (81.7783,361.5990) --
      (83.3141,362.1648) -- (83.8799,360.7098) -- (82.9100,358.9316) --
      (80.2426,356.9108) -- (79.1918,357.0725) -- (79.1918,358.2849) --
      cycle(77.7368,366.9338) -- (79.5151,370.0862) -- (80.7275,372.0261) --
      (79.2726,372.2686) -- (77.9793,371.0562) .. controls (77.9793,371.0562) and
      (77.2518,369.6012) .. (77.2518,369.1970) .. controls (77.2518,368.7929) and
      (77.2518,367.0146) .. (77.2518,367.0146) -- (77.7368,366.9338) -- cycle;

    \end{scope}
}

\usepackage{pgfplotstable}

\pgfplotstableset{
    color cells/.style={
        col sep=comma,
       string type,
       postproc cell content/.code={%
            \pgfkeysalso{@cell content=\rule{0cm}{2ex}%
                \pgfmathtruncatemacro\number{##1}%
                \ifnum\number<0
                    \cellcolor{red!-##1}##1
              \else 
                  \cellcolor{blue!##1}##1
              \fi
              }},
        every head row/.style={after row=\hline},
        every first column/.style={column type/.add={}{|}},
        columns/x/.style={
            column name={},
            postproc cell content/.code={}
        }
    }
}

    \makeatletter
    \def\multilimits@{\bgroup
  \Let@
  \restore@math@cr
  \default@tag
 \baselineskip\fontdimen10 \scriptfont\tw@
 \advance\baselineskip\fontdimen12 \scriptfont\tw@
 \lineskip\thr@@\fontdimen8 \scriptfont\thr@@
 \lineskiplimit\lineskip
 \vbox\bgroup\ialign\bgroup\hfil$\m@th\scriptstyle{##}$\hfil\crcr}
    \def\Sb{_\multilimits@}
    \def\endSb{\crcr\egroup\egroup\egroup}
\makeatother

\makeatletter
\DeclareRobustCommand*\cal{\@fontswitch\relax\mathcal}
\makeatother

\begin{document}

\bibliographystyle{IEEEtran}

\newcommand{\papertitle}{
Blind Community Detection from Low-rank Excitations of a Graph Filter
}

\newcommand{\paperabstract}{
This paper considers a new framework to  detect communities in a graph from the observation of signals at its nodes. We model the observed signals as noisy outputs of an unknown network process, represented as a graph filter that is excited by 
a set of unknown low-rank inputs/excitations. {Application scenarios of this model include diffusion dynamics, pricing experiments, and opinion dynamics.} Rather than learning the precise parameters of the 
graph itself, we aim at retrieving the community structure directly. 
The paper shows that communities can be detected by applying a spectral method
to the covariance matrix of graph signals.
Our analysis indicates that the community detection performance
depends on a `low-pass' property of the graph filter. 
We also show that the performance can be improved via
a low-rank matrix { plus sparse} decomposition method when the 
latent parameter vectors are known. 
Numerical experiments demonstrate that our approach 
is effective. 
}


\ifplainver

    \title{\papertitle}

    \author{
    Hoi-To Wai, Santiago Segarra, Asuman E. Ozdaglar, Anna Scaglione, Ali Jadbabaie\thanks{H.-T.~Wai is with the Department of SEEM, The Chinese University of Hong Kong, Shatin, Hong Kong. E-mail: \texttt{htwai@se.cuhk.edu.hk}. S.~Segarra is with Department of ECE, Rice University, TX, USA. E-mail: \texttt{segarra@rice.edu}. A.~E.~Ozdaglar is with LIDS, Massachusetts Institute of Technology, MA, USA. A.~Jadbabaie is with IDSS, Massachusetts Institute of Technology, MA, USA. E-mails: \texttt{\{asuman,jadbabai\}@mit.edu}. A.~Scaglione is with School of ECEE, Arizona State University, Tempe, AZ, USA. E-mail: \texttt{Anna.Scaglione@asu.edu}}
    }
    \maketitle
        \begin{abstract}
\paperabstract \end{abstract}

\else
    \title{\papertitle}

    \ifconfver \else {\linespread{1.1} \rm \fi

    \author{Hoi-To Wai, Santiago Segarra, Asuman E. Ozdaglar, Anna Scaglione, Ali Jadbabaie
    \thanks{This work is supported by  NSF CCF-BSF 1714672, an MIT IDSS seed fund, and Spanish MINECO TEC2013-41604-R. A preliminary version of this work has been presented at ICASSP 2018 in Calgary, Canada \cite{our_icassp}.}
        \thanks{H.-T.~Wai is with the Department of SEEM, The Chinese University of Hong Kong, Shatin, Hong Kong. E-mail: \texttt{htwai@se.cuhk.edu.hk}. S.~Segarra is with Department of ECE, Rice University, TX, USA. E-mail: \texttt{segarra@rice.edu}. A.~E.~Ozdaglar is with LIDS, Massachusetts Institute of Technology, MA, USA. A.~Jadbabaie is with IDSS, Massachusetts Institute of Technology, MA, USA. E-mails: \texttt{\{asuman,jadbabai\}@mit.edu}. A.~Scaglione is with School of ECEE, Arizona State University, Tempe, AZ, USA. E-mail: \texttt{Anna.Scaglione@asu.edu}.}}

    \maketitle

    \ifconfver \else
        \begin{center} \vspace*{-2\baselineskip}
        \end{center}
    \fi

    \begin{abstract}
\paperabstract \end{abstract}
    \begin{IEEEkeywords}\vspace{-0.0cm}
    community detection, graph signal processing, low-rank matrix recovery, spectral clustering
    \end{IEEEkeywords}
    \ifconfver \else \IEEEpeerreviewmaketitle} \fi
 \fi
\ifconfver \else
    \ifplainver \else
        \newpage
\fi \fi

\section{Introduction} \label{sec:intro}
The emerging field of \emph{network science} and availability of \emph{big data} 
have motivated researchers to extend signal processing techniques to the analysis
of signals defined on graphs, {propelling} a new area of research referred to as
\emph{graph signal processing} (GSP) \cite{sandryhaila2013discrete,EmergingFieldGSP,ortega2017graph}. 
As opposed to signals on time defined on a regular topology, the properties of \emph{graph signals} are 
intimately related to the generally irregular topology of the graph where they are defined.
The goal of GSP is to develop mathematical tools to leverage 
this topological structure in order to enhance 
our understanding of graph signals. 
A suitable way to  capture the graph's structure is via the 
so-called \emph{graph shift operator} (GSO), which is a matrix that reflects
the local connectivity of the graph and is a generalization of the time shift or delay operator in classical discrete signal processing \cite{sandryhaila2013discrete}. 
Admissible choices for the GSO include the graph's adjacency matrix and the Laplacian matrix. 
When the GSO is known, the algebraic and spectral characteristics of a given graph signal can be analyzed in an analogous way 
as in time-series analysis \cite{sandryhaila2013discrete}. Furthermore, signal processing tools 
such as sampling \cite{marquessampling,SamplingKovacevic}, interpolation \cite{romeroreconstruction,segarra2015reconstruction} and filtering \cite{segarrafilters,isufifilters} can be extended to the realm of graph signals.

This paper considers an \emph{inverse} 
problem in GSP where our focus is to infer information about the 
{GSO (or the graph)} from the observed graph signals. Naturally, \emph{graph or network inference}
is relevant to network and data science, 
and has been studied extensively.
Classical methods are based on partial correlations \cite{GLasso2008},
Gaussian graphical models \cite{pavez_laplacian_inference_icassp16}, 
and structural equation models \cite{shen2017kernel}, 
among others. 
Recently, GSP-based methods for graph inference 
have emerged, which tackle the problem as a system identification 
task. They postulate that the unknown graph is a structure encoded 
in the observed signals and the signals are 
obtained from observations of network 
dynamical processes defined on the graph \cite{BazerqueGeneNetworks,wai2016rids}.
Different assumptions are put forth in the literature to aid the graph topology inference, such as smoothness of the observed signals \cite{donglaplacian,Kalofolias2016inference_smoothAISTATS16,
chepuriinference},
richness of the inputs to the network process \cite{segarra2017network,segarrablind,shafipourinference,egilmez2018graph}, 
and partial knowledge of the network process \cite{wai2016active,shen2017kernel}.

A drawback common to the prior GSP work on graph inference
\cite{donglaplacian,Kalofolias2016inference_smoothAISTATS16,
chepuriinference,segarra2017network,segarrablind,shafipourinference,egilmez2018graph} is that
they require the observed graph signals to be
\emph{full-rank}. Equivalently, the signals observed are results of a 
network dynamical process excited by a set of input signals that  span
a space with the same dimension as  the number of nodes in the graph.
Such assumption can be unnecessarily stringent for a number of applications,
especially when the graph contains a large number of nodes. 
For example, whenever  graph inference experiments can only be performed 
by exciting a few nodes on the graph (such as rumor spreading initiated by a 
small number of sources and the gene perturbation experiments in \cite{marbach2012wisdom}); 
or the amount of data collected is limited due to cost and time constraints. 

{
Oftentimes, inferring the entire graph structure is only the first step since the ultimate  goal is to obtain \emph{interpretable} information from the set of graph signals. 
To this end, a feature that is often sought in network science is the \emph{community structure} \cite{fortunato2010community} that offers a coarse description of graphs.
For this task, applying conventional methods necessitates a \emph{two-step} procedure  which comprises of a  graph learning and a community detection step.
This paper departs from the conventional methods by developing a \emph{direct analysis} framework to recover the communities based on the observation of graph signals. 
We consider a setting where the observations are graph signals modeled as the outputs of an unknown  network process represented by a graph filter. Such signal model can be applicable to observations from, e.g., diffusion dynamics, pricing experiments in consumer networks \cite{candogan2012optimal,candogan2018latent}, and DeGroot dynamics \cite{degroot1974reaching} with stubborn agents. 
In addition, unlike the prior works on graph learning, we allow the excitations to the graph filter to be \emph{low-rank}. This is a challenging yet practical scenario as we demonstrate later.

We propose and analyze two \emph{blind community detection} ({\sf BlindCD}) methods that do not require learning the graph topology nor knowing the dynamics governing the generation of graph signals explicitly. 
The first method applies spectral clustering on the sampled covariance matrix, which is akin to a common heuristics used in data clustering, e.g., \cite{zha2002spectral}. 
Here our contribution lies in showing when sampled covariance carries information about the communities. 
Under a mild assumption that the underlying graph filter is \emph{low-pass} with the GSO taken as graph Laplacian, we show that the covariance matrix of observed graph signals is a \emph{sketch} of the Laplacian matrix that retains coarse topological features of the graph, like communities. 
We quantify the \emph{suboptimality} of the {\sf BlindCD} method compared to the minimizer of a convex relaxation of the {\sf RatioCut} objective defined on the actual graph Laplacian. 
Our result helps in justifying the successful application of such heuristics on real data.
Furthermore, the theoretical analysis of {\sf BlindCD} identifies the key bottleneck in the spectral method applied to some GSP models.  
This leads to the development of our second method, called boosted {\sf BlindCD}. 
The method works under an additional assumption that the latent parameter vectors are available and \emph{boosts} the performance of the first method by leveraging a low-rank plus sparse structure in the linear transformation between excitations and observed graph signals. 
Performance bound is also analyzed for this method.

The organization of this paper is as follows. In Section~\ref{sec:prelim}, we introduce   notations by describing the graph model and a formal definition for communities on graph. Section~\ref{sec:model} presents the GSP signal model with real world examples. In Section~\ref{sec:bcd} and \ref{sec:bbcd}, we describe and analyze the proposed {\sf BlindCD} method and its boosted version. In Section~\ref{sec:num}, we present numerical results on synthetic and real data to validate our findings.}\vspace{.1cm}

\textbf{Notation} --- 
We use boldface  lower-case (\resp  upper-case) letters to denote vectors  (\resp matrices).
For a vector ${\bm x}$, the notation $x_i$ denotes its $i$th element
and we use $\| {\bm x} \|_2$ to denote the standard Euclidean norm.
For a matrix ${\bm X}$, the notation $X_{ij}$ denotes its $(i,j)$th element
whereas $[{\bm X}]_{i,:}$ denotes its $i$th row vector
and $[{\bm X}]_{{\cal I},:}$ denotes the collection of its row vectors in ${\cal I}$.
Also, ${\cal R}( {\bm X} ) \subseteq \RR^N$ denotes
the range space of ${\bm X} \in \RR^{N \times M}$. 
Moreover, $\|{\bm X}\|_{\rm F}$ (\resp $\| {\bm X} \|_2$) 
denotes the Frobenius norm (\resp spectral norm). 
For a symmetric matrix ${\bm E}$, $\beta_i({\bm E})$ denotes its $i$th largest eigenvalue. For a  matrix ${\bm M} \in \RR^{P \times N}$, $\sigma_i({\bm M})$ denotes its $i$th largest singular value and $[{\bm M}]_K$ denotes its rank $K$ approximation.
Moreover,  ${\bm M}$  admits the partition
${\bm M} = [ {\bm M}_K~{\bm M}_{N-K} ]$  where  ${\bm M}_K$ (\resp ${\bm M}_{N - K}$) denotes the matrix consisting of the 
\emph{left-most} $K$ (\resp \emph{right-most} $N-K$) columns of ${\bm M}$.
Similarly, ${\bm m} \in \RR^N$ is partitioned into ${\bm m} = [{\bm m}_K; {\bm m}_{N-K}]$, where ${\bm m}_K$ (\resp ${\bm m}_{N-K}$) consists of its \emph{top} $K$ (\resp \emph{bottom} $N-K$) elements. 
For any integer $K$, we denote $[K] \eqdef \{1,...,K\}$.


\section{Preliminaries}\label{sec:prelim}


\subsection{Graph Signal Processing}\label{sec:gsp}
Consider an undirected graph $G = (V,E, {\bm A})$ with $N$ nodes such that 
$V = [N] \eqdef \{1,...,N\}$ and $E \subseteq V \times V$ is the set of edges
where $(i,i) \notin E$  for all $i$. 
The graph $G$ is also associated with a symmetric and weighted adjacency matrix ${\bm A}
\in \RR_+^{N \times N}$ 
such that $A_{ij} = A_{ji} > 0$ if and only if $(i,j) \in E$.
The graph Laplacian matrix for $G$ is defined as 
${\bm L} \eqdef {\bm D} - {\bm A}$,
where ${\bm D} \eqdef {\rm Diag} ( {\bm A} {\bf 1} )$ is a diagonal matrix
containing the weighted degrees of $G$. 
As ${\bm L}$ is symmetric and positive semidefinite,
it admits the following eigendecomposition
\beq \label{eq:svd}
{\bm L} = {\bm V} \bm{\Lambda} {\bm V}^\top \eqs,
\eeq
where $\bm{\Lambda} = {\rm Diag}( [\lambda_1, ..., \lambda_N] )$
and $\lambda_i$ is sorted in \emph{ascending order} such that
$0 = \lambda_1 \leq \lambda_2 \leq \cdots \leq \lambda_N$. 

A graph signal is defined as a function on the nodes of $G$, 
$f : V \rightarrow \RR$, and can be equivalently represented as a vector ${\bm x} \eqdef
[ x_1,x_2,...,x_N ] \in \RR^N$, where $x_i$ is the signal value at the $i$th node.
The graph is endowed with a graph shift operator (GSO) 
that is set as the graph Laplacian ${\bm L}$. Note that 
it is also possible to define alternative GSOs such as
the adjacency matrix ${\bm A}$ and its normalized versions; 
see \cite{sandryhaila2013discrete} for an overview on the subject, { yet the analysis result in this paper may  differ slightly for the latter cases}.
Having defined the GSO, the graph Fourier transform (GFT) \cite{sandryhaila2013discrete}
of ${\bm x}$ is given by
\beq
\tilde{\bm x} \eqdef {\bm V}^\top {\bm x} \eqs.
\eeq
The vector $\tilde{\bm x}$ is called the \emph{frequency} domain representation of ${\bm x}$
with respect to (w.r.t.) the GSO ${\bm L}$ \cite{sandryhaila2013discrete,ortega2017graph}. 

The GSO can be used to define linear graph filters. These are linear graph
signal operators that can be expressed as matrix polynomials on ${\bm L}$:
\beq \label{eq:filsvd}
{\cal H}( {\bm L} ) \eqdef \sum_{ t = 0 }^{T_{\sf d}-1} h_{t} {\bm L}^t 
= {\bm V} \left( \sum_{t= 0}^{T_{\sf d}-1}  h_t \bm{\Lambda}^t \right) {\bm V}^\top \eqs,
\eeq
where $T_{\sf d}$ is the \emph{order} of the graph filter. 
Note that by the Cayley-Hamilton theorem, any matrix polynomial (even of infinite degree)
can be represented using the form \eqref{eq:filsvd} with $T \leq N$. 
For a given { excitation} graph signal ${\bm x} \in \RR^N$, the output 
of the filter is simply ${\bm y} = {\cal H} ({\bm L}) {\bm x}$, and
carries the classical interpretation of being a linear combination of shifted versions
of the input. 
The graph filter 
${\cal H}({\bm L})$ 
may also be represented by its frequency response $\tilde{\bm h}$, 
defined as 
\beq \label{eq:tildeh} \textstyle
\tilde{h}_i \eqdef h( \lambda_i ) = \sum_{t=0}^{T_{\sf d}-1} h_t \lambda_i^t \eqs.
\eeq
We denominate the polynomial 
$h( \lambda ) \eqdef \sum_{t=0}^{T_{\sf d}-1} h_t \lambda^t$
as the \emph{generating function} of the graph filter.
From \eqref{eq:filsvd} it follows that the frequency representations
of the input and the output of a filter are related by
\beq
\tilde{\bm x} = \tilde{\bm h} \odot \tilde{\bm z} \eqs,
\eeq
where $\odot$ denotes the element-wise product.
This is analogous to the convolution theorem for time signals.
{In Section~\ref{sec:model}, we utilize GSP to model the relationship between the observed data and the unknown graph $G$.}

\subsection{Community Structure and its Detection} \label{sec:cd}
{Intuitively, a community on the graph $G$ is a subset of nodes, ${\cal C}_k^\star \subseteq V$, that induces a densely connected subgraph while loosely connected with nodes not in ${\cal C}_k^\star$. 
To formally describe a community structure, in this paper we refer to the common notion of \emph{ratio-cut} \cite{fortunato2010community} that measures the total \emph{cut} weight across the boundary between a \emph{disjoint} partition of $G=(V,E, {\bm A})$.
In particular, for any disjoint $K$ partition of $V$, \ie $V = {\cal C}_1 \cup ... \cup {\cal C}_K$,  define the function:
\beq \label{eq:ratiocut}
{\sf RatioCut}( {\cal C}_1, ..., {\cal C}_K ) 
\eqdef 
\sum_{ k=1 }^K \frac{ 1 } { | {\cal C}_k | } \sum_{i \in {\cal C}_k} \sum_{j \notin {\cal C}_k} A_{ij}.
\eeq
Throughout this paper, we assume that there are $K$ \emph{non-overlapping communities} in $G$ as given by ${\cal C}_1^\star, ..., {\cal C}_K^\star$, where the latter is a minimizer to the ratio-cut function and it results in a small objective value. For instance,  
\beq
\delta^\star \eqdef {\sf RatioCut}( {\cal C}_1^\star, ..., {\cal C}_K^\star ) \leq {\sf RatioCut}( {\cal C}_1, ..., {\cal C}_K ) 
\eeq
where $\delta^\star \ll 1$ indicates that the graph has $K$ communities.

Having defined the above notion, the \emph{community detection} problem is solved by \emph{minimizing} \eqref{eq:ratiocut} with the given number of communities $K$ and  graph adjacency matrix ${\bm A}$. However, the ratio-cut minimization problem is combinatorial and difficult to solve. As such,}
a popular remedy is to apply a convex relaxation -- a method known as the \emph{spectral clustering} \cite{abbe2017community,vonluxvurgspectral}. 
To describe the method, let us define the left-$K$ eigenmatrix of the graph Laplacian ${\bm L}$ as
\beq
{\bm V}_K \eqdef \big( {\bm v}_1~{\bm v}_2~\cdots~{\bm v}_K \big) \in \RR^{N \times K} \eqs,
\eeq
where ${\bm v}_i$ is the $i$th eigenvector of ${\bm L}$ corresponding 
to the $i$th eigenvalue $\lambda_i$ [cf.~\eqref{eq:svd}]. 
The $K$-means method \cite{kmeans} is applied on the row vectors of ${\bm V}_K$,
which seeks a partition ${\cal C}_1,...,{\cal C}_K$ that minimizes the distance
of each row vector to their respective means. The spectral clustering minimizes
\beq \label{eq:obj}
F( {\cal C}_1, ..., {\cal C}_K ) \eqdef \sum_{k=1}^K \sum_{i \in {\cal C}_k} \Big\| 
{\bm v}_i^{\rm row}  - \frac{1}{|{\cal C}_k|} \sum_{j \in {\cal C}_k} {\bm v}_j^{\rm row} \Big\|_2^2 \eqs,
\eeq
where ${\bm v}_j^{\rm row} \eqdef [ {\bm V}_K ]_{j,:}$ is the $j$th row vector of ${\bm V}_K$. 

For general $K$, \cite{kumar2004simple}  proposed a polynomial-time algorithm
that finds an $(1+\epsilon)$-optimal solution, $\tilde{\cal C}_1, ..., \tilde{\cal C}_K$, 
to the $K$-means problem \eqref{eq:obj}
satisfying
\beq \label{eq:approx}
F( \tilde{\cal C}_1, ..., \tilde{\cal C}_K ) \leq (1 + \epsilon) \min_{ {\cal C}_1, ..., {\cal C}_K \subseteq V} F( {\cal C}_1, ..., {\cal C}_K ) \eqs,
\eeq
under some statistical assumptions on $\{ {\bm v}_i^{\rm row} \}_{i=1}^N$. {The spectral clustering method is shown to be effective both in theory and in practice. In particular, when $K=2$ and the graph of interest is drawn from a stochastic block model ({\sf SBM}) satisfying certain spectral gap conditions, the spectral method exactly recovers the ground truth clusters in the {\sf SBM} when $N \rightarrow \infty$ [which also gives a minimizer to \eqref{eq:ratiocut}], see  \cite{abbe2017community}.}


{
\section{Graph Signal Model} \label{sec:model}
Consider a graph signal ${\bm y}^\ell \in \RR^N$ defined on the graph $G$ described in Section~\ref{sec:gsp}. The graph signal is obtained by exciting the graph filter ${\cal H}( {\bm L} )$ with an excitation ${\bm x}^\ell  \in \RR^N$, 
\beq \label{eq:data_collect}
{\bm y}^\ell = {\cal H} ({\bm L}) {\bm x}^\ell + {\bm w}^\ell,~\ell = 1,...,L  \eqs,
\eeq
where ${\bm w}^\ell \in \RR^N$ includes both the modeling and measurement 
error in data collection. We assume that ${\bm w}^\ell$ is zero mean and sub-Gaussian with
$\EE[ {\bm w}^\ell ( {\bm w}^\ell)^\top ] = \sigma_w^2 {\bm I}$. 
Consider a low-rank excitation setting where $\{ {\bm x}^\ell \}_{\ell=1}^L$ belong to an $R$-dimensional subspace of $\RR^N$. Assume $K \leq R \ll N$, where $K$ is the number of communities specified in Section~\ref{sec:cd}. Let ${\bm B} \in \RR^{N \times R}$ and 
\beq \label{eq:lowrank_obs}
{\bm x}^\ell = {\bm B} {\bm z}^\ell ,
\eeq
where ${\bm z}^\ell \in \RR^R$ is a latent parameter vector controlling the excitation signal.
Under this model, the sampled covariance matrix of $\{ {\bm y}^\ell \}_{\ell=1}^L$ is low rank with at most rank $R$.
As mentioned, under such setting it is difficult to reconstruct ${\bm L}$ from $\{ {\bm y}^\ell \}_{\ell=1}^L$ using the existing methods \cite{donglaplacian,Kalofolias2016inference_smoothAISTATS16,
chepuriinference,segarra2017network,segarrablind,shafipourinference}.

Before discussing the proposed methods for inferring communities from $\{ {\bm y}^\ell \}_{\ell=1}^L$ in Section~\ref{sec:bcd} and \ref{sec:bbcd}, let us justify the model \eqref{eq:data_collect}, \eqref{eq:lowrank_obs} with three motivating examples. 


\subsection{Example 1:  Diffusion Dynamics} \label{sec:diff}
The first example describes graph signals resulting from a diffusion process. 
For example, this model is commonly applied to temperatures within a geographical region \cite{egilmez2018graph}. Under this model, each node in the graph of Section~\ref{sec:gsp} is a location and the weights $A_{ij} = A_{ji}$ represent the strengths of relative influence between $i$ and $j$ such that $\sum_{j=1}^N A_{ij} = 1$ for $i=1,...,N$.

The $\ell$th sample graph signal obtained is the result of a diffusion over $T$ steps, described as
\beq
\begin{split}
{\bm y}^\ell & = ((1-\alpha) {\bm I} + \alpha {\bm A})^T {\bm x}^\ell + {\bm w}^\ell \\
& = ( {\bm I} - \alpha {\bm L} )^T {\bm x}^\ell + {\bm w}^\ell,
\end{split}
\eeq
where $\alpha \in (0,1)$ is the speed of the diffusion process. As $( {\bm I} - \alpha {\bm L} )^T$ is a polynomial of the graph's Laplacian, we observe that ${\bm y}^\ell$ is an output of a graph filter \eqref{eq:data_collect}.

On the other hand, the excitation signal ${\bm x}^\ell$ may model the changes in temperature in the region due to a weather condition. The number of modes of temperature changes maybe limited, e.g., a typical hurricane in North America affects the east coast of the US. This effect can be captured by having 
a tall matrix ${\bm B}$, \ie the excitation lies in a low-dimensional space. The columns of ${\bm B}$ represents the potential modes on which weather conditions may affect the region. 

\subsection{Example 2: Pricing Experiments in Consumers' Game} \label{sec:price}
This example is concerned with graph signals obtained as the equilibrium consumption levels of a consumers' game subject to pricing experiments~\cite{candogan2012optimal,candogan2018latent}. 
Here, the graph described in Section~\ref{sec:gsp} represents a network of $N$ agents where ${A}_{ij} = A_{ji} \geq 0$ is the influence strength between agents $i$ and $j$. 
We assume that ${\bm A} {\bf 1} = c {\bf 1}$ such that each agent experiences the same
level of influence from the others.


It has been suggested in \cite{candogan2018latent} that conducting a set of pricing experiments and observing the equilibrium behavior of agents can unveil the influence network between agents. 
Let $\ell$ be the index of a pricing experiment. Agent $i$ chooses to  consume $y_i$ units 
of a product depending on (i) the price of the product $p_i^\ell$ and (ii) the consumption
levels of other agents who are neighbors of him/her in the network, weighted by the influence strength $A_{ij}$. The consumption
level $y_i$ is determined by maximizing the utility
\beq \label{eq:util}
u_i ( y_i, {\bm y}_{-i}, p_i^\ell ) \eqdef a y_i - \frac{b}{2} y_i^2 + y_i \sum_{j=1}^N A_{ij} y_j - p_i^\ell  y_i \eqs,
\eeq
where ${\bm y}_{-i} \eqdef (y_j)_{j \neq i}$ and $a, b \geq 0$ are model parameters. 
As the utility function above depends on ${\bm y}_{-i}$, the equilibrium consumption level for the $i$th agent can be solved by the following network game:
\beq \label{eq:game} \textstyle
y_i^\ell = \argmax_{ y_i \in \RR_+ }~u_i ( y_i, {\bm y}_{-i}^\ell, p_i^\ell ),~~\forall~i \eqs.
\eeq
Under the conditions that $b > \sum_{j=1}^N A_{ij}$ and $a > p_i^\ell$,
the equilibrium to the above game is unique \cite{candogan2012optimal} and it satisfies 
\beq \label{eq:equi}
{\bm y}^\ell = ( b {\bm I} - {\bm A} )^{-1} (a {\bf 1} - {\bm p}^\ell ) \eqs.
\eeq
Removing the mean from ${\bm y}^\ell$ gives the graph signal:
\beq \label{eq:equi_mean} \textstyle
\tilde{\bm y}^\ell \eqdef  \frac{1}{L} \sum_{\tau=1}^L {\bm y}^\tau - {\bm y}^\ell 
= (b {\bm I} - {\bm A})^{-1} \tilde{\bm p}^\ell \eqs,
\eeq 
where $\tilde{\bm p}^\ell \eqdef {\bm p}^\ell - (1/L)\sum_{l=1}^L {\bm p}^l$ can be interpreted as a vector of discounts to agents
during the $\ell$th pricing experiment. 

In fact, \eqref{eq:equi_mean} can be interpreted as 
a filtered graph signal as in \eqref{eq:data_collect} by recognizing 
$\tilde{\bm p}^\ell$ as the excitation signal and 
$\tilde{\bm y}^\ell$ as the observed graph signal. 
Since $b > c$ and ${\bm A} {\bf 1} = c{\bf 1}$,  
\beq
\begin{split}
& ( b {\bm I}- {\bm A} )^{-1} 
= \frac{1}{b-c} \sum_{t=0}^\infty \big( \frac{1}{b-c} {\bm L} \big)^t 
\eqs,
\end{split}
\eeq
which is a matrix polynomial in ${\bm L}$.
This shows that the linear operator $( b {\bm I} - {\bm A} )^{-1}$ is indeed
a graph filter. 

Next, we study the types of discounts offered in the pricing experiment. 
A practical case is that due to the limitation of market, the pricing experiments only control the prices on $R$ agents, while the prices of the rest are unchanged
across experiments. This gives rise to a low-rank structure
for the excitation signal.
Note that $\tilde{\bm p}^\ell = {\bm B} {\bm z}^\ell$ holds with
\beq \label{eq:b_consume}
[ {\bm B} ]_{ {\cal I}, : } = {\bm I}, [{\bm B}]_{[N] \setminus {\cal I},:} = {\bm 0} \eqs,
\eeq
where ${\cal I} \subset [N]$ is the index set
of $R$ agents whom prices are controlled,
and ${\bm z}^\ell \in \RR^R$ is simply a vector of the price  variations from the mean. The latter can be assumed as known in a controlled experiment setting.
The discount offered in the $\ell$th experiment is a special case of low-rank excitation.

\subsection{Example 3: DeGroot Dynamics with Stubborn Agents} \label{sec:degroot}
The last example is related to a social network with $N$ agents where the graph signals are opinions sampled from the agents on different topics, e.g., votes casted by  Senators on different topics \cite{wu2018estimating}.
The network is represented by a \emph{directed} graph $G = (V,E, {\bm A})$ such 
that $A_{ij} \geq 0$ captures the amount of `trust' that agent $i$ has on agent $j$. 
The agents are influenced by $R$ \emph{stubborn} agents
in the sense that their opinions are not influenced by the others \cite{acemoglu2010spread,yildiz2013binary,jia2015opinion}. 

Consider the discussions on the $\ell$th topic, the agents exchange opinions 
according to the DeGroot opinion dynamics \cite{degroot1974reaching} --- 
let $y_i^\ell (\tau)$ (\resp $z_j^\ell$)
be the opinion of the $i$th agent (\resp $j$th stubborn agent) at time $\tau$, 
e.g., $y_i^\ell (\tau) \in [0,1]$ represents the probability for agent $i$ to agree, we have 
\beq \label{eq:stubborn}
{\bm y}^\ell(\tau+1) = {\bm A} {\bm y}^\ell(\tau) + {\bm B} {\bm z}^\ell,~\tau=1,2,... \eqs,
\eeq
where ${\bm B} \in \RR^{N \times R}$ is a weight matrix 
describing the bipartite graph that connects the stubborn agents to the agents in $G$.
We assume that the concatenated matrix is stochastic 
such that  $[ {\bm A}, {\bm B} ] {\bf 1} = {\bf 1}$
and therefore the updated opinions are  
convex combinations of the opinions of neighboring
agents;
see \cite{wai2016active} for detailed description on the model. 
Note that it is possible to estimate the latent parameter ${\bm z}^\ell$ as well since the latter represents the opinions of stubborn agents.

Let us focus on
the \emph{steady-state} opinions, \ie the opinions when $\tau \rightarrow \infty$.
Under mild assumptions, it holds  \cite{wai2016active,yildiz2010computing}
\beq \label{eq:steady_stub}
\begin{split}
{\bm y}^\ell&  \eqdef \lim_{\tau \rightarrow \infty} {\bm y}^\ell(\tau) = ({\bm I} - {\bm A})^{-1} {\bm B} {\bm z}^\ell \\
& = ( {\rm Diag}( {\bf 1} - {\bm A} {\bf 1} ) + {\bm L} )^{-1} {\bm B} {\bm z}^\ell \\
& \approx c^{-1} ({\bm I} + c^{-1} {\bm L} )^{-1} {\bm B} {\bm z}^\ell \eqs,
\end{split}
\eeq
where the last approximation holds when
there exists $c > 0$ such that $c {\bf 1} \approx {\bf 1} - {\bm A} {\bf 1} = {\bm B} {\bf 1}$, e.g., when the out-degrees of the stubborn agents are almost the same. 
From~\eqref{eq:steady_stub} it follows that the steady state opinions is a special case of \eqref{eq:data_collect}, \eqref{eq:lowrank_obs}.}

\section{Blind Community Detection} \label{sec:bcd}
{ 
We study the \emph{blind community detection} problem, whose goal is to infer a disjoint partition of the nodes $V$ that corresponds to the communities, ${\cal C}_1^\star, ..., {\cal C}_K^\star$, in the graph $G = (V,E,{\bm A})$ as defined in Section~\ref{sec:cd}, when the only given inputs are the observed graph signals $\{ {\bm y}^\ell \}_{\ell=1}^L$ [cf.~\eqref{eq:data_collect}, \eqref{eq:lowrank_obs}] and the desired number of communities $K$.
Only in this section, we assume that the latent parameter vector ${\bm z}^\ell$ is a random, zero-mean, sub-Gaussian vector with $\EE[ {\bm z}^\ell ({\bm z}^\ell )^\top ] = {\bm I}$.}
The covariance matrix of ${\bm y}^\ell$ is given by
\beq \label{eq:cy}
\begin{split}
{\bm C}_y & \eqdef \EE[ {\bm y}^\ell ( {\bm y}^\ell )^\top ] = {\cal H}( {\bm L} ){\bm B} {\bm B}^\top {\cal H}^\top ({\bm L}) + \sigma_w^2 {\bm I} \eqs. 
\end{split}
\eeq
We also denote by $\overline{{\bm C}}_y \eqdef {\cal H}( {\bm L} ) {\bm B} {\bm B}^\top {\cal H}^\top ({\bm L})$ the covariance of ${\bm y}^\ell$ in the absence of measurement error. Observe that
\beq \label{eq:sketchB}
{\cal H}( {\bm L} ) {\bm B} 
= {\bm V} {\rm Diag} ( \tilde{\bm h} ) {\bm V}^\top {\bm B} \eqs,
\eeq
which is due to \eqref{eq:filsvd}, \eqref{eq:tildeh}.
We can interpret ${\cal H}( {\bm L} ) {\bm B}$ as a \emph{sketch} of the graph filter ${\cal H}({\bm L})$, where ${\bm B}$ is a {sketch matrix} { that compresses} the right dimension from $N$ to $R$. 

{To perform blind community detection based on $\{ {\bm y}^\ell \}_{\ell=1}^L$, let us gain intuition by considering the scenario when the noise is small ($\sigma_w^2 \approx 0$), the first $K$ elements in $\tilde{\bm h}$ are non-zero which have larger magnitudes than the rest of elements, and the columns of ${\bm B}$  span  the same space as 
${\rm span} \{ {\bm v}_{1}, ..., {\bm v}_K \}$.}
In this scenario, from \eqref{eq:cy} and \eqref{eq:sketchB}, we observe that
${\bm V}_K$ can be estimated (up to a rotation) by simply obtaining the top-$K$
eigenvectors of ${\bm C}_y$. 
{ This intuition suggests that we can detect communities by applying spectral clustering on ${\bm C}_y$, similar to the one applied to the Laplacian ${\bm L}$ in Section~\ref{sec:cd}. 
The proposed {\sf BlindCD} method is summarized in Algorithm~\ref{alg:bcd}.

The computation complexity of {\sf BlindCD} is dominated by   covariance estimation and eigenvalue decomposition in Line~\ref{cd:sample}-\ref{cd:evd}, which costs ${\cal O}( N^2 (L+K) )$ FLOPS for large $N$.	
This is significantly less complex than a two-step procedure using a sophisticated graph learning step, e.g., \cite{segarra2017network}. In addition to estimating the covariance, the latter requires a linear program with ${\cal O}(N^2)$ variables and constraints. This learning step entails a total complexity of ${\cal O}( N^2L + N^7 \log \epsilon_{\sf acc}^{-1} )$ FLOPS with the interior point method in \cite{ben2001lectures}\footnote{In practice, the said linear program can be solved efficiently with a tailor-made solver such as \cite{shafi2019}.}, where $\epsilon_{\sf acc}>0$ is the accuracy.

Similar method to the {\sf BlindCD} method have been proposed in the data clustering literature \cite{zha2002spectral}, offering a simple interpretation of ${\bm C}_y$ as the similarity graph between nodes. We provide a different interpretation here. Precisely, we view ${\bm C}_y$ as a spectral sketch of the Laplacian ${\bm L}$ and analyze the performance of {\sf BlindCD} as an indirect algorithm to \emph{approximately} find the  ground truth communities in ${\bm L}$.}

\algsetup{indent=1em}
\begin{algorithm}[t]
	\caption{Blind Community Detection ({\sf BlindCD}).}\label{alg:bcd}
	\begin{algorithmic}[1]
		\STATE \textbf{Input}: Graph signals $\{{\bm y}^\ell\}_{\ell=1}^L$; desired number of communities $K$.
		\STATE \label{cd:sample}  Compute the sample covariance $\widehat{\bm C}_y$ as
		\beq \label{eq:chat} \textstyle
		\widehat{\bm C}_y = (1/L) \sum_{\ell=1}^L {\bm y}^\ell ({\bm y}^\ell )^\top \eqs.\vspace{-.4cm}
		\eeq
		\STATE \label{cd:evd} Find the top-$K$ eigenvectors of $\widehat{\bm C}_y$ (with the eigenvalues sorted in \emph{descending} order). Denote the set of eigenvectors as $\widehat{\bm V}_K \in \RR^{N \times K}$.
		\STATE \label{cd:kmeans} Apply the $K$-means method, which seeks to optimize \vspace{-.1cm}
		\beq \label{eq:obj_k}
		\min_{ {\cal C}_1, ..., {\cal C}_K \subseteq V }~ \sum_{k=1}^K \sum_{i \in {\cal C}_k} \Big\| {\hat{\bm v}_i}^{\rm row} - \frac{1}{|{\cal C}_k|} \sum_{j \in {\cal C}_k} {\hat{\bm v}_j}^{\rm row} \Big\|_2^2 \eqs, \vspace{-.1cm}
		\eeq 
		where $\hat{\bm v}_i^{\rm row} \eqdef [\widehat{\bm V}_K]_{i,:} \in \RR^K$.
		\STATE \textbf{Output}: $K$ communities $\hat{\cal C}_1, ..., \hat{\cal C}_K$.
	\end{algorithmic} 
\end{algorithm}


\subsection{Low-pass Graph Filters}\label{sec:lpf}
Following \eqref{eq:sketchB} and the ensuing discussion, 
the performance of {\sf BlindCD} depends on $\tilde{\bm h}$, the frequency
response of the graph filter.
In particular, a desirable situation would be one where $\tilde{\bm h}$ contains
only significant entries over the first $K$ elements; in this way,
the graph filter ${\cal H}({\bm L})$ is approximately rank $K$ and retains
all the eigenvectors required for spectral clustering. 
To quantify the above conditions, we formally introduce the notion of a 
\emph{low-pass graph filter} (LPGF) as follows.

\begin{Def} \label{def:lpgf}
A graph filter ${\cal H}( {\bm L})$ is a $(K,\eta)$-LPGF
if 
\beq \label{eq:def}
\eta \eqdef \frac{ \max \big\{ | \tilde{h}_{K+1} |, ...., |\tilde{h}_N| \big\} }{ 
\min \big\{ | \tilde{h}_1 |, ...., |\tilde{h}_K| \big\} } < 1 \eqs,
\eeq
where $\tilde{h}_i$ is defined in \eqref{eq:tildeh}.  
The LPGF is \emph{ideal} if $\eta = 0$. 
\end{Def}
Note that a small $\eta$ implies a `good' LPGF, 
since $\eta \ll 1$ implies that most of the energy is concentrated in the first $K$ 
frequency bins of the graph filter. 
In fact, as we show later in Section~\ref{sec:theory}, the low-pass coefficient $\eta$ 
plays an important role in the performance of {\sf BlindCD}.

We now survey a few graph filter designs that are LPGF
and comment on their low-pass 
coefficients $\eta$. 
\begin{Exa} \label{ex:fil1}
Consider the filter order $T_{\sf d} < \infty$ and
\beq
{\cal H}_1 ({\bm L}) = ( {\bm I} - \alpha {\bm L} )^{T_{\sf d}-1},~\alpha \in (0,1/\lambda_N) .
\eeq
This filter models a discrete time diffusion process after $(T_{\sf d}-1)$ time instances 
on the graph
\cite{Tsitsiklis1984}. In particular, 
\beq\label{E:eta_example_1}
\eta_1 = \left( \frac{ 1 - \alpha \lambda_{K+1} }{ 1 - \alpha \lambda_K } \right)^{T_{\sf d}-1} \eqs.
\eeq
Observe that the coefficient $\eta_1$ improves exponentially with  $T_{\sf d}$.
\end{Exa}
\begin{Exa} \label{ex:fil3} Consider
\beq \label{eq:iir}
{\cal H}_2 ({\bm L}) = ( {\bm I} +  c^{-1} {\bm L} )^{-1} \eqs,
\eeq
for some $c > 0$.
This filter is analogous to a single-pole infinite impulse response (IIR) filter 
in classical signal processing. Its low-pass coefficient can be bounded as
\beq\label{E:eta_example_3}
\eta_2 = \frac{ 1 + c^{-1} \lambda_{K} }{ 1 + c^{-1} \lambda_{K+1} } = 1 - c^{-1} \!~ \frac{  \lambda_{K+1} - \lambda_K }{1 + c^{-1} \lambda_{K+1}}  \eqs.
\eeq
Observe that the coefficient $\eta_2 \approx 1$ for $\lambda_{K+1} \gg 1$ or $c \gg 1$.  
\end{Exa}
{Example~\ref{ex:fil1} is related to the diffusion dynamics in Section~\ref{sec:diff}, while Example~\ref{ex:fil3} is related to the consumers' game and opinion dynamics in Sections~\ref{sec:price} and~\ref{sec:degroot}.}
For further reference, an overview of graph filters and their relevant network processes can be found in \cite{sandryhaila2013discrete,ortega2017graph}. 

We conclude this subsection by characterizing the low-pass coefficient $\eta$
from the properties of the generating function $h(\lambda)$.
To simplify the analysis, we consider the class of filters such that $h(\lambda)$ 
satisfies the following assumption.
\begin{Assumption} \label{ass:gen}
The generating 
function $h(\lambda)$ is non-negative and non-increasing for all $\lambda \geq 0$.
\end{Assumption}
Note that Assumption~\ref{ass:gen} holds for 
the graph filters in Examples~1 and 2. 
The following observation gives a bound on $\eta$ using the 
first and second order derivatives of $h(\lambda)$.
\begin{Observation} \label{obs:gf}
Suppose that Assumption~\ref{ass:gen} holds and that $h(\lambda)$ is
$L_h$-smooth and $\mu_h$-strongly convex for $ \lambda \in [\lambda_K,\lambda_{K+1}]$,
where $0 \leq \mu_h \leq L_h$.
Then, the graph filter ${\cal H}({\bm L})$ is a $(K,\eta)$-LPFG with
\beq \label{eq:obs1}
\begin{split}
& \eta \leq 1 - \frac{ 1 }{\tilde{h}_K} \Big( \frac{\mu_h}{2} \Delta \lambda_K^2 -  h'(\lambda_{K+1}) \Delta \lambda_K \Big) \eqs, \\
& \eta \geq 1 - \frac{ 1 }{\tilde{h}_K} \Big( \frac{L_h}{2} \Delta \lambda_K^2 - h'(\lambda_{K+1}) \Delta \lambda_K \Big) \eqs,
\end{split}
\eeq
where $\Delta \lambda_K \eqdef \lambda_{K+1} - \lambda_K$ is the spectral gap of ${\bm L}$.
\end{Observation}
The observation can be verified using the definitions of $L_h$-smooth and $\mu_h$-strongly convex 
functions \cite{bertsekas1999nonlinear}.
Note that Assumption~\ref{ass:gen} implies that $h(\lambda)$ is convex
and the derivative $h'(\lambda_{K+1})$ is non-positive.  
Consequently, the upper bound on $\eta$ depends on the spectral gap 
$\Delta \lambda_K$ and the magnitude of $\tilde{h}_K$. 
In particular, {for a \emph{constant spectral gap}}, a small $\tilde{h}_K$ leads to $\eta \approx 0$
and thus a good LPGF.




\subsection{Performance Analysis} \label{sec:theory}
{ This subsection shows that under the GSP model \eqref{eq:data_collect}, \eqref{eq:lowrank_obs} and using Definition~\ref{def:lpgf}, we can bound the `suboptimality' of the communities obtained by {\sf BlindCD} compared to the `optimal' ones found using spectral clustering on ${\bm L}$ [cf.~Section~\ref{sec:cd}]. Together with recent advances in the theoretical analysis of spectral clustering \cite{abbe2017community}, this result allows us to quantify the accuracy of {\sf BlindCD} to perform \emph{blind community detection} and provides new insights on how to improve its performance.} 

To { proceed}, first let us take the $K$-means objective function $F(\cdot)$ in \eqref{eq:obj} constructed from eigenvectors of ${\bm L}$ as our performance metric. Let us denote 
\beq \label{eq:fstar}
F^\star \eqdef \min_{{\cal C}_1,..., {\cal C}_K \subseteq V} F( {\cal C}_1,..., {\cal C}_K ) \eqs
\eeq
as the optimal objective value. 
Furthermore, $\widehat{\bm C}_y$ is the sampled covariance of $\{ {\bm y}^\ell \}_{\ell=1}^L$ and $\overline{\bm C}_y$ is the covariance of ${\bm y}^\ell$ in the absence of noise.
The ensuing performance guarantee follows:\begin{Theorem}\label{T:main}
	Under the following conditions: 
	\begin{enumerate}
	\item ${\cal H}({\bm L})$ is a $(K, \eta)$-LPGF [cf.~Definition~\ref{def:lpgf}],
	\item $\mathrm{rank}({\bm V}_K \mathrm{diag}(\tilde{\bm h}_K) {\bm V}_K^\top {\bm B} {\bm Q}_K)= K$, where ${\bm Q}_K$ is the top-$K$ right singular vector of ${\cal H}({\bm L}){\bm B}$.
	\item $\mathrm{rank}({\cal H}({\bm L}) {\bm B} ) \geq K$,
        \item There exists $\delta > 0$ such that 
        \beq \label{eq:thm_5}
        \delta \eqdef \beta_K(\overline{\bm C}_y) - \beta_{K+1}(\overline{\bm C}_y) - \| \widehat{\bm C}_y - \overline{\bm C}_y \|_2 > 0 \eqs,
        \eeq 
where $\beta_K( \overline{\bm C}_y )$ is the $K$th largest eigenvalue of 
        $\overline{\bm C}_y$. 
        \end{enumerate}
	{For any $\epsilon > 0$, if the partition $\hat{\cal C}_1,...\hat{\cal C}_K$ found by {\sf BlindCD} is a $(1+\epsilon)$-optimal solution\footnote{This means that the objective value obtained is at most $(1+\epsilon)$ times the optimal value. See \cite{kumar2004simple} for a polynomial-time algorithm achieving this.} to problem \eqref{eq:obj_k}}, then, 
	\beq \label{E:main_theorem}
	\begin{split}
	& \sqrt{F( \hat{\cal C}_1, ..., \hat{\cal C}_K )}  -  \sqrt{(1+\epsilon) F^\star} \\
	& \leq (2+\epsilon) \sqrt{2K} \left( \sqrt{ \frac{\gamma^2}{1 + \gamma^2} } + \frac{  \| \widehat{\bm C}_y - \overline{\bm C}_y \|_2 }{\delta} \right) \eqs,
	 \end{split}
	\eeq 
	where $\gamma$ is bounded by
	\begin{equation}\label{E:gamma}
		\gamma \, \leq \, \eta  \!~ \| {\bm V}_{N-K}^\top {\bm B} {\bm Q}_K \|_2 \!~ \| ({\bm V}_{K}^\top {\bm B} {\bm Q}_K)^{-1} \|_2 \eqs.
	\end{equation}
\end{Theorem}
The proof (inspired by \cite{boutsidis2015spectral_a}, also see \cite{tremblay2016compressive}) can be found in Appendix~\ref{app:thm1}.
{
Condition 1) requires that the graph filter involved is an LPGF. This natural requisite imposes that the frequency response must be higher for those eigenvectors that capture the community structure in the graph.
Conditions 2) and 3) are technical requirements implying that the rank $R$ of the excitation matrix ${\bm B}$ cannot be smaller than the number of clusters $K$ that we are trying to recover.
Lastly, condition 4) imposes a restriction on the distance between the true covariance $\overline{\bm C}_y$ and the observed one $\widehat{\bm C}_y$.  
This condition may be violated if the spectral gap $\beta_K(\overline{\bm C}_y) - \beta_{K+1}(\overline{\bm C}_y)$  is small or, relying on Lemma~\ref{lem:conc}, if the noise power $\sigma^2_w$ is large.}

Moreover, Eq.~\eqref{E:main_theorem} in Theorem~\ref{T:main} bounds the optimality gap for the communities found applying {\sf BlindCD} compared to $F^\star$ in \eqref{eq:fstar}. {We first observe that the performance decreases when the number of communities $K$ increases, which is natural.}
This bound 
consists of the sum of two contributions.
The first term is {a function of} $\gamma$, which in turn depends on the low-pass coefficient $\eta$ of the LPGF involved as well as {the alignment 
between the matrices ${\bm B} {\bm Q}_K$ and ${\bm V}_{N-K}$}. 
From \eqref{E:gamma}, the recovered communities are more accurate when: 1) the LPGF is close to ideal ($\eta \approx 0$) and 2)  the distortion induced by ${\bm B}$ on the relevant eigenvectors {${\bm V}_{K}$} is minimal.
The second term in \eqref{E:main_theorem} depends on the distance between $\widehat{\bm C}_y$ and $\overline{\bm C}_y$, capturing the combined effect of noise in the observations (via $\sigma_w^2$) as well as the finite sample size. 
To further control this term, if we define $\bm{\Delta} \eqdef \widehat{\bm C}_y - \overline{\bm C}_y$, the next result follows.
\begin{Lemma} \cite[Remark 5.6.3, Exercise 5.6.4]{vershynin2017high} \label{lem:conc}
Suppose that \emph{i)} 
${\bm y}^1,... , {\bm y}^L$ are independent, and 
\emph{ii)} they are bounded almost surely with $\| {\bm y}^\ell \|_2 \leq {\blue Y}$.
Let the effective rank of ${\bm C}_y$ 
be $r \eqdef {\rm Tr}({\bm C}_y) / \| {\bm C}_y \|_2$, 
then for every $c > 0$ with probability at least $1-c$, one has that
\beq
\| \bm{\Delta} \|_2 \leq \sigma_w + C \Big( \sqrt{\frac{{\blue Y}^2 r \log( N/c ) }{L}} + \frac{{\blue Y}^2 r \log( N/c ) }{L} \Big) \eqs,
\eeq
for some constant $C$ that is independent of $N,r,L,c$, and $\sigma_y$. 
\end{Lemma}
Condition ii) in Lemma~\ref{lem:conc} is satisfied if ${\bm y}^\ell$ is sub-Gaussian and
$N \gg 1$. 
From Lemma \ref{lem:conc} it follows that the error  
converges to $\sigma_w$ at the rate of ${\cal O}(\sqrt{r K^2 \log(N) /L})$. 
For our model, it can be verified that 
$r \approx R \ll N$, where $R$ is the rank of ${\bm B}$ 
and the sampling complexity 
is significantly reduced compared to a signal model with full-rank excitations.

In a nutshell, Theorem~\ref{T:main} illustrates the effects that the observation noise, the finite number of observations, 
and the low-pass structure of the filter have on the suboptimality of the communities obtained.
As
 discussed above, the low-pass coefficient $\eta$ plays an important role
in the performance of {\sf BlindCD}. While $\eta$ is determined by the dynamics that generates the graph signals $\{ {\bm y}^\ell \}_{\ell=1}^L$,
it is possible to improve this coefficient, as described in the next section.



\section{Boosted Blind Community Detection} \label{sec:bbcd}
{The performance analysis in the previous section shows that the performance of {\sf BlindCD} depends on the low-pass filter coefficient $\eta$. While it is impossible to change the graph filter that generates the data, this section presents a `boosting' technique that extracts an improved low-pass filtered component,  \ie one with a smaller  $\eta$, from the observed graph signals.} 
For the application of the boosting technique, 
we shall work with low-pass graph filters satisfying Assumption~\ref{ass:gen} 
and consider a data model where, 
{apart from the access to the graph signals ${\bm y}^\ell \in \RR^N$ 
we also have access to the latent parameter vector ${\bm z}^\ell \in \RR^R$ [cf.~\eqref{eq:data_collect}, \eqref{eq:lowrank_obs}]}. 
{This scenario can be justified in the example of pricing experiments [cf.~Section~\ref{sec:price}] when the price discounts are directly controlled by the seller attempting to estimate the network; or in the example of DeGroot dynamics [cf.~Section~\ref{sec:degroot}] where the latent parameter vectors are the opinions of the stubborn agents.}

First,
the input-output pairs $\{ {\bm z}^\ell, {\bm y}^\ell \}_{\ell=1}^L$ enable us
to estimate the $N \times R$ matrix ${\cal H}( {\bm L} ) {\bm B}$ via
the least square estimator
\beq \label{eq:lse} 
\bm{\mathcal{H}}^\star \in \argmin_{ \widehat{\bm{\mathcal{H}}} \in \RR^{N \times R} }~\frac{1}{L} 
\sum_{\ell=1}^L \left\| {\bm y}^\ell - \widehat{\bm{\mathcal{H}}} {\bm z}^\ell \right\|_2^2 \eqs,
\eeq
where the solution is unique when $L \geq R$ and $\{ {\bm z}^\ell \}_{\ell=1}^L$  spans $\RR^R$. 
Importantly, we note the decomposition:
\beq \label{eq:decom}
{\cal H}( {\bm L} ) {\bm B} = \widetilde{\cal H}({\bm L}) {\bm B}  + \tilde{h}_N {\bm B} \eqs,
\eeq
where  
\beq
\widetilde{\cal H}({\bm L}) \eqdef {\cal H}( {\bm L} ) - \tilde{h}_N {\bm I} 
\eeq
is a graph filter with the generating function $
\tilde{h}(\lambda) = h(\lambda) - \tilde{h}_N$. 
{ The graph filter $\widetilde{\cal H}({\bm L})$ is called a \emph{boosted 
LPGF} as it has a smaller low-pass coefficient, denoted by $\tilde{\eta}$, 
than the low-pass coefficient of the original ${\cal H}({\bm L})$.}
This can be seen since 
(i) the magnitude of the boosted $K$th frequency response is reduced
to $\tilde{h}_K - \tilde{h}_N$; (ii)
the first and second order derivatives of $\tilde{h}(\lambda)$ are the same
as $h(\lambda)$. Applying Observation~\ref{obs:gf} 
it follows that 
$\widetilde{\cal H}({\bm L})$ has a smaller low-pass coefficient $\tilde{\eta}$
by replacing $\tilde{h}_K$ by $\tilde{h}_K - \tilde{h}_N$ in \eqref{eq:obs1}.
Concretely, we observe the example. 
\begin{Exa} (Boosted single-pole IIR filter).
Consider  
\beq
{\cal H}_3 ({\bm L}) \eqdef {\cal H}_2({\bm L}) - (1+ c^{-1}\lambda_N)^{-1} {\bm I} \eqs,
\eeq 
where ${\cal H}_2({\bm L})$ was defined in \eqref{eq:iir}
and we note that $\tilde{h}_N = (1+c^{-1}\lambda_N)^{-1}$. We have
\beq \notag
\eta_3 = \frac{\lambda_N - \lambda_{K+1}}{\lambda_N - \lambda_K} \!~ \frac{ 1 + c^{-1} \lambda_{K} }{ 1 + c^{-1} \lambda_{K+1} } = \left( \frac{\lambda_N - \lambda_{K+1}}{\lambda_N - \lambda_K} \right) \eta_2 \eqs.
\eeq
{ It follows that} $\eta_3 \ll \eta_2$ whenever $\lambda_{K+1} \gg \lambda_K$.
\end{Exa} 

In general, the discussion above shows that it is possible to 
reduce the low-pass coefficient $\eta$ significantly
by adjusting the constant level of the frequency responses 
in graph filters. 
As a result, applying spectral clustering based on the top-$K$ left singular vectors 
of $\widetilde{\cal H}({\bm L}){\bm B}$ will return a more accurate
community detection result. 

In order to estimate $\widetilde{\cal H}({\bm L}){\bm B}$ from $\bm{\mathcal{H}}^*$ as in \eqref{eq:decom},
one needs, in principle, to have access to ${\bm B}$ and the frequency response
$\tilde{h}_N$.
However, our goal is to obtain a boosting effect in the absence of knowledge about ${\bm B}$ and $\tilde{h}_N$.
A key towards achieving this goal is to notice 
that $\widetilde{\cal H}({\bm L}) {\bm B}$ 
is close to a rank-$K$ matrix since $\widetilde{\cal H}({\bm L})$ 
has a small low-pass coefficient $\tilde{\eta}$. 
Hence, for $R > K$, it follows  from \eqref{eq:decom} that 
$\bm{\mathcal{H}}^\star$ can be decomposed into a low-rank matrix 
and a scaled version of the sketch matrix ${\bm B}$. 
This motivates us to consider the noisy matrix decomposition problem proposed
in \cite{agarwal2012noisy}:
\beq \label{eq:lowrank} 
\begin{array}{rl}
\ds \min_{ \widehat{\bm{\mathcal{S}}}, \widehat{\bm B} \in \RR^{N \times R} } & 
\ds \frac{1}{2} \| {\bm{\mathcal{H}}}^\star - \widehat{\bm{\mathcal{S}}} - \widehat{\bm B} \|_{\rm F}^2 + \kappa \| \widehat{\bm{\mathcal{S}}} \|_{\sigma,1} + \rho g \big( \widehat{\bm B} \big) \\
{\rm s.t.} & g^\star ( \widehat{\bm{\mathcal{S}}} ) \leq \alpha \eqs, \vspace{-.4cm}
\end{array}\vspace{.2cm}
\eeq
where $\| \widehat{\bm{\mathcal{S}}}\|_{\sigma,1}$ is the trace  norm 
of the matrix $\widehat{\bm{\mathcal{S}}}$, $\bm{\mathcal{H}}^\star$ is a solution to \eqref{eq:lse}, 
$\alpha, \kappa, \rho > 0$ are predefined parameters, 
$g(\cdot)$ is a decomposable regularizer of $\widehat{\bm B}$,
which is a norm chosen according to the prior knowledge on the unknown
sketch matrix ${\bm B}$ and $g^\star(\cdot)$ is its dual norm.
A few examples for choices of $g(\cdot)$ are listed below.
\vspace{0.1cm}
\begin{itemize}
\item \emph{Localized excitation}: We set
\beq \label{E:g_1}
g_1(\widehat{\bm B}) = \| {\rm vec} ( \widehat{\bm B} ) \|_1,~g_1^\star(\widehat{\bm{\mathcal{S}}} ) = \| {\rm vec} ( \widehat{\bm{\mathcal{S}}} ) \|_\infty \eqs.
\eeq
This regularization forces the solution $\widehat{\bm B}^\star$ to \eqref{eq:lowrank}
to be an element-wise sparse matrix. This corresponds to the scenario where
each element of the latent variables in ${\bm z}^\ell$ excites only a few of the nodes
in our graph.
\item \emph{Small number of excited nodes}: Let $\widehat{\bm b}_i^{\rm row}$ 
be the $i$th row vector of $\widehat{\bm B}$. We then set
\beq  \label{E:g_2}
g_2( \widehat{\bm B} ) = \sum_{i=1}^N \| \widehat{\bm b}_i^{\rm row} \|_2,~g_2^\star(\widehat{\bm{\mathcal{S}}} ) = \max_{i=1,....,N} \| \hat{\bm s}_i^{\rm row} \|_2 \eqs.
\eeq
This regularization is motivated by the group-sparsity formulation in \cite{huang2011learning}
which forces the solution $\widehat{\bm B}^\star$ to \eqref{eq:lowrank} to be row-sparse.
Notice that this is relevant when 
the graph filter is excited on a small number of nodes.
\item \emph{Small perturbation}: We set 
\beq \label{E:g_3}
g_3(\widehat{\bm B}) = \| \widehat{\bm B} \|_{\rm F},~g_3^\star(\widehat{\bm{\mathcal{S}}}) = \| \widehat{\bm{\mathcal{S}}} \|_{\rm F} \eqs.
\eeq
This regularization models each entry of $\tilde{h}_N {\bm B}$ as a Gaussian random
variable of small, identical variance. This can   be used when there is no prior 
knowledge on ${\bm B}$. 
\end{itemize}
\vspace{0.1cm}
\noindent Notice that for every choice of the regularizer $g(\cdot)$ discussed, 
\eqref{eq:lowrank} is a convex problem that can be solved 
in polynomial time. 
Let the optimal solution to \eqref{eq:lowrank} be $\widehat{\bm{\mathcal{S}}}^\star, \widehat{\bm B}^\star$. We apply spectral method 
on $\widehat{\bm{\mathcal{S}}}^\star$ based on its
top-$K$ left singular vectors. The boosted {\sf BlindCD}
method is overviewed in Algorithm~\ref{alg:bbcd}. 



\algsetup{indent=1em}
\begin{algorithm}[t]
	\caption{Boosted {\sf BlindCD} method.}\label{alg:bbcd}
	\begin{algorithmic}[1]
		\STATE \textbf{Input}: Graph signals and excitation signals $\{{\bm y}^\ell, {\bm z}^\ell\}_{\ell=1}^L$; desired number of communities $K$.
		\STATE Solve the convex optimization problem \eqref{eq:lowrank}
		and denote its solution as $(\widehat{\bm{\mathcal{S}}}^\star, \widehat{\bm B}^\star)$. 
		\STATE Find the top-$K$ \emph{left} singular vectors of $\widehat{\bm{\mathcal{S}}}^\star$ and denote the set of singular vectors as $\widetilde{\bm S}_K \in \RR^{N \times K}$.
		\STATE Apply the $K$-means method on the row vectors of $\widetilde{\bm S}_K$. 
		\STATE \textbf{Output}: $K$ communities $\tilde{\cal C}_1, ..., \tilde{\cal C}_K$.
	\end{algorithmic} \vspace{-.1cm}
\end{algorithm}

\subsection{Performance Analysis}
This section analyzes the performance of the \emph{boosted} {\sf BlindCD} method, mimicking the ideas in Section~\ref{sec:theory}.
Due to the space limitation, 
we focus on the special case where ${\bm B}$ is sparse and select $g_1(\widehat{\bm B})$ in \eqref{E:g_1} when solving \eqref{eq:lowrank}. 

Our first step towards deriving a theoretical bound for the performance of 
boosted {\sf BlindCD}
is to characterize the estimation error of ${\cal H}({\bm L)} {\bm B}$
when solving \eqref{eq:lse},
defined as 
\beq
\bm{\mathcal{E}} \eqdef \bm{\mathcal{H}}^\star - {\cal H}({\bm L)} {\bm B} \eqs.
\vspace{-.8cm}
\eeq 
\begin{Lemma} \label{prop:err1}
Suppose that $L \geq R$, $\{ {\bm z}^\ell \}_{\ell=1}^L$  spans $\RR^R$, and $\| {\bm w}^\ell ({\bm z}^\ell)^\top \| < \infty$ almost surely.
For every $c>0$ and with probability at least $1-2c$, it holds that
\beq
\|  \bm{\mathcal{E}} \|_2 = {\cal O} \left( \frac{ \sigma_w \log((N+R)/c)}{\sqrt{L}}\right) \eqs. 
\eeq
\end{Lemma}
The proof can be found in Appendix~\ref{app:lse}.
Lemma~\ref{prop:err1} captures the expected behavior of a vanishing 
estimation error when $L \to \infty$.
Next, we show that $\widehat{\bm L}^\star$ from \eqref{eq:lowrank}
is close to $\tilde{{\cal H}}({\bm L}) {\bm B}$ by leveraging the fact that the latter is approximately rank-$K$.\begin{Lemma}\cite[Corollary 1]{agarwal2012noisy} \label{prop:lowrank}
Consider problem \eqref{eq:lowrank} with
\beq \label{eq:cor_cond} \begin{split} 
& \kappa \geq 4 \| \bm{\mathcal{E}} \|_2,~\rho \geq 4 \Big( \frac{\alpha}{\sqrt{NR}} + \| {\rm vec}( \bm{\mathcal{E}} ) \|_\infty \Big) \eqs, \\
& \alpha \geq \sqrt{NR} \!~ \| {\rm vec}( \widetilde{\cal H}( {\bm L} ) {\bm B} ) \|_\infty \eqs.
\end{split}
\eeq
Let $R \geq K$. 
There exists constants $c_1, c_2$ such that
\beq \label{eq:fro_err} \begin{split}
& \| \widehat{\bm{\mathcal{S}}}^\star - \widetilde{\cal H}({\bm L}) {\bm B}  \|_{\rm F}^2 + 
\| \widehat{\bm B}^\star - \tilde{h}_N {\bm B} \|_{\rm F}^2 \leq \\
& c_1 \kappa^2 \Big( K + \frac{1}{\kappa} \sum_{j=K+1}^R  \sigma_j ( \widetilde{\cal H}({\bm L}) {\bm B} ) \Big) + c_2 \rho^2 \| {\rm vec}( {\bm B} ) \|_0.
\end{split}
\eeq
\end{Lemma}
The term
$\sum_{j=K+1}^R  \sigma_j ( \widetilde{\cal H}({\bm L)} {\bm B} )$ is negligible
when $\widetilde{\cal H}({\bm L}){\bm B}$ is approximately rank-$K$.
Therefore,
the implication is that the distance between $\widehat{\bm{\mathcal{S}}}^\star$ 
and $\widetilde{\cal H}({\bm L}) {\bm B}$ can be bounded by the sum of two terms --- one 
that is dependent on $\bm{\mathcal{E}}$,
and one that is dependent on $\alpha / \sqrt{NR}$. 
Overall, it shows that the error reduces when the excitation rank 
$R$ and number of observations $L$ increases. 
{On the other hand, \eqref{eq:cor_cond} suggests that one should set $\kappa = c_1 / \sqrt{L}$, $\rho = c_2 / \sqrt{RL}$ in \eqref{eq:lowrank} for some $c_1,c_2$ for the optimal performance.} 

Having established these results, the boosted {\sf BlindCD} method is an approximation
of {\sf BlindCD} operating on the boosted LPGF $\widetilde{\cal H}({\bm L}) {\bm B}$. 
Next, we define  the SVD of $\tilde{{\cal H}}({\bm L}) {\bm B}$
as $\widetilde{\bm V} \widetilde{\bm{\Sigma}} \widetilde{\bm Q}^\top$ and analyze
the performance of the boosted {\sf BlindCD}  
through a minor modification of Theorem~\ref{T:main}.\begin{Corollary} \label{cor:bbcd}
Suppose that Conditions~1 to 3 in Theorem~\ref{T:main} are met when 
replacing ${\bm Q}_K$ by $\widetilde{\bm Q}_K$ and ${\cal H}({\bm L})$ by $\widetilde{\cal H}({\bm L})$. Let 
$\widetilde{\bm{\Delta}} \eqdef \widehat{\bm{\mathcal{S}}}^\star - \widetilde{\cal H}({\bm L}) {\bm B}$
and assume that
\beq \label{eq:deltatil}
\widetilde{\delta} \eqdef \sigma_K ( \widetilde{\cal H}({\bm L}) {\bm B} ) - \sigma_{K+1} ( \widetilde{\cal H}({\bm L}) {\bm B} ) - \| \widetilde{\bm{\Delta}} \|_2 > 0 \eqs.
\eeq
If Step~4 in the boosted {\sf BlindCD} method finds an $(1+\epsilon)$ optimal solution to the $K$-means problem, where $\epsilon >0$, then, 
	\beq \label{E:main_theorem2}
	\begin{split}
	 \sqrt{F( \tilde{\cal C}_1, ..., \tilde{\cal C}_K )}  & - \sqrt{(1+\epsilon) F^\star} \leq  \\
	&  (2+\epsilon) \sqrt{2K} \left( \sqrt{ \frac{\tilde\gamma^2}{1 + \tilde\gamma^2} } + \frac{ \| \widetilde{\bm{\Delta}} \|_2 }{ \tilde{\delta} } \right) \eqs,
	 \end{split}
	\eeq 
where $F(\cdot)$, $F^\star$ are defined in \eqref{eq:obj}, \eqref{eq:fstar}, respectively, and 
\beq \label{eq:bdgamma}
\tilde{\gamma} \leq \tilde{\eta}  \!~ \| {\bm V}_{N-K}^\top {\bm B} \widetilde{\bm Q}_K \|_2 \!~ \| ({\bm V}_{K}^\top {\bm B} \widetilde{\bm Q}_K)^{-1} \|_2 \eqs,
\eeq
where $\tilde{\eta}$ is the low-pass coefficient of the boosted LPGF
$\widetilde{\cal H}({\bm L}) {\bm B}$.
\end{Corollary}
The proof of Corollary~\ref{cor:bbcd} can be found in Appendix~\ref{app:cor}. 
We see that the performance of the boosted {\sf BlindCD} method 
depends on $\tilde{\eta}$, the low-pass coefficient of the boosted LPGF. {
As $\tilde{\eta} \ll \eta$ due to our prior discussions, it is anticipated that the boosted method achieves a much better performance, especially when the original LPGF is not markedly low-pass.}

While the bound in \eqref{E:main_theorem2} is similar to that in Theorem~\ref{T:main},
we observe that 
applying Lemma~\ref{prop:err1} and Lemma~\ref{prop:lowrank} yields
\beq \notag \label{E:rate_delta_tilde}
\| \widetilde{\bm{\Delta}} \|_2 \leq \| \widetilde{\bm{\Delta}} \|_{\rm F} = {\cal O} \left( \sigma_{K+1} ( \widetilde{\cal H}({\bm L}) {\bm B} ) + \frac{1}{ \sqrt{L}} + \frac{1}{\sqrt{NR}} \right) .
\eeq
From the definition of $\tilde{\delta}$
we have 
\beq \label{eq:fin_fin}
\frac{ \| \widetilde{\bm{\Delta}} \|_2 }{ \tilde{\delta} } = {\cal O} \left(  \frac{ \sigma_{K+1} ( \widetilde{\cal H}({\bm L}) {\bm B} ) + 1/\sqrt{L} + 1/ \sqrt{NR}  }{ \sigma_K ( \widetilde{\cal H}({\bm L}) {\bm B} ) - C \sigma_{K+1} ( \widetilde{\cal H}({\bm L}) {\bm B} ) } \right) ,
\eeq
for some constant $C$.
Substituting \eqref{eq:fin_fin} into \eqref{E:main_theorem2}
shows that the sub-optimality of boosted {\sf BlindCD} can be minimized 
when 1) the spectral gap for the sketched matrix
$\widetilde{\cal H}( {\bm L} ) {\bm B}$, 2) the number of samples $L$, and 3)
the excitation rank $R$, are large.
We remark that it is possible to undertake analogous performance analysis for the other proposed regularizers on ${\bm B}$ [cf.~\eqref{E:g_2} and~\eqref{E:g_3}]. For example, this can be done using \cite{negahban2012unified} and replacing Lemma~\ref{prop:lowrank} with the corresponding result. 
These extensions, however, are beyond the scope of the current paper.

\section{Numerical Examples}\label{sec:num}
To illustrate the efficacy of the {\sf BlindCD} methods, we study
three application examples that pertain to consensus dynamics, 
consumer networks, and social networks. Numerical examples will be given for these applications, { which were introduced in Sections~\ref{sec:diff} through~\ref{sec:degroot}.}

Unless otherwise specified, the graphs used in the simulations 
will be generated  according to a \emph{stochastic block 
model} (SBM) \cite{holland1983stochastic}, denoted by $G \sim {\sf SBM}(N,K,a,b)$, such that $G$ has $N$ nodes,
$K$ equal-sized non-overlapping communities and the intra (\resp inter)
community connectivity probability is $a \in [0,1]$ (\resp $b \in [0,1]$). 
The weights on the graph, $A_{ij}$, are set to $1$ if $(i,j) \in E$ and $0$ otherwise. 
We use the ground truth community membership in generating
the {\sf SBM} graphs when evaluating the accuracies. 
The error rate is given by
\beq \textstyle
P_e \eqdef \EE \left[ \frac{1}{N} \min_{ \pi : [K] \rightarrow [K] } \sum_{i=1}^N \mathbbm{1}_{ \pi(c_i) \neq c_i^{\rm true} } \right] \eqs,
\eeq
and the above is approximated via Monte-Carlo simulations, 
where $\mathbbm{1}_{{\cal E}}$ is an indicator function for the event ${\cal E}$, 
$\pi : [K] \rightarrow [K]$ is 
a permutation function and $c_i \in [K]$ (\resp $c_i^{\rm true}$) 
is the detected (\resp true) community membership of node $i$. 

\subsection{Diffusion Dynamics}
We first evaluate the performance of {\sf BlindCD} using  
 graph signals generated according to the observation model 
in \eqref{eq:data_collect}. 
We focus on  the diffusion dynamics in Section~\ref{sec:diff}. 
We perform Monte-Carlo simulations to evaluate the community detection performance
on random graphs. 
In this example, the SBM graphs generated are 
$G \sim {\sf SBM}(N,K,8 \log N / N, \log N / N)$ with $N=150$ and $K=3$.
We simulate a scenario where the graph filter is excited on only $R$ 
nodes. In this case, the sketch matrix ${\bm B}$ is generated by first 
picking $R$ rows uniformly from the $N$ available rows, and the elements
in each selected rows are set to one uniformly with probability $p_b = 0.5$. 
For 
the boosted {\sf BlindCD} method, 
we test the formulation of \eqref{eq:lowrank} with regularizers 
$g_1(\widehat{\bm B})$ and $g_2(\widehat{\bm B})$ [cf.~\eqref{E:g_1} and \eqref{E:g_2}] by  
setting $\kappa = 2 / \sqrt{L}$ and $\rho = 0.5 / \sqrt{RL}$. 
The variance of observation noise is $\sigma_w^2 = 10^{-2}$
and each element of ${\bm z}^\ell$ is generated independently as
$[ {\bm z}^\ell ]_i \sim {\cal U}[-1,1]$.

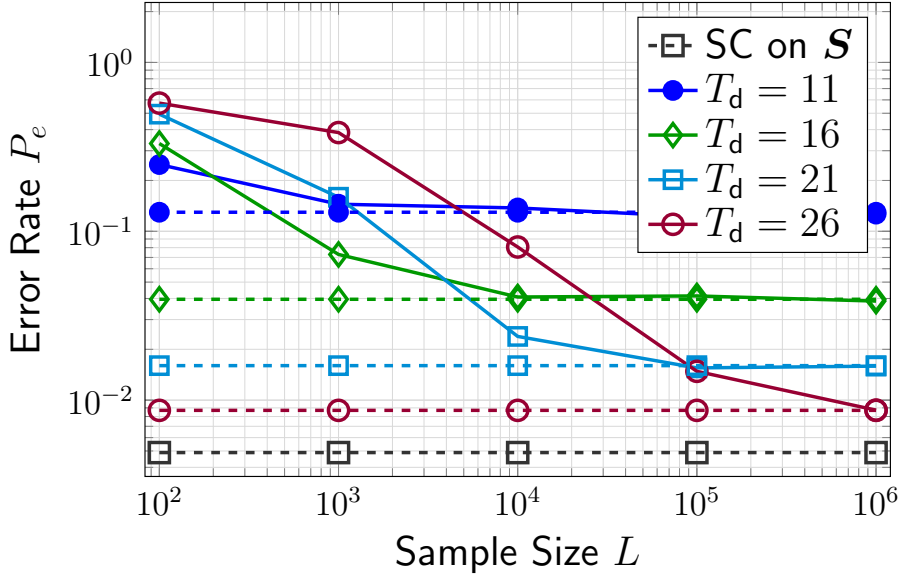
\begin{figure}[t]
\centering
\resizebox{.75\linewidth}{!}
{ \sf 
\begin{tikzpicture}
\pgfplotsset{ scale only axis,
    width=0.55\textwidth,height=0.35\textwidth,
    grid=both,grid style={line width=.01pt, draw=gray!30},
    legend cell align=left, legend style={legend pos=north east,font=\Large},
    xlabel={\Large Sample Size $L$},
    enlarge x limits=0.02,
    enlarge y limits=0.02, ymin = 4e-3, ymax = 2, xmin = 1e2, xmax = 1e6,
}

\pgfplotsset{every tick label/.append style={font=\large}}

\begin{loglogaxis}[ylabel={\Large Error Rate $P_e$}]
\addplot[black!80, dashed, very thick, mark options={solid,mark size=3.5}, mark=square] 
      table[x index=0, y index=1, col sep=comma] {./Results_Tikz/EG1_N150K3R15_L25_VarSamp2_Avg_err_S.csv};
      \addlegendentry{SC on ${\bm S}$};
\addplot[blue, very thick, mark options={solid,mark size=3}, mark=*] 
      table[x index=0, y index=1, col sep=comma] {./Results_Tikz/EG1_N150K3R15_L10_VarSamp2_err_CY.csv};
      \addlegendentry{$T_{\sf d}=11$};
\addplot[green!60!black, very thick, mark options={solid,mark size=4}, mark=diamond] 
      table[x index=0, y index=1, col sep=comma] {./Results_Tikz/EG1_N150K3R15_L15_VarSamp2_err_CY.csv};
      \addlegendentry{$T_{\sf d}=16$};
\addplot[asublue, very thick, mark options={solid,mark size=3}, mark=square] 
      table[x index=0, y index=1, col sep=comma] {./Results_Tikz/EG1_N150K3R15_L20_VarSamp2_err_CY.csv};
      \addlegendentry{$T_{\sf d}=21$};
\addplot[asured, very thick, mark options={solid,mark size=3.5}, mark=o] 
      table[x index=0, y index=1, col sep=comma] {./Results_Tikz/EG1_N150K3R15_L25_VarSamp2_err_CY.csv};
      \addlegendentry{$T_{\sf d}=26$};
\addplot[blue, dashed, very thick, mark options={solid,mark size=3}, mark=*] 
      table[x index=0, y index=1, col sep=comma] {./Results_Tikz/EG1_N150K3R15_L10_VarSamp2_Avg_err_CYperf.csv};
\addplot[green!60!black, dashed, very thick, mark options={solid,mark size=4}, mark=diamond] 
      table[x index=0, y index=1, col sep=comma] {./Results_Tikz/EG1_N150K3R15_L15_VarSamp2_Avg_err_CYperf.csv};
\addplot[asublue, dashed, very thick, mark options={solid,mark size=3}, mark=square] 
      table[x index=0, y index=1, col sep=comma] {./Results_Tikz/EG1_N150K3R15_L20_VarSamp2_Avg_err_CYperf.csv};
\addplot[asured, dashed, very thick, mark options={solid,mark size=3.5}, mark=o] 
      table[x index=0, y index=1, col sep=comma] {./Results_Tikz/EG1_N150K3R15_L25_VarSamp2_Avg_err_CYperf.csv};
\end{loglogaxis}

\end{tikzpicture}
} \vspace{-.2cm}
\caption{\textbf{Community detection performance versus sample size $L$}. We consider graphs generated as $G \sim {\sf SBM}(N , K,8 \log N /N, \log N /N)$
with $N=150,K=3$ and fix the excitation rank at $R=15$. The solid (\resp dashed) lines show
the performance of {\sf BlindCD} on the sampled \emph{output} covariance $\widehat{\bm C}_y$
(\resp true and noiseless covariance ${\overline{\bm C}_y}$).}
\label{fig:sample}\vspace{-.2cm}
\end{figure}

The first example examines the effect of the graph filter's low-pass coefficient $\eta$
and sample size $L$ on the performance of {\sf BlindCD}.  
In particular, Fig.~\ref{fig:sample} shows the performance of community detection 
for different filter orders $T_{\sf d}$ against the number of samples $L$ 
accrued.
Notice that the low-pass coefficient $\eta$ decreases with the filter order $T_{\sf d}$ [cf.~\eqref{E:eta_example_1}].  
As such, we observe that the performance  
improves with $T_{\sf d}$. The error rate approaches that achieved by applying spectral clustering on the
actual ${\bm L}$. 
An interesting observation is that for sample covariances, 
as $T_{\sf d}$ increases, the sample size $L$ required to reach the performance of 
noiseless covariance also increases. 
This can be explained with the condition \eqref{eq:thm_5} 
in Theorem~\ref{T:main}. In particular, as $T_{\sf d}$ increases, 
the absolute value of 
$\beta_K ( {\overline{\bm C}_y} ) 
- \beta_{K+1} ( {\overline{\bm C}_y} )$
decreases, therefore restricting $\| \widehat{\bm C}_y - \overline{\bm C}_y \|_2$ to be smaller [cf.~\eqref{eq:thm_5}]. 
The latter is satisfied when the number of samples accrued is sufficiently large. 

\begin{figure}[t]
\centering
\resizebox{.75\linewidth}{!}
{ \sf 
\begin{tikzpicture}
\pgfplotsset{ scale only axis,
    width=0.55\textwidth,height=0.35\textwidth,
    grid=both,grid style={line width=.01pt, draw=gray!30},
    legend cell align=left, legend style={opacity = 0.8,legend pos=north east,font=\large},
    xlabel={\Large Excitation Rank $R$},
    enlarge x limits=0.02,
    enlarge y limits=0.02, ymin = 3e-3, ymax = 1.2
}

\pgfplotsset{every tick label/.append style={font=\large}}

\begin{semilogyaxis}[ylabel={\Large Error Rate $P_e$}]
\addplot[black!80, dashed, very thick, mark options={solid,mark size=3.5}, mark=square] 
      table[x index=0, y index=1, col sep=comma] {./Results_Tikz/EG1_N150K3_L15_VarR_Avg_err_S.csv};
      \addlegendentry{SC on ${\bm S}$};
\addplot[green!60!black, very thick, mark options={solid,mark size=4}, mark=diamond] 
      table[x index=0, y index=1, col sep=comma] {./Results_Tikz/EG1_N150K3_L15_VarR2_err_CYperf.csv};
      \addlegendentry{BlindCD on $\overline{\bm C}_y$};
\addplot[blue, very thick, mark options={solid,mark size=4}, mark=o] 
      table[x index=0, y index=1, col sep=comma] {./Results_Tikz/EG1_N150K3_L15_VarR2_err_CY.csv};
      \addlegendentry{BlindCD on $\hat{\bm C}_y$};
\addplot[red, very thick, mark options={solid,mark size=4}, mark=triangle] 
      table[x index=0, y index=1, col sep=comma] {./Results_Tikz/EG1_N150K3_L15_VarR2_err_Shat.csv};
      \addlegendentry{Boosted w/ $g_1(\hat{\bm B})$};
\addplot[asuorange, very thick, mark options={solid,mark size=4}, mark=square] 
      table[x index=0, y index=1, col sep=comma] {./Results_Tikz/EG1_N150K3_L15_VarR2_err_Shat2.csv};
      \addlegendentry{Boosted w/ $g_2(\hat{\bm B})$};
\addplot[purple, very thick, mark options={solid,mark size=4}, mark=*] 
      table[x index=0, y index=1, col sep=comma] {./Results_Tikz/Revised_EG1_L15_err_SpecTemp.csv};
      \addlegendentry{2-step w/ \cite{segarra2017network}};
\end{semilogyaxis}

\end{tikzpicture}
} \vspace{-.2cm}
\caption{\textbf{Community detection performance versus excitation rank $R$}.
We consider graphs generated as $G \sim {\sf SBM}(N , K,8 \log N /N, \log N /N)$
with $N=150,K=3$ and fixed $T_{\sf d}=16$, $L=10^3$.} \label{fig:varR}\vspace{-.2cm}
\end{figure}
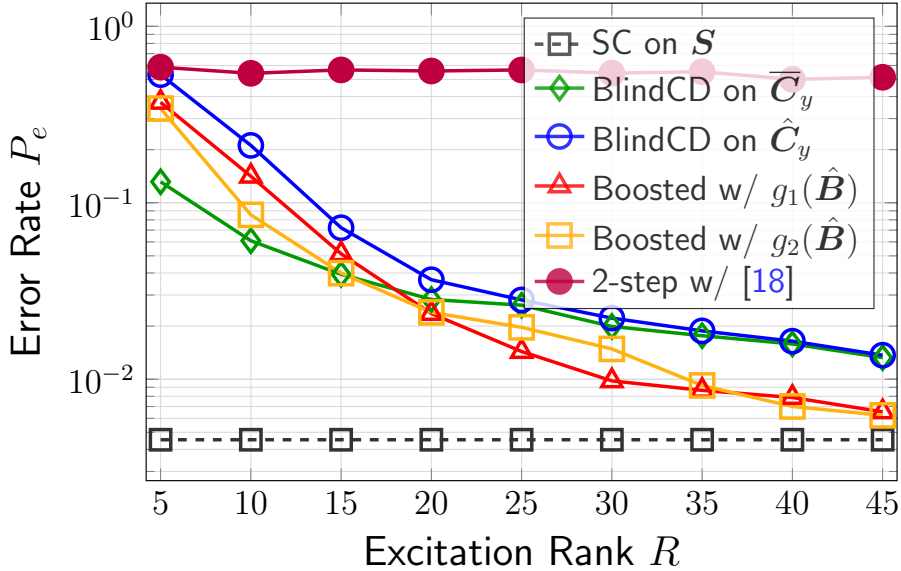

The second example shows the effect of the excitation rank $R$. 
The results are shown in Fig.~\ref{fig:varR} where we have fixed $L=10^3$
and $T_{\sf d}=16$.  
{In this example, we have compared the performance of {\sf BlindCD} to a 2-step procedure which uses \cite{segarra2017network} (with efficient implementation in \cite{shafi2019}) to recover the GSO, then it applies spectral clustering on the recovered GSO to detect communities.
For {\sf BlindCD}, we observe that the performance improves with the rank $R$, 
while the 2-step procedure performs poorly\footnote{The 2-step method with \cite{segarra2017network} provides accurate result only when $R \geq 100$.}.  
As predicted by Corollary~\ref{cor:bbcd},
the boosting technique enhances the performance of {\sf BlindCD}.}

\begin{figure}[t]
\centering
\includegraphics[width=0.7 \linewidth]{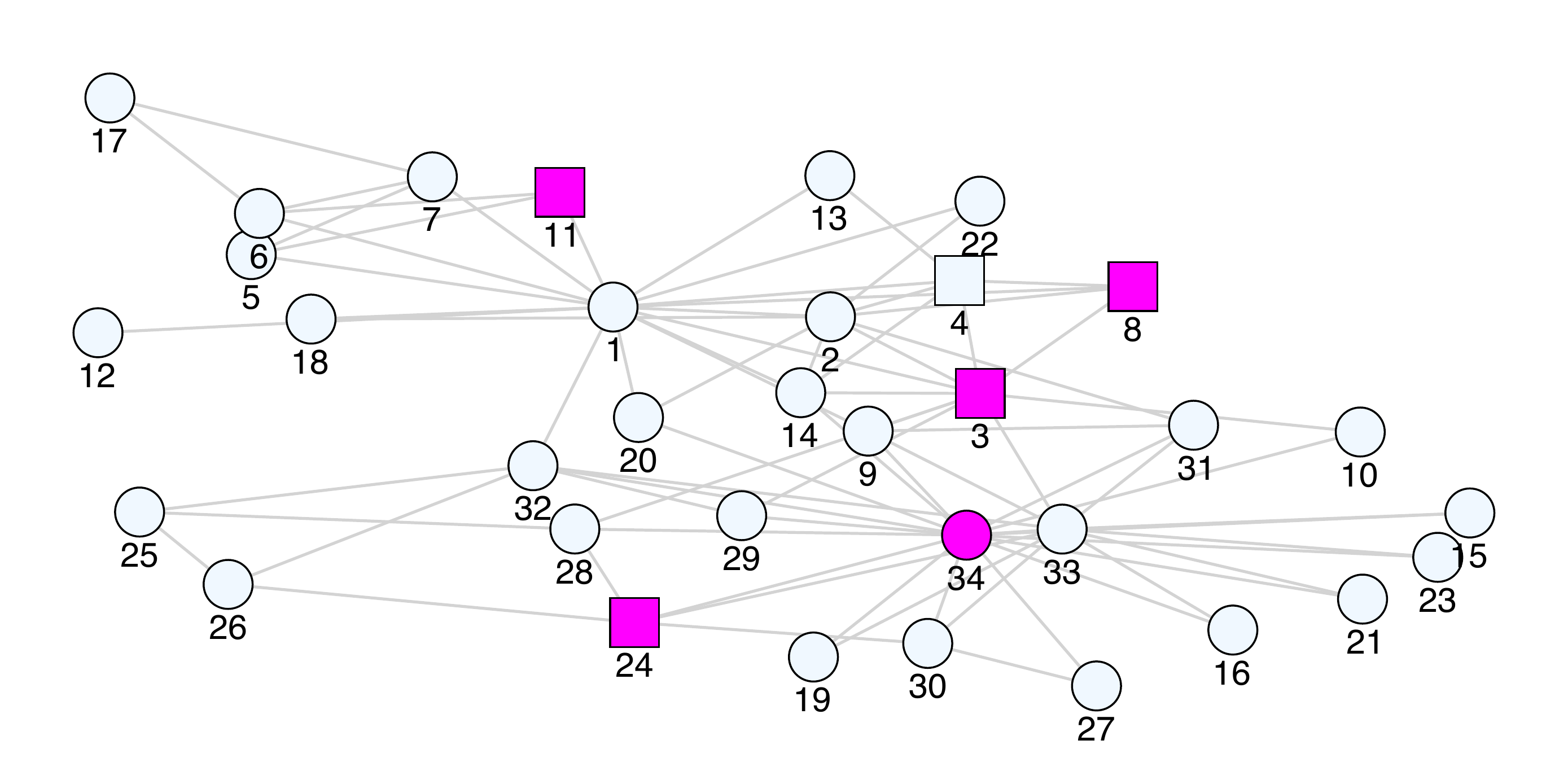}\\[-.3cm]
{\sf \scriptsize (a) Zachary's karate club network. Highlighted nodes in magenta are the actual sites of excitations, while   nodes marked as rectangles are the detected sites of excitations using boosted BlindCD. The only mismatches were node `4' and `34'.}\\[.2cm]

\begin{tabular}{cc}\hspace{-.5cm}\includegraphics[width=.375\linewidth]{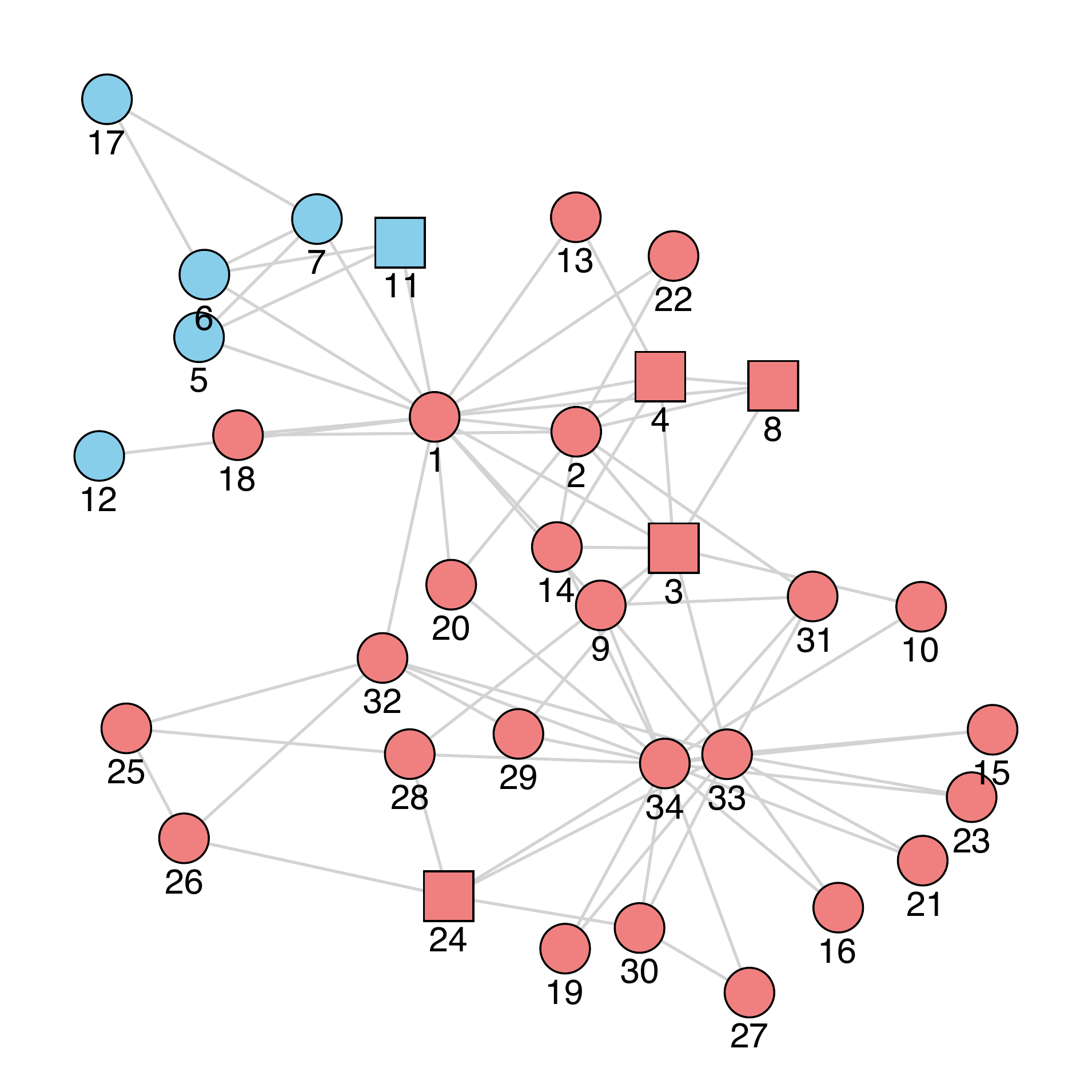}&\hspace{-.5cm}
\includegraphics[width=.375\linewidth]{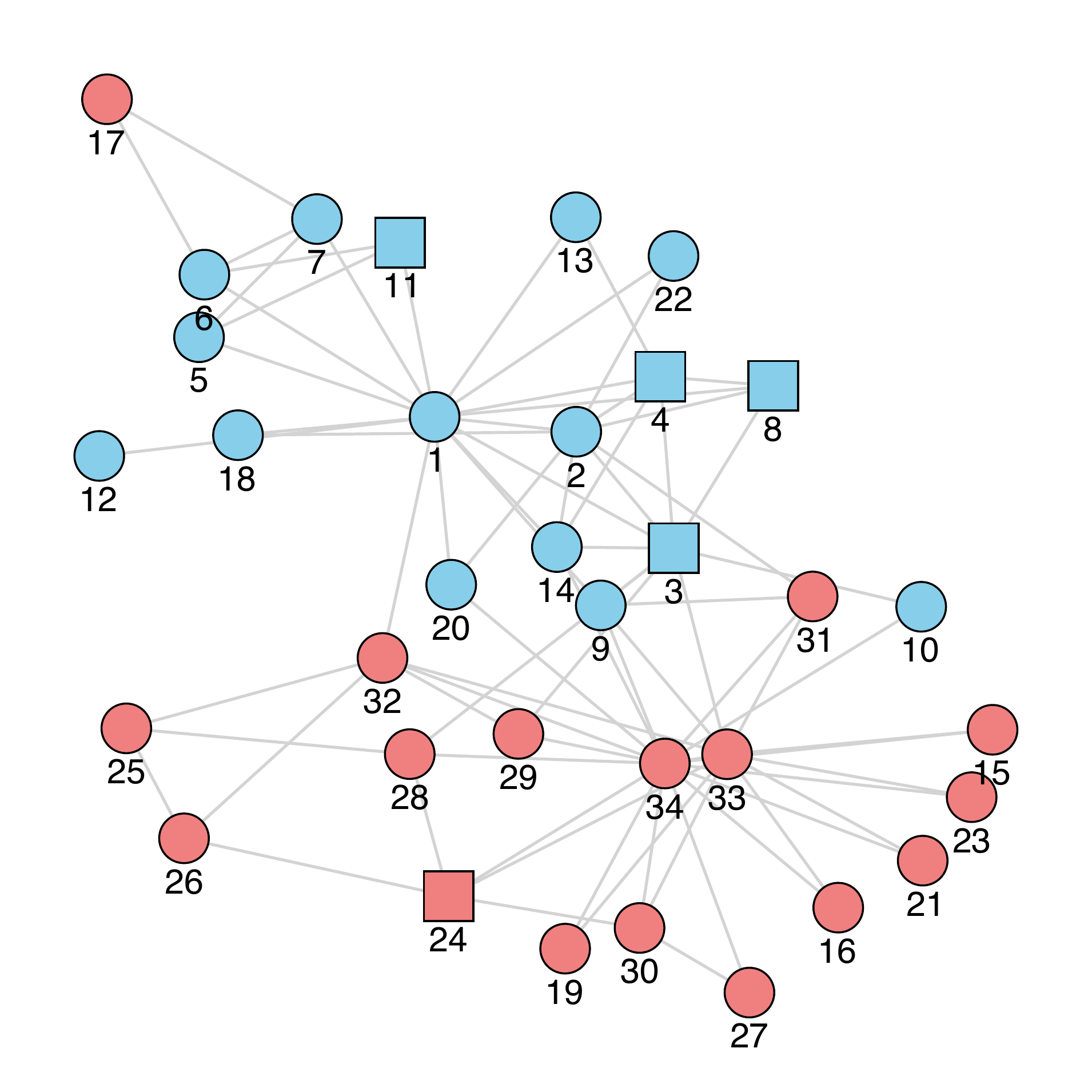}\\[-.3cm]
{\sf \scriptsize (b) {\sf BlindCD} result} & {\sf \scriptsize (c) {Boosted} {\sf BlindCD} result}
\end{tabular}\\[.1cm]
\caption{\textbf{Experiments on Zachary's karate club network}. The network consists of $N=34$ nodes and (approximately) $K=2$ communities. The graph filter models 
a diffusion dynamics with an order of $T_{\sf d} = 6$ and the graph signals observed
have only rank $R=5$ as only $5$ nodes are injected with input signals.
The bottom plots show the result of both {\sf BlindCD} methods.} \label{fig:zachary} \vspace{-.4cm}
\end{figure}

The example in Fig.~\ref{fig:zachary} shows 
the performance of an instance of {\sf BlindCD} on the Zachary's
Karate Club network when the graph signals are generated from
the diffusion dynamics. 
To capitalize on the benefit of the \emph{boosting} technique, 
we consider a scenario with a filter order of $T_{\sf d}=6$, observation
rank of $R=5$ (the graph is excited on just $5$ nodes) and we observe $L=10^3$ noisy samples of the graph signals. 
Observe that the low-pass coefficient for the filter may be close to $1$ 
as $T_{\sf d}$ is small. This explains the poorer performance of
{\sf BlindCD} in Fig.~\ref{fig:zachary}.{\sf (b)}. 
The boosted {\sf BlindCD}, instead, delivers good performance 
as it identifies the two communities in the network except for 
a miss-classification of agent $17$. 
Through sorting the row sums of the estimated $\widehat{\bm B}$, we also detected the sites of the excitations, as shown
in the Fig.~\ref{fig:zachary}.a.  


\subsection{Network Dynamics Models}
We describe applications of our  {\sf BlindCD} methods 
on detecting communities in consumer and social networks, where the models have been studied in Sections~\ref{sec:price} and~\ref{sec:degroot}.  

\begin{figure}[t]
\centering
\includegraphics[width=.375\linewidth]{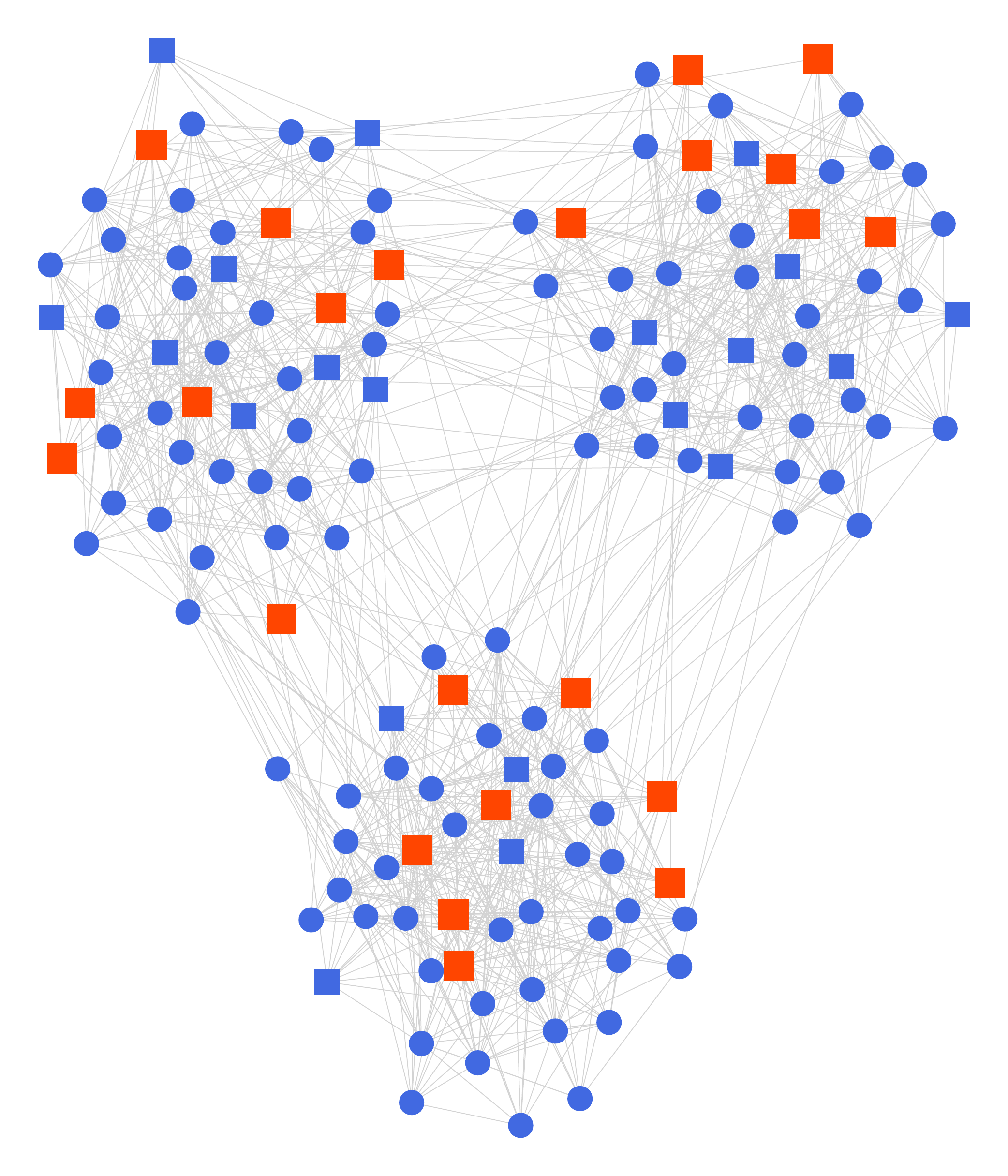}\includegraphics[width=.375\linewidth]{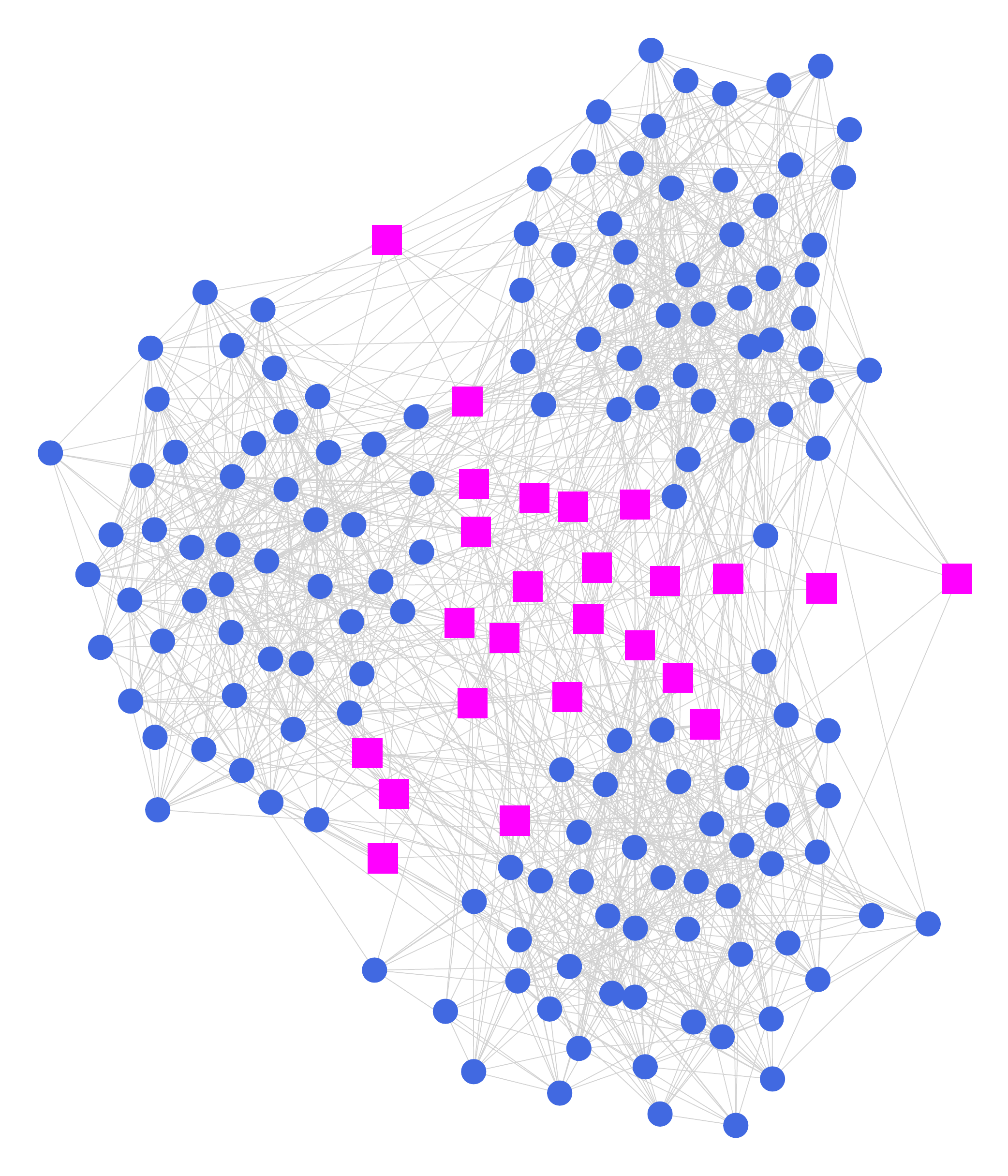}\vspace{-.3cm}
\caption{\textbf{Snapshots of set-ups for case studies on network dynamics}. (Left) A consumers network, where the highlighted nodes are the agents that the pricing experiments
were performed on. (Right) A social network, where the highlighted nodes are the stubborn agents. Both networks are generated according to ${\sf SBM}(150,3,8\log N / N, \log N /(2N))$.}\vspace{-.3cm}
\label{fig:setup}
\end{figure}

In the Monte-Carlo simulations below, we generate the graphs as 
$G \sim {\sf SBM}(150,3,8 \log N /N , \log N / (2N) )$, $N=150$.
For the \emph{consumer} games, ${\bm A}$ is  
taken as the binary adjacency matrix of $G$
and ${\bm B}$
is chosen as in \eqref{eq:b_consume} 
where  the set of affected agents ${\cal I}$ is selected uniformly.
Furthermore, in the utility \eqref{eq:util}, 
we set $b = 2 \|{\bm A} {\bf 1}\|_\infty$
and $a = 2 \max_\ell \| {\bm p}^\ell \|_\infty$ such that the 
equilibrium always satisfies \eqref{eq:equi}. 
For the \emph{social} networks, 
we first generate the support of ${\bm B}$ as a sparse bipartite 
graph with connectivity $2 \log N / N$, then the weights on ${\bm A}, {\bm B}$
are assigned uniformly such that all the rows in 
the concatenated matrix $[{\bm A}, {\bm B}]$
sum up to one. 
This models a setting where the stubborn agents are connected sparsely to the
others, \ie they are located at the \emph{periphery of the communities}. 
Note the support of ${\bm A}$ is symmetric with $A_{ij} \neq 0 \Leftrightarrow A_{ji} \neq 0$. 
Snapshots of the set-ups for both networks are found in Fig.~\ref{fig:setup}.

Despite the similarity to the previous examples, it is important to note that for the social networks, the Laplacian matrix ${\bm L}$ can be  asymmetric. 
Nevertheless, we anticipate that {\sf BlindCD} would
work in this case provided that ${\bm L}$ is approximately symmetric.
This symmetry in ${\bm L}$ is consistent with assuming that trust in social networks is of mutual nature\footnote{Additional numerical experiments
show that, on a directed graph where the trusts are not mutual, the {\sf BlindCD} method recovers the same sets of nodes that are discovered by performing spectral clustering on the \emph{eigenvectors} of ${\bm L}$. We omit these interesting results here since 
their interpretation requires a different notion of community for directed graphs, e.g.,
see \cite{malliaros2013clustering}. Further investigations on this subject 
are left for future work.}.

The consumption levels and 
steady-state opinions 
can both be generated from the graph filter in Example~\ref{ex:fil3}. 
The difference between the two cases rests on 
the design of the sketch matrix ${\bm B}$. 
In the following, we fix the number of samples at $L=10^4$
with a noise variance of 
$\sigma_w^2 = (10^{-1} / b^2)^2$ 
for consumer games
and $\sigma_w^2 = 10^{-2}$ for social networks.
For the boosted {\sf BlindCD} method, we set $\kappa = 2 / \sqrt{L}, \rho = 4 /\sqrt{RL}$ 
for consumer games and $\kappa = 2 / \sqrt{L}, \rho = 1/\sqrt{RL}$ for
social networks; and we test the formulation of \eqref{eq:lowrank} with the 
regularizer $g_1(\widehat{\bm B})$. 
{For the social network, we   included a comparison to a 2-step procedure 
which first recovers the graph topology using \cite{wai2016active}, 
and then applying spectral clustering on the inferred topology.}

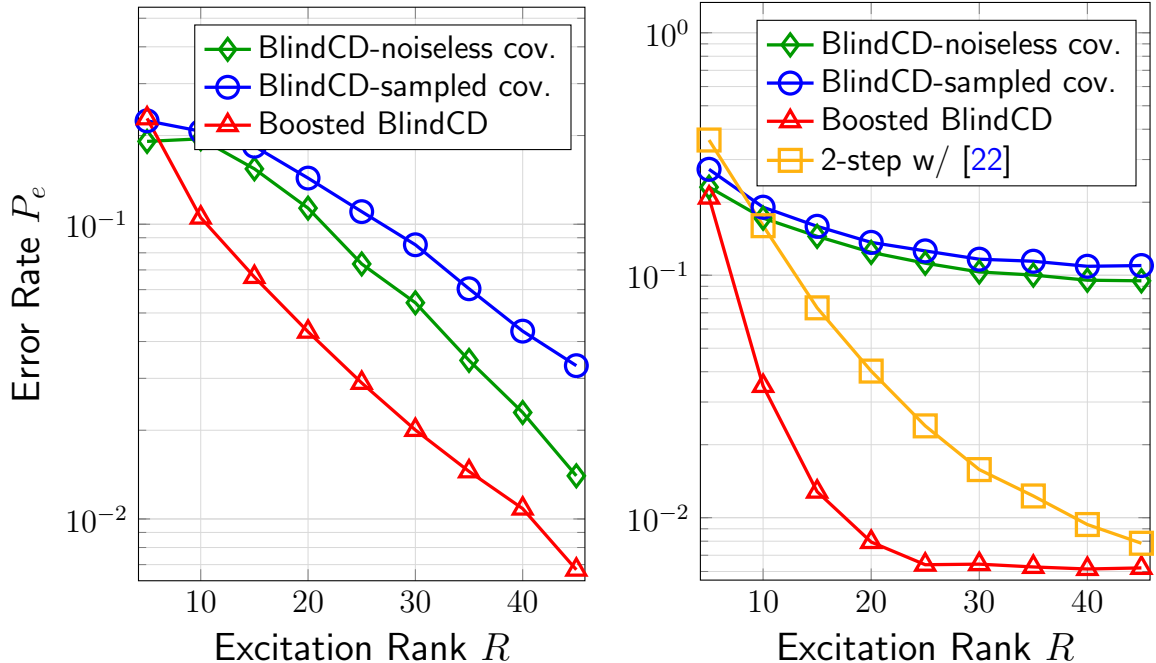
\begin{figure}[t]
\centering
\resizebox{.95\linewidth}{!}{\resizebox{.49\linewidth}{!}
{ \sf 
\begin{tikzpicture}
\pgfplotsset{ scale only axis,
    width=0.35\textwidth,height=0.45\textwidth,
    grid=both,grid style={line width=.01pt, draw=gray!30},
    legend cell align=left, legend style={legend pos=north east,font=\large},
    xlabel={\Large Excitation Rank $R$},
    enlarge x limits=0.02,
    enlarge y limits=0.02, ymax = 0.5
}

\pgfplotsset{every tick label/.append style={font=\large}}

\begin{semilogyaxis}[ylabel={\Large Error Rate $P_e$}]
\addplot[green!60!black, very thick, mark options={solid,mark size=4}, mark=diamond] 
      table[x index=0, y index=1, col sep=comma] {./Results_Tikz/EG2_N150K3_VarR2_err_CYperf.csv};
      \addlegendentry{BlindCD-noiseless cov.};
\addplot[blue, very thick, mark options={solid,mark size=4}, mark=o] 
      table[x index=0, y index=1, col sep=comma] {./Results_Tikz/EG2_N150K3_VarR2_err_CY.csv};
      \addlegendentry{BlindCD-sampled cov.};
\addplot[red, very thick, mark options={solid,mark size=4}, mark=triangle] 
      table[x index=0, y index=1, col sep=comma] {./Results_Tikz/EG2_N150K3_VarR2_err_Shat.csv};
      \addlegendentry{Boosted BlindCD};
\end{semilogyaxis}

\end{tikzpicture}
} 
\resizebox{.45\linewidth}{!}
{ \sf 
\begin{tikzpicture}
\pgfplotsset{ scale only axis,
    width=0.35\textwidth,height=0.45\textwidth,
    grid=both,grid style={line width=.01pt, draw=gray!30},
    legend cell align=left, legend style={legend pos=north east,font=\large},
    xlabel={\Large Excitation Rank $R$},
    enlarge x limits=0.02,
    enlarge y limits=0.02, ymax = 1.2
}

\pgfplotsset{every tick label/.append style={font=\large}}

\begin{semilogyaxis}
\addplot[green!60!black, very thick, mark options={solid,mark size=4}, mark=diamond] 
      table[x index=0, y index=1, col sep=comma] {./Results_Tikz/EG3_N150K3_VarR2_err_CYperf.csv};
      \addlegendentry{BlindCD-noiseless cov.};
\addplot[blue, very thick, mark options={solid,mark size=4}, mark=o] 
      table[x index=0, y index=1, col sep=comma] {./Results_Tikz/EG3_N150K3_VarR2_err_CY.csv};
      \addlegendentry{BlindCD-sampled cov.};
\addplot[red, very thick, mark options={solid,mark size=4}, mark=triangle] 
      table[x index=0, y index=1, col sep=comma] {./Results_Tikz/EG3_N150K3_VarR2_err_Shat.csv};
      \addlegendentry{Boosted BlindCD};
\addplot[asuorange, very thick, mark options={solid,mark size=4}, mark=square] 
      table[x index=0, y index=1, col sep=comma] {./Results_Tikz/EG3_N150K3_VarR2_err_act.csv};
      \addlegendentry{2-step w/\cite{wai2016active}};
\end{semilogyaxis}

\end{tikzpicture}
} \vspace{-.3cm}}\vspace{-.2cm}
\caption{\textbf{Community detection performance on cases of network dynamics}. (Left) Pricing
experiments on consumers network. (Right) Opinion dynamics with stubborn agents on 
social networks.
In both cases, we consider networks generated as $G \sim {\sf SBM}(N,3,8\log N /N, \log N / (2N))$, $N=150$.}\vspace{-.4cm} \label{fig:cases}
\end{figure}

The results of our numerical experiments are shown in Fig.~\ref{fig:cases}, where
we compare the community detection performance as the excitation rank $R$
increases in both systems. 
Similar to the previous experiment in Fig.~\ref{fig:varR}, 
for both cases we observe that the performance 
improves with $R$  and the boosted {\sf BlindCD} method
delivers the best performance consistently. 
Overall,   the performance improvement with boosted 
{\sf BlindCD} is greater than in the previous example 
 [cf.~Fig.~\ref{fig:varR}]. 
The reason behind this is the fact that the IIR graph filter 
has a poor low-pass coefficient depending on the parameter $c \gg 1$ for the
scenario we have considered. 
Another observation is that the community detection performance 
of the un-boosted {\sf BlindCD}
saturates at $R \approx 25$ for the opinion dynamics experiments
while it continues to improve with $R$ for pricing ones.
This is due to the different model used for the sketch matrix ${\bm B}$. 
In particular, for the pricing experiments, ${\bm B}$ is 
merely a sub-matrix of the identity matrix. Recall from Theorem~\ref{T:main} 
that the performance of {\sf BlindCD} depend on the product
$\| {\bm V}_{N-K}^\top {\bm B} {\bm Q}_K \|_2 \| ( {\bm V}_K^\top {\bm B} {\bm Q}_K )^{-1} \|_2$, which is anticipated
to decrease since ${\bm B}$ approaches a permutation of ${\bm I}$ as $R$ approaches $N$, yielding
a better performance. The same observation does not apply for opinion dynamics
as the sketch
matrix does not approximate the identity matrix as $R$ grows.

\begin{figure}[t]
\centering
\includegraphics[width=0.7\linewidth]{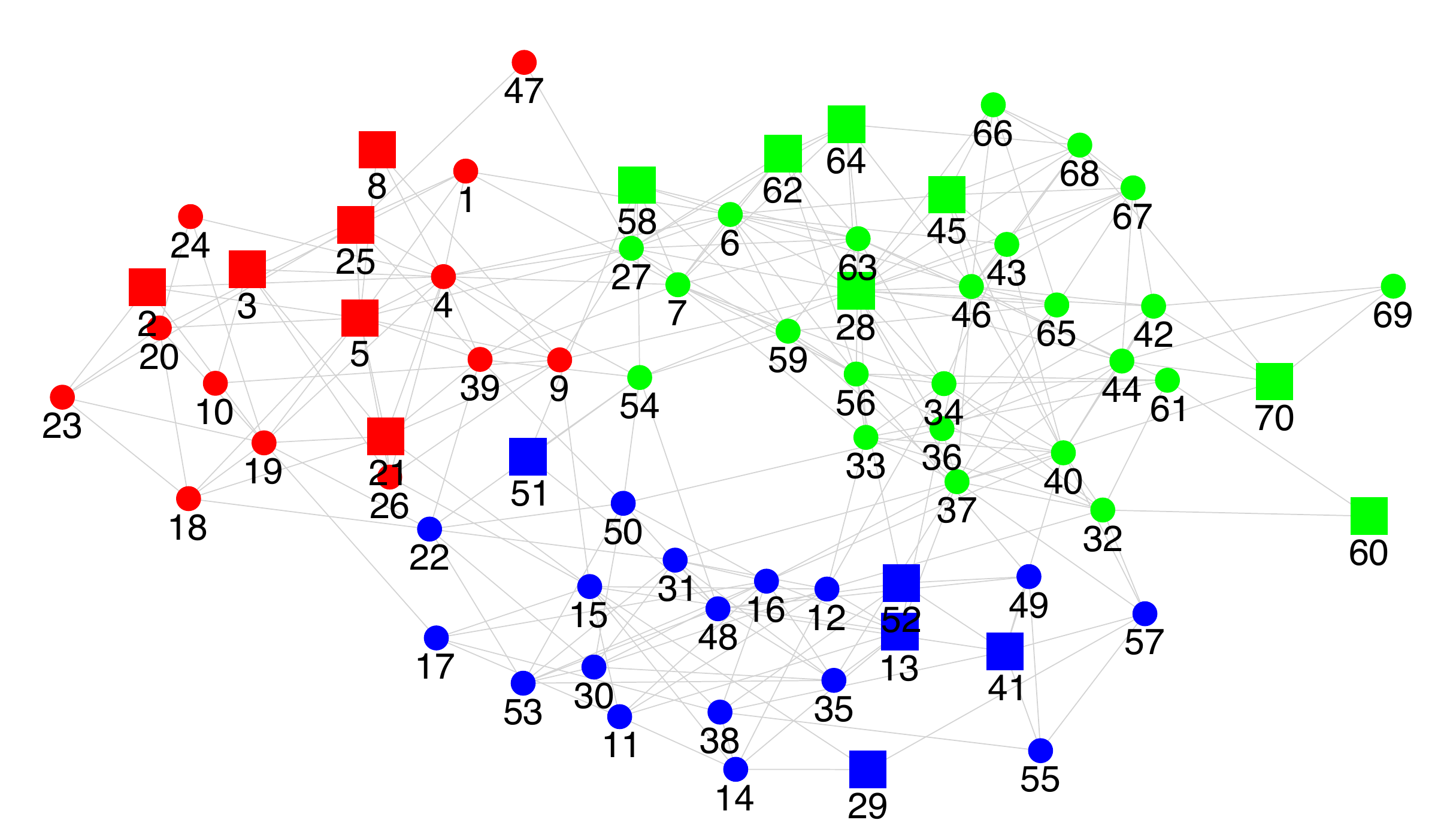}\\[-.1cm]
{\sf \scriptsize (a) \texttt{Highschool} network. The pricing experiments modify prices for the agents marked with \emph{square}. The above clustering on the network is computed from the true Laplacian matrix ${\bm L}$, which has a RatioCut of $3.618$.}\\[.0cm]

\begin{tabular}{cc}\hspace{-.2cm}\includegraphics[width=.365\linewidth]{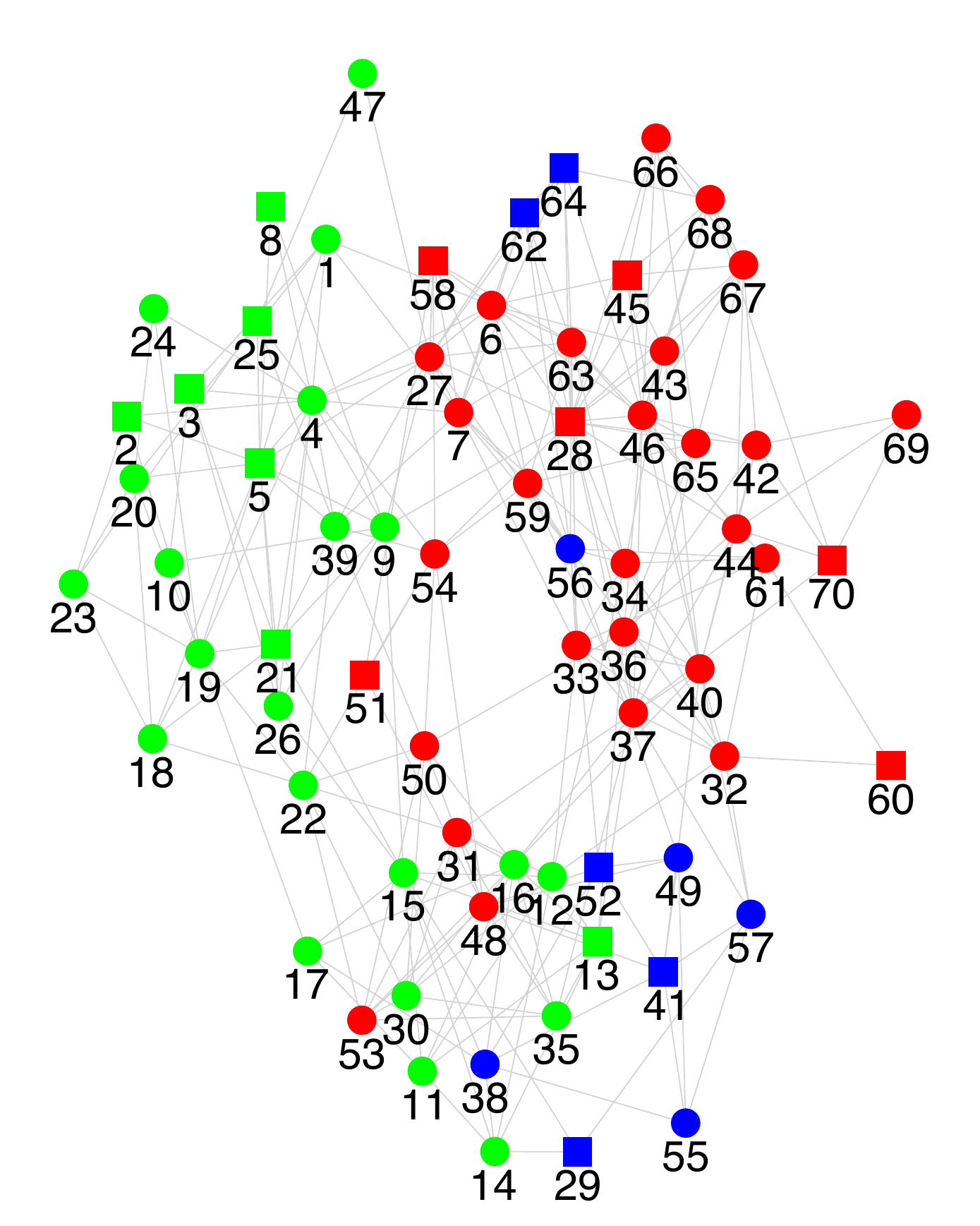}\hspace{-.5cm}&\hspace{-.5cm}
\includegraphics[width=.365\linewidth]{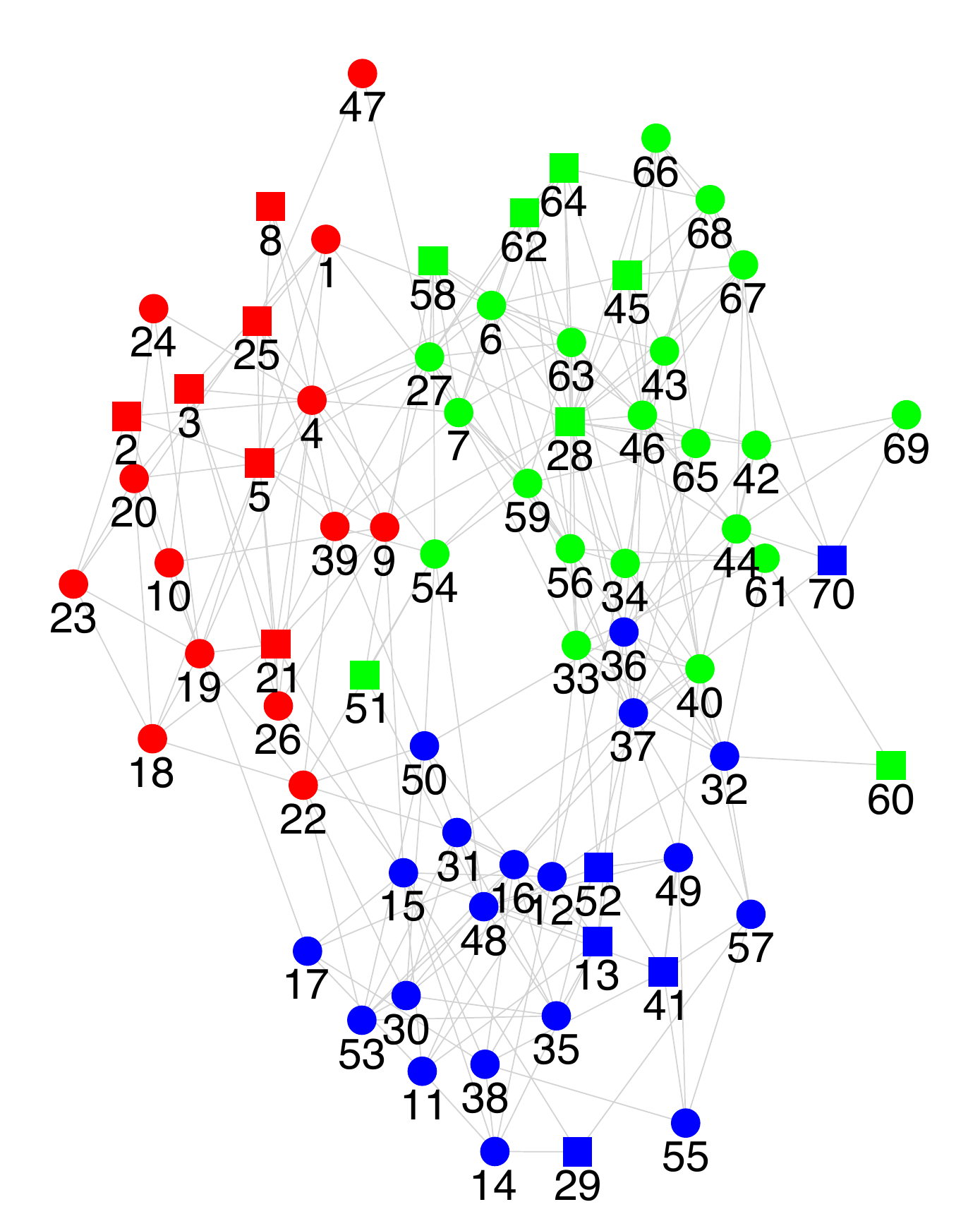}\\
{\sf \scriptsize (b) {\sf BlindCD}, RatioCut$=6.769$.} \hspace{-.25cm}&\hspace{-.25cm} {\sf \scriptsize (c) {Boosted} {\sf BlindCD}, RatioCut$=4.701$.}
\end{tabular}
\caption{\textbf{Pricing experiments on the \texttt{Highschool} network}. The network comprises of $N=70$ agents and (approximately) $K=3$ communities.
The utility parameters in \eqref{eq:util} are $a = R$ and $b = 2 \| {\bm A} {\bf 1} \|_\infty$. The network's equilibrium consumption
levels are collected for $L=10^3$ samples, each observed with a noise variance of 
$\sigma_w^2 = 10^{-2}/b^4$.
The pricing experiments are conducted through controlling the prices for $R=18$ agents.}\vspace{-.4cm}  \label{fig:price}
\end{figure}

\begin{figure}[t]
\centering \vspace{-.2cm}
\includegraphics[width=0.7 \linewidth]{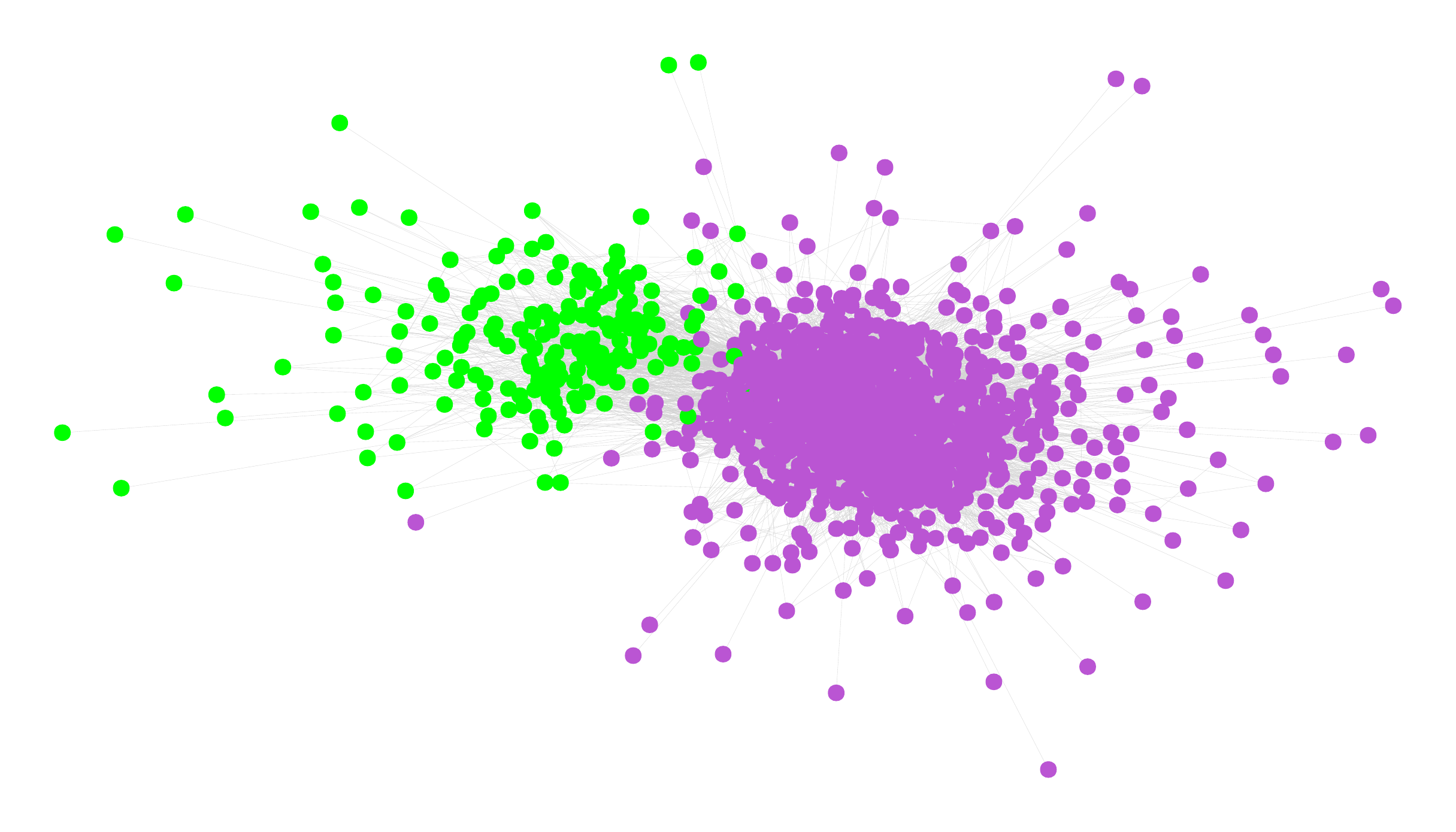} \vspace{-.8cm}
\begin{flushleft}{\sf \scriptsize (a) \texttt{ReedCollege} network. The clustering above is found by applying spectral clustering on the true ${\bm L}$. The obtained RatioCut is $0.1419$ w.r.t.~${\bm A}$.}\end{flushleft}\vspace{-.2cm}
\includegraphics[width=0.7 \linewidth]{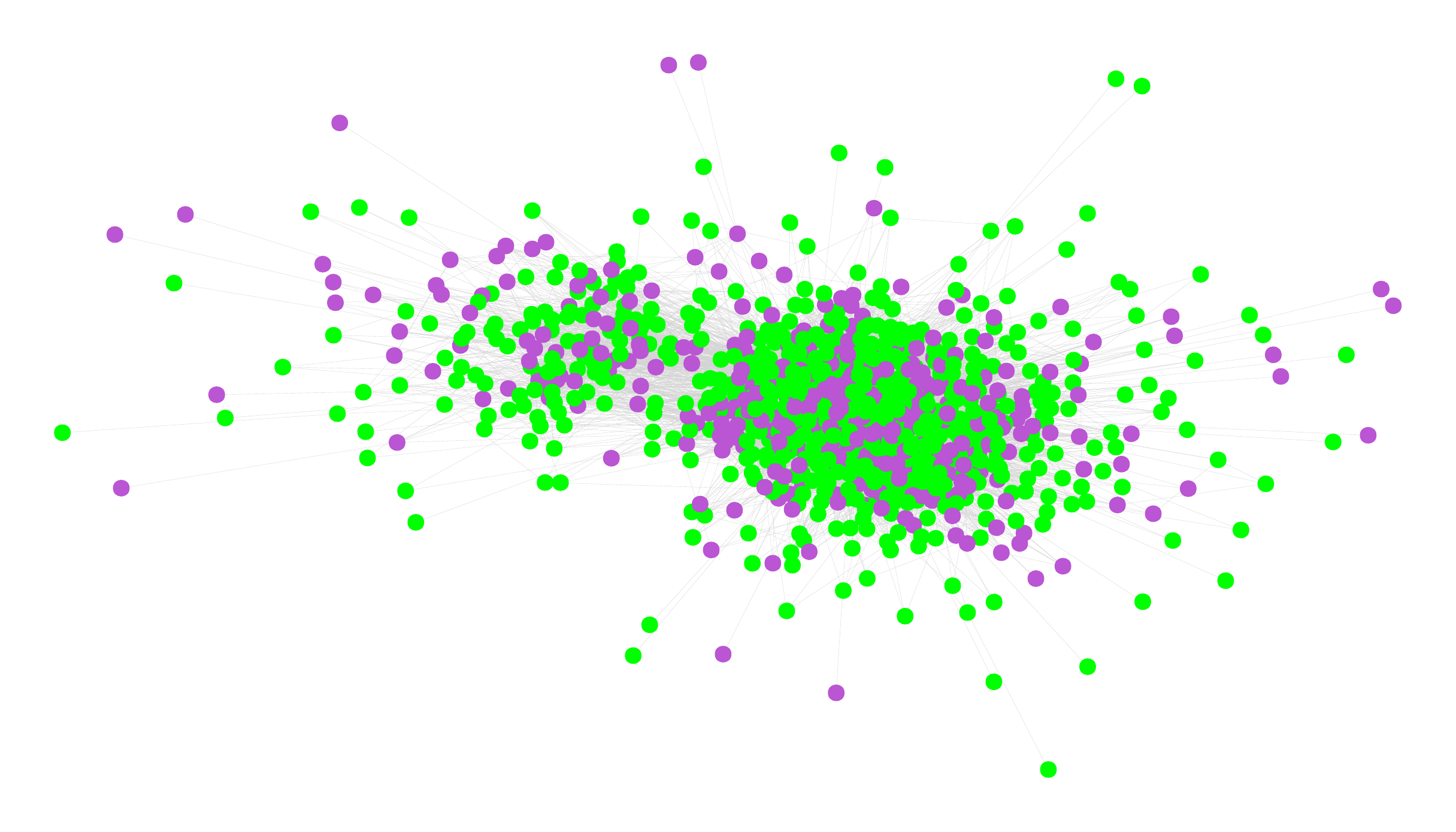} \vspace{-.8cm}
\begin{flushleft}{\sf \scriptsize (b) Applying BlindCD method. The obtained RatioCut is $0.8953$ {w.r.t.~${\bm A}$}.}\end{flushleft}\vspace{-.2cm}
\includegraphics[width=0.7 \linewidth]{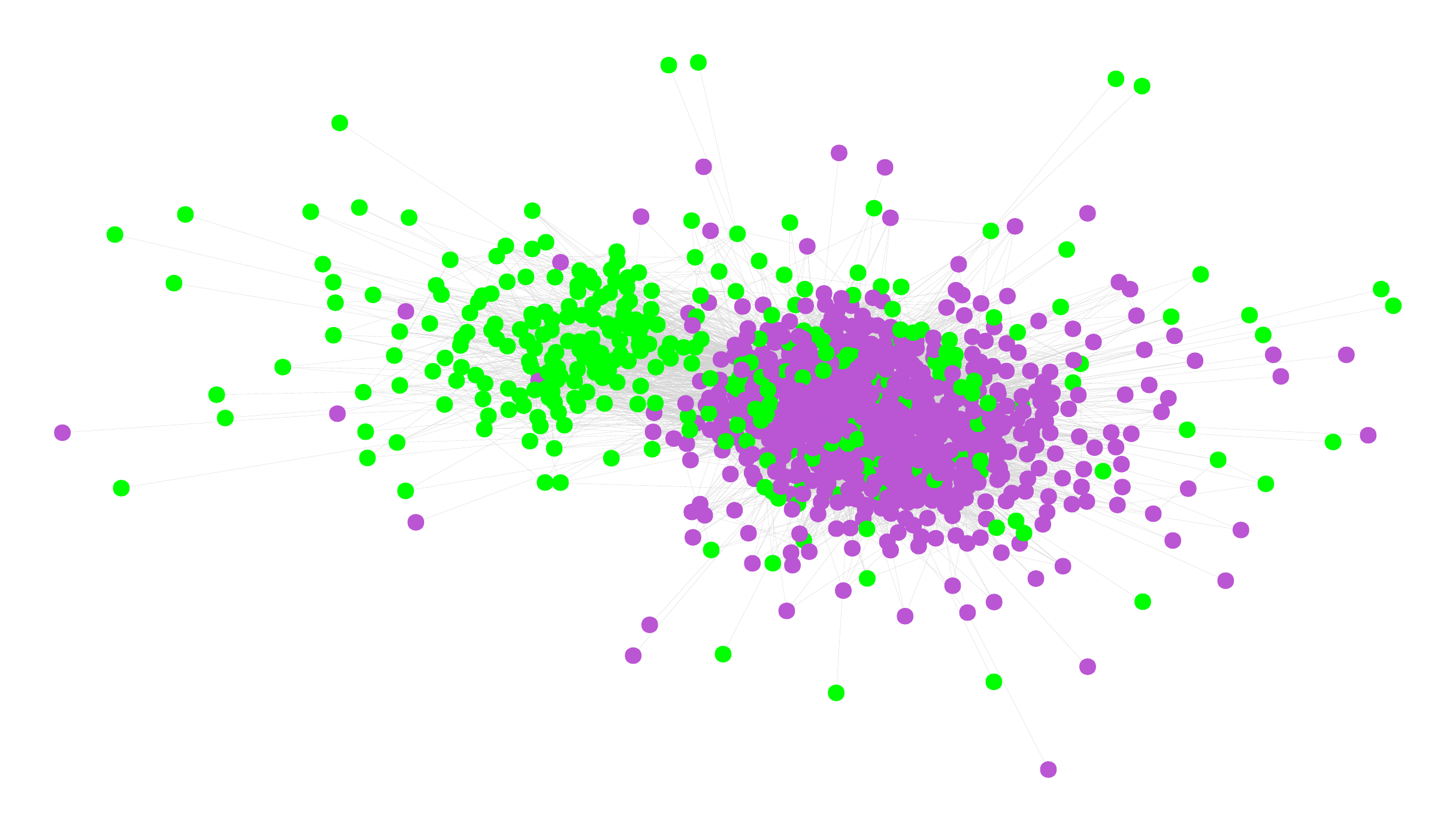} \vspace{-.8cm}
\begin{flushleft}{\sf \scriptsize (c) Applying Boosted BlindCD method. The obtained RatioCut is $0.5249$ {w.r.t.~${\bm A}$}.}\end{flushleft}\vspace{-.3cm}
\caption{\textbf{Opinion dynamics experiments on \texttt{ReedCollege} network.} The network comprises of $N=962$ agents and (approximately) $K=2$ communities.
The network's steady-state opinions are collected 
with $L=10^4$ samples, each observed with a noise variance of 
$\sigma_w^2 = 10^{-2}$.
There are $R=150$ stubborn agents in the experiments, which 
are connected to the social network
through a random bi-partite graph with connectivity $\log N / N$.}\vspace{-.2cm}\label{fig:stubborn}
\end{figure}

We then illustrate an application on real network topologies for the two network dynamics.
Fig.~\ref{fig:price} shows an example of simulated 
pricing experiments on the network
\texttt{highschool} from \cite{coleman1964introduction}, which is a friendship network 
between $N = 70$ high school
students with $|E| = 273$ undirected edges. 
On the other hand, Fig.~\ref{fig:stubborn} shows 
the case study for opinion dynamics 
on the Facebook network of \texttt{ReedCollege} \cite{traud2012social},
which is a friendship network with $N=962$ college students with $|E| = 18,812$ 
undirected edges, and we influence the network using $R=150$ stubborn agents.
To handle the high dimensionality, we 
  applied the fast algorithm from \cite{aravkin2014} to solve
the robust PCA problem in \eqref{eq:lowrank}.
In both cases, we observe that the boosted 
{\sf BlindCD} method recovers the communities in the networks, as evidenced
from the illustrations and the ratio-cut scores.\vspace{-.2cm}

{
\subsection{Application to US Senate Rollcall Records}
We consider applying the {\sf BlindCD} methods
to the US Senate rollcall record on \url{https://voteview.com} for the 110th congress. 
The dataset contains 657 rollcalls during the period from 2007 to 2009. To represent the opinions of the states during a rollcall, we consider the votes from the two Senators of a state by counting a `yay' as 1, while a `nay' or `absent' is counted as 0.  
By treating each state as a node on a graph with $50$ nodes, this results in $L=657$ samples of graph signals with values $\{0,1,2\}$. 
As argued in~\cite{wu2018estimating}, the rollcall data may be modeled as the equilibrium of an opinion dynamics process with stubborn agents. Therefore, we selected $4$ states -- Massachusetts (MA), New York (NY), Alabama (AL), Louisiana (LA), which are the most liberal/conservative states \cite{pew}, as the `stubborn' states modeled in Section~\ref{sec:degroot}. We then apply the {\sf BlindCD} methods to detect communities for the remaining $N=46$ states. For the boosted method, we use the sparse regularizer $g_1( \widehat{\bm B})$ to promote sparsity in the $\widehat{\bm B}^\star$ component of the solution. 

Fig.~\ref{fig:senate} shows the $K=2$ communities detected using the proposed methods. We observe that the boosted {\sf BlindCD} method  successfully identifies Maine to be in the same community as Texas, where both states were controlled by Republicans in this congress. Fig.~\ref{fig:bmat} shows the inferred $\widehat{\bm B}^\star$ matrix modeled in Section~\ref{sec:degroot}, where we labeled the rows as the stubborn states and the columns as the regular states. A large number in the table indicates strong influence from the stubborn to regular state. We observe consistent results, e.g., NY (\resp LA) positively influencing Illinois (\resp Arkansas) as both are Democrat (\resp Republican) states in this congress; NY is negatively influencing Idaho (Republican in this congress).}\vspace{-.2cm}

\begin{figure}[t]
\begin{center}
\resizebox{.49\linewidth}{!}{
\begin{tikzpicture}
\tikzset{set state val/.style args={#1/#2}{#1={fill=black!#2}}}
\tikzset{set state val/.list={MA/20,NY/20,AL/20,LA/20}}
\tikzset{set state val/.style args={#1}{#1={fill=red!60}}}
\tikzset{set state val/.list={NH,KS,SD,GA,MS,NC,SC,TX,KY,OK,TN,AZ,ID,UT,WY,AK}}
\USA[every state={draw=white, ultra thick, fill=blue!60}]
\end{tikzpicture}}
\resizebox{.49\linewidth}{!}{
\begin{tikzpicture}
\tikzset{set state val/.style args={#1/#2}{#1={fill=black!#2}}}
\tikzset{set state val/.list={MA/20,NY/20,AL/20,LA/20}}
\tikzset{set state val/.style args={#1}{#1={fill=red!60}}}
\tikzset{set state val/.list={      ME 
     ,NH 
     ,IN 
     ,KS 
     ,MO 
     ,NE 
     ,SD 
     ,FL 
     ,GA 
     ,MS 
     ,NC 
     ,SC 
     ,TX 
     ,KY 
     ,OK 
     ,TN 
     ,AZ 
     ,CO 
     ,ID 
     ,UT 
     ,WY 
     ,AK  }}
\USA[every state={draw=white, ultra thick, fill=blue!60}]
\end{tikzpicture}}
\end{center}
\vspace{-.2cm}
\caption{Applying {\sf BlindCD} methods on the 110th US Senate Rollcall records. The states marked in {\color{red}red}/{\color{blue}blue} are found to be in different communities; while the states marked in gray are marked as the `stubborn' states as explained in the text. (Left) Results of {\sf BlindCD}. (Right) Results of boosted {\sf BlindCD}.} \label{fig:senate}\vspace{-.4cm}
\end{figure}
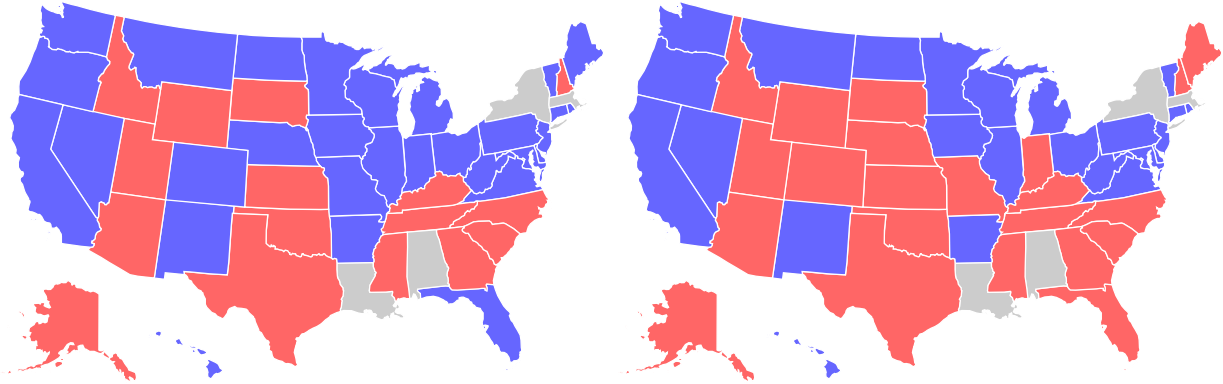

\begin{figure}[t]
\centering
\resizebox{1\linewidth}{!}
{\sf\begin{filecontents*}{./SenateData/bstar1.csv}
x, CT, ME, NH, RI, VT, DE, NJ, PA, IL, IN, MI, OH, WI, IA, KS, MN
MA,0,0,0,9,4,0,0,8,0,0,0,5,0,0,0,0
NY,11,3,0,5,4,0,14,0,29,9,6,0,0,0,0,0
AL,0,0,31,0,0,0,0,0,0,0,0,0,0,5,5,0
LA,0,0,0,-2,0,0,0,0,0,2,13,0,0,0,0,0
\end{filecontents*}
\pgfplotstabletypeset[color cells]{./SenateData/bstar1.csv}}\vspace{.2cm}
\resizebox{1\linewidth}{!}
{\sf\begin{filecontents*}{./SenateData/bstar4.csv}
x, TN, WV, AZ, CO, ID, MT, NV, NM, UT, WY, CA, OR, WA, AK, HI
MA,2,8,0,4,6,0,0,0,2,0,0,6,0,0,0
NY,0,0,0,0,-9,0,0,0,0,0,10,0,0,0,0
AL,19,0,7,0,19,0,18,0,28,34,0,0,0,9,0
LA,0,5,0,0,0,28,0,0,0,0,0,0,0,0,0
\end{filecontents*}
\pgfplotstabletypeset[color cells]{./SenateData/bstar4.csv}}\vspace{.2cm}
\resizebox{1\linewidth}{!}
{\sf\begin{filecontents*}{./SenateData/bstar5.csv}
x, OK, MO, NE, ND, SD, VA, AR, FL, GA, MS, NC, SC, TX, KY, MD
MA,-7,6,0,0,1,0,1,0,0,-2,0,0,0,0,0
NY,0,0,0,0,0,0,0,8,0,0,0,0,0,0,8
AL,15,0,0,0,0,0,0,0,18,16,11,26,4,12,0
LA,0,19,11,20,2,0,34,5,0,5,0,0,5,0,0
\end{filecontents*}
\pgfplotstabletypeset[color cells]{./SenateData/bstar5.csv}}\vspace{.1cm}
\caption{Illustrating the $\widehat{\bm B}^\star$ matrix found with boosted {\sf BlindCD} method [cf.~\eqref{eq:stubborn},\eqref{eq:lowrank}]. Note that the values within the table have been rescaled.}\label{fig:bmat}\vspace{-.4cm}
\end{figure}

\section{Conclusions}\vspace{-.1cm}
This paper proposes two \emph{blind} community detection   
methods for inferring community structure from graph signals.
We consider 
a challenging and realistic setting where the observed graph signals are outcomes
of a graph filter with low-rank excitations.
The {\sf BlindCD} methods rely on an intrinsic \emph{low-pass} property
of the graph filters that models the network dynamics.
This property holds for common network processes and the accuracy of {\sf BlindCD} is analyzed by viewing the graph signals as sketches of the graph filters.
We propose a boosting technique to improve the performance of {\sf BlindCD}.
The technique leverages the latent `low-rank plus sparse' structure
related to the graph signals. 
Extensive numerical experiments  verify our findings.\vspace{-.2cm}

\ifplainver
\appendix
\else
\appendices 
\fi

\section{Proof of Theorem~\ref{T:main}} \label{app:thm1}
To simplify the notations while proving the theorem, let us define 
the following indicator matrices for the communities.
Firstly, the matrix $\widehat{\bm X} \in \RR^{N \times K}$ is associated with 
the communities $\{ \hat{\cal C}_1,..., \hat{\cal C}_K \}$ found with 
{\sf BlindCD} and defined as
\beq \label{eq:indicatorX}
\hat{X}_{ij} \eqdef \begin{cases}
{1 / \sqrt{| \hat{\cal C}_j |}}, & \text{if}~i \in \hat{\cal C}_j, \\
0, & \text{otherwise}. 
\end{cases}
\eeq 
We have
\[
\| \widehat{\bm V}_K - \widehat{\bm X} \widehat{\bm X}^\top \widehat{\bm V}_K \|_{\rm F}^2 = 
\sum_{k=1}^K \sum_{i \in \hat{\cal C}_k} \Big\| \hat{\bm v}_i^{\rm row} - \frac{1}{|\hat{\cal C}_k|} \sum_{j \in \hat{\cal C}_k} \hat{\bm v}_j^{\rm row} \Big\|_2^2  .
\]
Define ${\cal X}$ as the set of all possible indicator matrices of partitions.
Using Condition~1 in Theorem~\ref{T:main}, we have that
\beq \label{eq:xhat} \begin{split}
\| \widehat{\bm V}_K - \widehat{\bm X} \widehat{\bm X}^\top \widehat{\bm V}_K \|_{\rm F}^2 & \leq (1 + \epsilon)
\min_{ {\bm X} \in {\cal X} } \| \widehat{\bm V}_K - {\bm X} {\bm X}^\top \widehat{\bm V}_K \|_{\rm F}^2 \\
& \hspace{-.8cm} \leq (1+\epsilon) \| \widehat{\bm V}_K - {\bm X}^\star ({\bm X}^\star)^\top \widehat{\bm V}_K \|_{\rm F}^2 \eqs,\\[-.3cm]
\end{split}
\eeq
where we have defined ${\bm X}^\star \in \RR^{N \times K}$ 
by replacing $\hat{\cal C}_i$ in \eqref{eq:indicatorX} with ${\cal C}_i^\star$
such that $ {\cal C}_1^\star, \ldots, {\cal C}_K^\star$
is an optimal set of communities found by minimizing $F( {\cal C}_1,..., {\cal C}_K )$
[cf.~\eqref{eq:obj}].
On the other hand, by the definition,
\beq \label{eq:xstar} \begin{split}
\| {\bm V}_K - {\bm X}^\star ({\bm X}^\star)^\top {\bm V}_K \|_{\rm F}^2 & =
\min_{ {\bm X} \in {\cal X} } \| {\bm V}_K - {\bm X} {\bm X}^\top {\bm V}_K \|_{\rm F}^2 \\
& \hspace{-.5cm} = \min_{ {\cal C}_1,..., {\cal C}_K } F( {\cal C}_1,..., {\cal C}_K ) = F^\star \eqs, \vspace{-.4cm}
\end{split}\vspace{.2cm}
\eeq
and furthermore 
$\| {\bm V}_K - \widehat{\bm X} \widehat{\bm X}^\top {\bm V}_K \|_{\rm F}^2 = F( \hat{\cal C}_1, ..., \hat{\cal C}_K )$.

Define the error matrix as
${\bm E}  \eqdef {\bm V}_K {\bm V}_K^\top - \widehat{\bm V}_K \widehat{\bm V}_K^\top$. 
We observe the following chain of inequalities:
\beq \begin{split}
& \| {\bm V}_K - \widehat{\bm X} \widehat{\bm X}^\top {\bm V}_K \|_{\rm F} 
= \| 
( {\bm I} -  \widehat{\bm X} \widehat{\bm X}^\top ) ( \widehat{\bm V}_K \widehat{\bm V}_K^\top + {\bm E} ) \|_{\rm F} \\
& \leq \| ( {\bm I} -  \widehat{\bm X} \widehat{\bm X}^\top ) \widehat{\bm V}_K \widehat{\bm V}_K^\top \|_{\rm F} + 
\| ( {\bm I} -  \widehat{\bm X} \widehat{\bm X}^\top ) {\bm E} \|_{\rm F} \\
& \leq \| ( {\bm I} -  \widehat{\bm X} \widehat{\bm X}^\top ) \widehat{\bm V}_K \widehat{\bm V}_K^\top \|_{\rm F} + 
\| {\bm E} \|_{\rm F} \eqs,\\[-.5cm]
\end{split}
\eeq
where 
the first equality is due to ${\bm V}_K^\top {\bm V}_K = {\bm I}$ and
the last inequality is due to $ {\bm I} - \widehat{\bm X} \widehat{\bm X}^\top $
is a projection matrix.
Using \eqref{eq:xhat}, we have that
\beq \label{eq:thm1end} \begin{split}
& 
\| ( {\bm I} -  \widehat{\bm X} \widehat{\bm X}^\top ) \widehat{\bm V}_K \widehat{\bm V}_K^\top \|_{\rm F} + 
\| {\bm E} \|_{\rm F} \\
& \leq 
\sqrt{1 + \epsilon} \| ( {\bm I} -  {\bm X}^\star ({\bm X}^\star)^\top ) ( {\bm V}_K {\bm V}_K^\top - {\bm E} ) \|_{\rm F} + 
\| {\bm E} \|_{\rm F} \\
& \leq \sqrt{1 + \epsilon} \| ( {\bm I} -  {\bm X}^\star ({\bm X}^\star)^\top ) {\bm V}_K {\bm V}_K^\top \|_{\rm F} \\
& \hspace{1cm} + \sqrt{1 + \epsilon} \| ( {\bm I} -  {\bm X}^\star ({\bm X}^\star)^\top ) {\bm E} \|_{\rm F} + \| {\bm E} \|_{\rm F} \\
& \leq \sqrt{ 1 + \epsilon } \| ( {\bm I} -  {\bm X}^\star ({\bm X}^\star)^\top ) {\bm V}_K {\bm V}_K^\top \|_{\rm F} + (2 + \epsilon) \| {\bm E} \|_{\rm F} \\
& =  \sqrt{(1 + \epsilon) F^\star} + (2 + \epsilon)  \| {\bm E} \|_{\rm F} \eqs,
\end{split}
\eeq
where we have used the fact   ${\bm I} -  {\bm X}^\star ({\bm X}^\star)^\top$
is a projection matrix and $\sqrt{1+\epsilon} \leq 1 + \epsilon$
in the third inequality. 
The final step is to bound $\| {\bm E} \|_{\rm F}$, where
we rely on the following results.
\begin{Lemma} \label{fact:spectral} \cite[Lemma 7]{boutsidis2015spectral_a}
For any ${\bm A}, {\bm B} \in \RR^{N \times K}$ with $N \geq K$ and ${\bm A}^\top {\bm A} = {\bm B}^\top {\bm B} = {\bm I}$, it holds that
\beq
\| {\bm A} {\bm A}^\top - {\bm B} {\bm B}^\top \|_{\rm F}^2 \leq 2K \| 
{\bm A} {\bm A}^\top - {\bm B} {\bm B}^\top \|_{2}^2 \eqs. \vspace{-.2cm}
\eeq 
\end{Lemma}
\begin{Prop} \label{prop:sketch}
Under Conditions 2 to 4 in Theorem~\ref{T:main}, we have  
\beq \label{eq:err_1}
\| \overline{\bm V}_K \overline{\bm V}_K^\top - {\bm V}_K {\bm V}_K^\top \|_2^2 
= (1+\gamma^2)^{-1} \gamma^2 \eqs,
\eeq
where the columns of $\overline{\bm V}_K$ are the top $K$ eigenvectors of $\overline{\bm C}_y$ and
$\gamma$ is bounded as stated in \eqref{E:gamma}. \vspace{-.2cm}
\end{Prop}
\begin{Prop} \label{prop:coverr}
Under Condition~5 in Theorem~\ref{T:main}, it holds that
\beq \label{eq:ineq}
\| \overline{\bm V}_K \overline{\bm V}_K^\top - \widehat{\bm V}_K \widehat{\bm V}_K^\top \|_2 \leq \| \widehat{\bm C}_y - \overline{\bm C}_y \|_2 / \delta \eqs.
\eeq
\end{Prop}
The proofs of the propositions can be found in the subsections A and B
of this appendix.
Applying Lemma~\ref{fact:spectral} we obtain that 
\beq
\| {\bm E} \|_{\rm F} \leq \sqrt{2K} \| {\bm V}_K {\bm V}_K^\top - \widehat{\bm V}_K \widehat{\bm V}_K^\top \|_2 \eqs.
\eeq
Combining \eqref{eq:err_1}, \eqref{eq:ineq} and using the triangle inequality
yields
\beq \notag \begin{split}
& \sqrt{F( \hat{\cal C}_1, ..., \hat{\cal C}_K )}  = \| ( {\bm I} -  \widehat{\bm X} \widehat{\bm X}^\top ) {\bm V}_K {\bm V}_K^\top \|_{\rm F} \\
& \leq \sqrt{(1 + \epsilon)  F^\star} + (2+\epsilon) \sqrt{2K} \Big(
\sqrt{\frac{ \gamma^2 }{ 1 + \gamma^2 }} + \frac{ \| \widehat{\bm C}_y - \overline{\bm C}_y \|_2 }{\delta } \Big),
\end{split}
\eeq
concluding the proof.

\subsection{Proof of Proposition~\ref{prop:sketch}}
We begin our proof by establishing the relationships 
between $\overline{\bm V}_K, {\bm V}_K$ and the left singular vectors of 
${\cal H} ({\bm L}) {\bm B}$. 
Denote 
the rank-$K$ approximation to ${\cal H}({\bm L})$ as $[ {\cal H} ({\bm L}) ]_K \eqdef {\bm V}_K {\rm diag} (\tilde{\bm h}_K ) {\bm V}_K^\top$. 
This expression is valid due to the low pass property of ${\cal H}({\bm L})$. 
Define $\tilde{\bm B} \eqdef {\bm B} {\bm Q}_K$, 
we observe 
that
\beq \label{eq:range}
{\cal R} ( [ {\cal H} ({\bm L} ) ]_K ) = {\cal R} ( [ {\cal H} ({\bm L} ) ]_K \tilde{\bm B} ) \eqs,
\eeq 
which  is due to 
Condition~3 in Theorem~\ref{T:main} 
such that the linear transformation $\tilde{\bm B}$ does not modify 
the range space of $[{\cal H}({\bm L})]_K$. 
Similarly,  
$[ \overline{\bm C}_y]_K \!\eqdef\! \overline{\bm V}_K {\rm diag} ( \bm{\sigma}_K )^2 \overline{\bm V}_K^\top$
is the rank $K$ approximation to $\overline{\bm C}_y$.
We  observe the equivalences
\beq \label{eq:range2}
{\cal R} ( [ \overline{\bm C}_y]_K ) = {\cal R} ( [ {\cal H} ( {\bm L} ) {\bm B} ]_K )
= {\cal R} ( {\cal H} ( {\bm L} ) \tilde{\bm B} )
\eeq
where the last equality is due to  
${\cal H}({\bm L}) \tilde{\bm B} = {\cal H}({\bm L}) {\bm B} {\bm Q}_K = 
\overline{\bm V}_K {\rm diag} ( \bm{\sigma}_K )$, 
as we recall that the columns of ${\bm Q}_K$ are the
top $K$ right singular vectors of ${\cal H}({\bm L}) {\bm B}$. 
Furthermore, 
\beq \label{eq:orthogonal}
{\cal R} ( [ {\cal H}( {\bm L} ) ]_K \tilde{\bm B} ) 
~ \bot ~ {\cal R} ( ( {\cal H} ( {\bm L} ) - [{\cal H}({\bm L})]_{K} ) \tilde{\bm B} ) \eqs.
\eeq

Let the columns of $\widetilde{\bm V}_K$ and $\widetilde{\overline{\bm V}}_K$ 
be respectively the top-$K$ singular vectors of $[ {\cal H}( {\bm L} )]_K \tilde{\bm B}$
and ${\cal H}({\bm L}) \tilde{\bm B}$, 
therefore \eqref{eq:range} and \eqref{eq:range2} imply that 
${\bm V}_K {\bm V}_K^\top = \widetilde{\bm V}_K \widetilde{\bm V}_K^\top$
and $\overline{\bm V}_K \overline{\bm V}_K^\top = 
\widetilde{\overline{\bm V}}_K \widetilde{\overline{\bm V}}_K^\top$. 
Invoking \eqref{eq:orthogonal} with \cite[Lemma 8]{boutsidis2015spectral}
through setting ${\bf D} = {\cal H} ( {\bm L} ) \tilde{\bm B}$,
${\bf C} =  [ {\cal H}( {\bm L} ) ]_K \tilde{\bm B}$
and ${\bf E} = ( {\cal H}({\bm L}) - [ {\cal H}( {\bm L} ) ]_{K} ) \tilde{\bm B}$ therein, 
and applying \cite[Theorem 2.6.1]{golub},
we obtain that
\beq \label{eq:desired_exp}
\begin{split}
& \| {\bm V}_K {\bm V}_K^\top - \overline{\bm V}_K \overline{\bm V}_K^\top \|_2^2 = \| \widetilde{\bm V}_K \widetilde{\bm V}_K^\top - \widetilde{\overline{\bm V}}_K \widetilde{\overline{\bm V}}_K^\top \|_2^2 \\
& = 1 - \beta_K \Big( [ {\cal H}({\bm L} ) ]_K \tilde{\bm B} \bm{\Pi}^\dagger ([ {\cal H}({\bm L} ) ]_K \tilde{\bm B} )^\top \Big) \eqs,
\end{split}
\eeq
where we have defined
$\bm{\Pi} \eqdef ( {\cal H}({\bm L} ) \tilde{\bm B})^\top {\cal H}({\bm L}) \tilde{\bm B}$
and $\beta_K(\cdot)$ denotes the $K$th largest eigenvalue. 
Under Condition 4 in Theorem~\ref{T:main}, 
the $K \times K$ 
matrix $\bm{\Pi}$ 
is non-singular. We   observe the following chain of equalities
\beq
\begin{split}
& \beta_K \Big( [ {\cal H}({\bm L} ) ]_K \tilde{\bm B} \bm{\Pi}^{-1}   ([ {\cal H}({\bm L} ) ]_K \tilde{\bm B} )^\top \Big) \\
& = \beta_K \Big( {\rm diag} ( \tilde{\bm h}_K ) {\bm V}_K^\top \tilde{\bm B} 
\bm{\Pi}^{-1}
( {\rm diag} ( \tilde{\bm h}_K ) {\bm V}_K^\top \tilde{\bm B} )^\top \Big) \\
& = \frac{1}{ \beta_1 \Big( 
( {\rm diag} ( \tilde{\bm h}_K ) {\bm V}_K^\top \tilde{\bm B} )^{-\top}
\bm{\Pi}
( {\rm diag} ( \tilde{\bm h}_K ) {\bm V}_K^\top \tilde{\bm B} )^{-1} \Big) } \eqs,
\end{split}
\eeq
where the first equality is due to 
$\beta_K ( {\bm U} {\bm A} {\bm U}^\top ) = \beta_K ( {\bm A} )$
for any symmetric ${\bm A}$ and ${\bm U} \in \RR^{N \times K}$ with orthogonal columns, 
and the second equality follows since the argument in $\beta_K(\cdot)$ is of rank $K$.
Moreover, $\bm{\Pi}$ admits the decomposition
\beq
\begin{split}
\bm{\Pi} & = ( {\cal H}({\bm L}) \tilde{\bm B} )^\top {\cal H}({\bm L}) \tilde{\bm B} 
 = \tilde{\bm B}^\top {\cal H}({\bm L})^\top {\cal H}({\bm L}) \tilde{\bm B}
\\
& = \tilde{\bm B}^\top {\bm V}_{K} {\rm diag} ( \tilde{\bm h}_{K} )^2 {\bm V}_{K}^\top \tilde{\bm B} \\
& \hspace{0.5cm} + \tilde{\bm B}^\top {\bm V}_{N-K} {\rm diag} ( \tilde{\bm h}_{N-K} )^2 {\bm V}_{N-K}^\top \tilde{\bm B} \eqs.
\end{split}
\eeq
Thus, yielding that
\beq \notag
\begin{split}
& \beta_K \Big( [ {\cal H}({\bm L} ) ]_K \tilde{\bm B} \bm{\Pi}^{-1}   ([ {\cal H}({\bm L} ) ]_K \tilde{\bm B} )^\top \Big)  \\
& = \Big( 1 +  \beta_1 \Big( 
( {\rm diag} ( \tilde{\bm h}_K ) {\bm V}_K^\top \tilde{\bm B} )^{-\top} 
\tilde{\bm B}^\top {\bm V}_{N-K} \\
& \hspace{1cm} {\rm diag} ( \tilde{\bm h}_{N-K} )^2 {\bm V}_{N-K}^\top 
\tilde{\bm B} 
( {\rm diag} ( \tilde{\bm h}_K ) {\bm V}_K^\top \tilde{\bm B} )^{-1} \Big) \Big)^{-1} \\
& = \frac{1}{1 +  \| {\rm diag} ( \tilde{\bm h}_{N-K} ) {\bm V}_{N-K}^\top 
\tilde{\bm B}
( {\rm diag} ( \tilde{\bm h}_{K} ) {\bm V}_K^\top \tilde{\bm B} )^{-1} \|_2^2}\\
& = \Big( 1 + \gamma^2 \Big)^{-1} \eqs,
\end{split}
\eeq
where we have defined $\gamma$ such that
\beq \begin{split}
\gamma & \eqdef \| {\rm diag} ( \tilde{\bm h}_{N-K} ) {\bm V}_{N-K}^\top 
\tilde{\bm B}
( {\rm diag} ( \tilde{\bm h}_{K} ) {\bm V}_K^\top \tilde{\bm B} )^{-1} \|_2 \\
& \leq \eta \!~ \| {\bm V}_{N-K}^\top {\bm B} {\bm Q}_K \|_2 \| ({\bm V}_{K}^\top {\bm B} {\bm Q}_K)^{-1} \|_2 \eqs.
\end{split}
\eeq
Substituting the above into \eqref{eq:desired_exp} 
concludes the proof. 

\subsection{Proof of Proposition~\ref{prop:coverr}}
Denote the SVD of the sampled covariance as 
$\widehat{\bm C}_y = \widehat{\bm V} \widehat{\bm{\Sigma}} \widehat{\bm V}^\top$. 
The left hand side of \eqref{eq:ineq}
can be written as
\beq
\| \overline{\bm V}_K \overline{\bm V}_K^\top - \widehat{\bm V}_K \widehat{\bm V}_K^\top \|_2
= \| \widehat{\bm V}_{N-K}^\top \overline{\bm V}_K \|_2 \eqs,
\eeq
where the equality is due to \cite[Theorem 2.6.1]{golub}. 

Define   $\bm{\Delta} \eqdef \widehat{\bm C}_y - \overline{\bm C}_y$. 
Condition~5 in Theorem~\ref{T:main} implies that the largest eigenvalue
in $\hat{\bm{\Sigma}}_{N-K}$ will never exceed $\beta_K (\overline{\bm C}_y ) - \delta$
since
\beq \begin{split}
\beta_{\sf max} ( \hat{\bm{\Sigma}}_{N-K} ) = \beta_{K+1} ( \widehat{\bm C}_y ) &
\leq \beta_{K+1} ( \overline{\bm C}_y ) + \beta_1 ( \bm{\Delta} ) \\
& \leq \beta_{K+1} ( \overline{\bm C}_y ) + \| \bm{\Delta} \|_2 \eqs,
\end{split}
\eeq
where the first inequality is due to Weyl's inequality \cite{golub}. 
The perturbed matrix $\widehat{\bm C}_y$ thus satisfies the requirement of  
the Davis-Kahan's $\sin (\Theta)$ theorem \cite{dk70}
\beq
\| \widehat{\bm V}_{N-K}^\top \overline{\bm V}_K \|_2 \leq \delta^{-1} \| \widehat{\bm V}_{N-K}^\top 
\bm{\Delta} \overline{\bm V}_K \|_2 \eqs.
\eeq 
The inequality in \eqref{eq:ineq} is obtained by observing that both $\overline{\bm V}_K$ 
and $\widehat{\bm V}_{N-K}$ are orthogonal matrices.

\section{Proof of Lemma~\ref{prop:err1}} \label{app:lse}
Fix $1 \geq c > 0$. 
Under the conditions stated in the lemma, 
the least-squares optimization \eqref{eq:lse} admits a closed form solution
\beq\label{E:app_B_rhs}
\bm{\mathcal{H}}^\star - {\cal H}({\bm L}) {\bm B} =  \Big( \sum_{\ell=1}^L {\bm w}^\ell ({\bm z}^\ell)^\top \Big) \Big(
\sum_{\ell=1}^L {\bm z}^\ell ({\bm z}^\ell)^\top \Big)^{-1},
\eeq
where ${\bm w}^\ell$ was introduced in~\eqref{eq:data_collect}. 
Denoting the right hand side in \eqref{E:app_B_rhs} by $\bm{\mathcal{E}}$, we have that
\beq
\begin{split}
\| \bm{\mathcal{E}}\|_2 & = \left\| \Big( \frac{1}{L} \sum_{\ell=1}^L {\bm w}^\ell ({\bm z}^\ell)^\top \Big) \Big( \frac{1}{L}
\sum_{\ell=1}^L {\bm z}^\ell ({\bm z}^\ell)^\top \Big)^{-1} \right\|_2 \\
& \leq  \left\| \frac{1}{L} \sum_{\ell=1}^L {\bm w}^\ell ({\bm z}^\ell)^\top \right\|_2 
\left\| \Big( \frac{1}{L}
\sum_{\ell=1}^L {\bm z}^\ell ({\bm z}^\ell)^\top \Big)^{-1} \right\|_2.
\end{split}
\eeq
Observe that $\frac{1}{L}
\sum_{\ell=1}^L {\bm z}^\ell ({\bm z}^\ell)^\top$ converges to ${\bm I}$
such that with probability at least $1-c$, 
\beq
\left\| \frac{1}{L}
\sum_{\ell=1}^L {\bm z}^\ell ({\bm z}^\ell)^\top - {\bm I} \right\|_2 \leq C_0 \!~ \sqrt{ \frac{R \log(1/c)} {L} } \eqs,
\eeq
for some constant $C_0$.
Applying \cite[Proposition 2.1]{vershynin2012close} we get that 
\beq \notag \begin{split}
 \left\| \Big( \frac{1}{L}
\sum_{\ell=1}^L {\bm z}^\ell ({\bm z}^\ell)^\top \Big)^{-1} \right\|_2 & \leq \left( 1 - \Big\| \frac{1}{L}
\sum_{\ell=1}^L {\bm z}^\ell ({\bm z}^\ell)^\top - {\bm I} \Big\|_2 \right)^{-1} \\
& \leq \left( 1 - C_0 \!~ \sqrt{ \frac{R \log(1/c)} {L} } \right)^{-1} .
\end{split}
\eeq
On the other hand, observe that $\EE[ {\bm w}^\ell ({\bm z}^\ell)^\top ] = 0$ 
and $\| {\bm w}^\ell ({\bm z}^\ell)^\top \| \leq C_w$ almost surely. 
Applying
the matrix Bernstein's inequality \cite[Theorem 1.6]{tropp2012user} 
shows that with probability at least $1-c$ and for sufficiently large $L$,
\beq
\left\| \frac{1}{L} \sum_{\ell=1}^L {\bm w}^\ell ({\bm z}^\ell)^\top \right\|_2
\leq C_1 \!~ \sqrt{ \frac{ \sigma_w^2 \log( (N+R) / c ) }{L} } \eqs,
\eeq
for some constant $C_1$. Finally, with probability at least $1-2c$, 
\beq
\| \bm{\mathcal{E}} \|_2 \leq \frac{ C_1 \sqrt{ \sigma_w^2 \log( (N+R)/c ) } }{ \sqrt{L} - C_0 \sqrt{ R \log(1/c) } } = {\cal O}(\sigma_w/\sqrt{L}) \eqs. \vspace{-.1cm}
\eeq


\section{Proof of Corollary~\ref{cor:bbcd}} \label{app:cor}
Let $\widetilde{\bm V}_K$ and $\widetilde{\bm S}_K$ be 
the top $K$ left singular vectors of $\widetilde{\cal H}({\bm L}) {\bm B}$
and $\widehat{\bm{\mathcal{S}}}^\star$,
respectively. 
We can repeat the proof for Theorem~\ref{T:main} up to \eqref{eq:thm1end} 
by re-defining the error matrix ${\bm E}$ therein as 
$\widetilde{\bm E} = {\bm V}_K {\bm V}_K^\top - \widetilde{\bm S}_K \widetilde{\bm S}_K^\top$. 
This entails
\beq
\sqrt{F ( \tilde{\cal C}_1, ..., \tilde{\cal C}_K )} - \sqrt{(1+\epsilon) F^\star} \leq (2+\epsilon) \| \widetilde{\bm E} \|_{\rm F} \eqs.
\eeq
Next, we bound $\| \widetilde{\bm E} \|_{\rm F}$. Applying Lemma~\ref{fact:spectral}
and using the triangle inequality we get that
\beq \notag \begin{split}
\| \widetilde{\bm E} \|_{\rm F} & \leq \sqrt{2K} \| {\bm V}_K {\bm V}_K^\top - \widetilde{\bm S}_K \widetilde{\bm S}_K^\top \|_2 \\
& \hspace{-0.75cm} \leq \sqrt{2K} \big(  \| {\bm V}_K {\bm V}_K^\top - \widetilde{\bm V}_K \widetilde{\bm V}_K^\top \|_2 + \| \widetilde{\bm V}_K \widetilde{\bm V}_K^\top - \widetilde{\bm S}_K \widetilde{\bm S}_K^\top \|_2 \big),
\end{split}
\eeq 
 Proposition~\ref{prop:sketch} implies that
\beq
\| {\bm V}_K {\bm V}_K^\top - \widetilde{\bm V}_K \widetilde{\bm V}_K^\top \|_2 \leq
\sqrt{ \tilde{\gamma} / (1 + \tilde{\gamma} ) },
\eeq
{where $\tilde{\gamma}$ is bounded as in \eqref{eq:bdgamma}.}
Our remaining task is to bound 
$\| \widetilde{\bm V}_K \widetilde{\bm V}_K^\top - \widetilde{\bm S}_K \widetilde{\bm S}_K^\top \|_2$. 
Observe that
\beq
\| \widetilde{\bm V}_K  \widetilde{\bm V}_K^\top - \widetilde{\bm S}_K \widetilde{\bm S}_K^\top \|_2 = \| \widetilde{\bm S}_{R-K}^\top \widetilde{\bm V}_K \|_2 
\eeq
and 
\beq
\sigma_K( \widehat{\bm{\mathcal{S}}}^\star ) \geq \sigma_K ( \widetilde{\cal H}({\bm L}) {\bm B} ) - \| \widetilde{\bm{\Delta}} \|_2 \eqs,
\eeq
where we recalled the definition
$\widetilde{\bm{\Delta}} = \widehat{\bm{\mathcal{S}}}^\star - \widetilde{\cal H}({\bm L}) {\bm B}$
and applied the Weyl's inequality \cite{golub}. 
From \eqref{eq:deltatil}, we have that
\beq
\sigma_K ( \widetilde{\cal H}({\bm L}) {\bm B} ) - \| \widetilde{\bm{\Delta}} \|_2 = 
\sigma_{K+1} ( \widetilde{\cal H}({\bm L}) {\bm B} ) + \tilde{\delta} \eqs,
\eeq
with $\tilde{\delta} > 0$.
Finally, applying the Wedin theorem \cite{wedin1972perturbation} 
yields
\beq
\| \widetilde{\bm S}_{R-K}^\top \widetilde{\bm V}_K \|_2 \leq (\tilde{\delta} )^{-1} \!~ \| \widetilde{\bm{\Delta}} \|_2  \eqs.
\eeq

\bibliographystyle{IEEEtran}
{
\bibliography{sketch_GSP}
}


\end{document}